\documentclass[12pt]{article}
\usepackage{lscape}
\usepackage{geometry}
\usepackage[onehalfspacing]{setspace}
\usepackage{amssymb}
\usepackage{amsfonts}
\usepackage{graphicx}
\usepackage{amsmath}
\usepackage{bbm}
\usepackage{epstopdf}
\usepackage{natbib}
\usepackage{bibunits}
\usepackage{morefloats}
\usepackage{float}
\usepackage{comment}
\usepackage{pdflscape}
\usepackage{adjustbox}
\usepackage{subcaption}
\usepackage{longtable}
\usepackage{apalike}
\usepackage{xfrac}
\usepackage[dvipsnames,table]{xcolor}
\usepackage{etoolbox,siunitx}
\robustify\bfseries

\usepackage{hhline,multirow}
\usepackage{yfonts}

\usepackage{longtable}
\usepackage{threeparttablex}

\newcolumntype{d}[1]{D{.}{.}{#1}}

\usepackage{qtree}
\usepackage{multicol}
\usepackage{booktabs}
\usepackage{threeparttable}
\usepackage{tabularx}
\usepackage{dcolumn}
\newcolumntype{d}[1]{D..{#1}} 

\usepackage{siunitx}
\usepackage{mathpazo} 
\usepackage[scaled]{helvet} 
\usepackage{courier} 
\normalfont
\usepackage[T1]{fontenc}
\usepackage{listings}
\usepackage{epigraph}
\def\sym#1{\ifmmode^{#1}\else\(^{#1}\)\fi}

\usepackage[pdftex,bookmarks,colorlinks]{hyperref}
\hypersetup{
  colorlinks,
  citecolor=darkpowderblue,
  linkcolor=red,
  urlcolor=blue}
\usepackage{cleveref}

\DeclareMathOperator*{\argmin}{arg\,min}

\usepackage{etoolbox}

\definecolor{dukeblue}{rgb}{0.0, 0.0, 0.61}
\definecolor{darkred}{rgb}{0.8,0,0}

\makeatletter
\patchcmd{\epigraph}{\@epitext{#1}}{\itshape\@epitext{#1}}{}{}
\makeatother

\makeatletter
\def\munderbar#1{\underline{\sbox\tw@{$#1$}\dp\tw@\z@\box\tw@}}
\makeatother

\usepackage{graphicx}
\usepackage{wrapfig}

\usepackage{algorithm,algorithmic}

\usepackage{xspace}

\definecolor{bred}{RGB}{122, 0, 0}
\definecolor{darkpowderblue}{rgb}{0.0, 0.05, 0.5}

\definecolor{dpd}{rgb}{0.0, 0.05, 0.5}
{}

\newcommand{\acc}{Albacore$_\text{\scriptsize comps}$\xspace}
\newcommand{\acr}{Albacore$_\text{\scriptsize ranks}$\xspace}

\newcommand{\qacc}{Qualbacore$_\text{\scriptsize comps}$\xspace}
\newcommand{\qacr}{Qualbacore$_\text{\scriptsize ranks}$\xspace}
\newcommand{\gac}{Albacore$^\text{G}$\xspace}
\newcommand{\gacc}{Albacore$^\text{G}_\text{\scriptsize comps}$\xspace}
\newcommand{\gacr}{Albacore$^\text{G}_\text{\scriptsize ranks}$\xspace}

\makeatletter
\newcommand\halftiny{\@setfontsize\halftiny\@vipt\@viipt}
\newcommand\notsotiny{\@setfontsize\notsotiny{6.99}{9.2828}}
\makeatother

\usepackage{soul}

\renewenvironment{abstract}
 {\small
  \begin{center}
  \bfseries \abstractname\vspace{-.5em}\vspace{0pt}
  \end{center}
  \list{}{
        \setlength{\leftmargin}{1.8cm}    \setlength{\rightmargin}{\leftmargin}  }  \item\relax}
 {\endlist}
\begin{document}
\sloppy
\title{\vspace*{-1.4cm} \LARGE {\textbf{\color{bred} Maximally Forward-Looking Core Inflation}} \vspace{0.3cm}}
\small
\small 
\author{ \hspace{-1.15em} Philippe Goulet
Coulombe\thanks{%
Contact:  \href{mailto:p.gouletcoulombe@gmail.com}{\texttt{goulet\_coulombe.philippe@uqam.ca}}.  For helpful discussions,  we thank  Elena Bobeica,  Max Esser,  Fabio Rumler,  Mirjam Salish,  Dalibor Stevanovic,  Boyuan Zhang,  as well as participants at the IIF MacroFor seminar, the OeNB Freitagsseminar, internal seminars at the ECB,  CDPQ,  and UQAM.  For research assistance,  we thank Nicolas Harvie. The views expressed in this paper do not necessarily reflect those of the Oesterreichische Nationalbank or the Eurosystem. This draft: \today. The R package \texttt{assemblage} is available \href{https://github.com/christophebarrette/assemblage}{here}. }  \\
\textbf{\color{dpd} \texttt{\fontfamily{phv}\selectfont \footnotesize\normalsize \quad \quad \quad Universit\'{e} du Qu\'{e}bec \`{a} Montr\'{e}al} \quad \quad \quad}  \newline \smallskip 
\and Karin Klieber \\
 \textbf{\color{dpd} \texttt{\fontfamily{phv}\selectfont \footnotesize \normalsize \quad \quad \quad Oesterreichische Nationalbank} \quad \quad \quad}  \newline   \smallskip
\and Christophe Barrette \\
\textbf{\color{dpd} \texttt{\fontfamily{phv}\selectfont \footnotesize \normalsize \quad \quad \quad \enskip \phantom{..}Universit\'{e} du Qu\'{e}bec \`{a} Montr\'{e}al} \enskip \quad \quad \quad}
\newline   \smallskip
\and Maximilian G\"obel \\
\textbf{\color{dpd} \texttt{\fontfamily{phv}\selectfont \footnotesize \normalsize \quad \quad \quad \quad \enskip \quad Bocconi University} \quad \quad \quad \quad}  
}

\date{}

\setstretch{0.98}

\maketitle
\vspace{-1.3cm}
\textbf{
\large
\center 
 \vspace{2em}
 }
 \setstretch{1.2}

\begin{abstract}

\noindent Timely monetary policy decision-making requires timely core inflation measures.  We create a new core inflation series that is explicitly designed to succeed at that goal.  Precisely, we introduce the Assemblage Regression,  a generalized nonnegative ridge regression problem that optimizes the price index's subcomponent weights such that the aggregate is maximally predictive of future headline inflation.  Ordering subcomponents according to their rank in each period switches the algorithm to be learning supervised trimmed inflation---or,  put differently,  the maximally forward-looking summary statistic of the realized price changes distribution.  In an extensive out-of-sample forecasting experiment for the US and the euro area, we find substantial improvements for signaling medium-term inflation developments in both the pre- and post-Covid years.  Those coming from the supervised trimmed version are particularly striking, and are attributable to a highly asymmetric trimming which contrasts with conventional indicators.  We also find that this metric was indicating first upward pressures on inflation as early as mid-2020 and quickly captured the turning point in 2022.  We also consider extensions,  like assembling inflation from geographical regions, trimmed temporal aggregation,  and building core measures specialized for either upside or downside inflation risks. 


\end{abstract}

\thispagestyle{empty}



\clearpage


\clearpage 
\setcounter{page}{1}

\newgeometry{left=2 cm, right= 2 cm, top=2.3 cm, bottom=2.3 cm}

\section{Introduction}

Ideally,  monetary policy should be forward-looking.  However,  it can only be as forward-looking as the warning lights on the dashboard.  Given the extended delay between policy impulse and the economy's response,  the benefits of more timely inflation gauges cannot be understated.  Core inflation measures,  which distill noise and amplify signal,  are crucial inputs guiding monetary policy making,  investment strategies,  and other consequential economic decisions.  But what is noise,  what is signal,  and what kind of signal are we interested in? 

Core inflation measures abound, and all answer the above in their own way.   The most well-known are built from heuristics.  Some permanently exclude components (e.g., food and energy) that are known to be subject to large (mostly) transitory shocks that can obscure from the  deeper underlying trend \citep{gordon1975alternative}.    Others identify the components experiencing  the most extreme (positive or negative) growth in each month,  and exclude those before aggregating the remainder \citep{bryan1991median,BryanCecchetti1993}.   Then,  there are more formal alternatives based on factor models  where the key idea is that core inflation is a latent variable related to observable data through some assumed statistical structure,  and can be extracted as such \citep{StockWatson2016,BanburaBobeica2020}.  In a subsequent step,  core inflation series' empirical merits are assessed according to a variety of criteria.  One of the most desirable properties to support timely monetary policy decision-making is that the core series should be as indicative as possible of future inflation conditions \citep{clark2001core,Cogley2002}.   Thus,  from this perspective,  all the above qualify as unsupervised learning.  Model design and evaluating forward-looking qualities  are mostly independent affairs.   Significant volatility reduction with respect to headline inflation can come at the cost of the resulting product being a lagging indicator.  This paper proposes to focus on the forward-looking criterion,  and writes a simple supervised learning algorithm that delivers a core series that satisfies it.

\vskip 0.15cm
{\sc \noindent \textbf{Assemblage Regression}.} We abide by the "train for what you aim" principle.  {Our first proposition assembles components so that the resulting aggregate is maximally predictive of the hard indicator that the central bank intends to bring (or keep) at 2\%. }
This is achieved through a generalized nonnegative ridge regression where the dependent variable is future headline inflation and the regressors  inflation subcomponents.  At aggregated levels,  where components are scarce and (additional) regularization is expendable,  this is constrained nonnegative least squares. \textcolor{black}{However,  regularization is necessary when considering highly disaggregated levels,  an environment where our penalized regression is most free to discover non-trivial predictability patterns.}  Our second offering extracts the \textit{supervised} summary statistics of the realized distribution of price growth rates that best achieve the same objective.   This is done by running the assemblage regression and replacing components data by the order statistics time series of the price growth rates distribution.   \textcolor{black}{Coefficients then reflect which part of the distribution gets trimmed out, which is retained, and, in the second case,  what are  the respective weights.} We name the newly constructed indicator Albacore,  for \textbf{a}daptive \textbf{l}earning-\textbf{ba}sed \textbf{core} inflation,  which summarizes both the philosophy (the construction should adapt to the objective) and the methodology.  As we will see,  both \acc and \acr (respectively,  weighting in components and rank space) will reaffirm parts of the common wisdom and bring new stylized facts to the table.

\vskip 0.15cm
{\sc \noindent \textbf{Related Work}.} Econometrically,  assemblage regression's closest relatives come from the regression-based forecast combination literature \citep{wang2023forecast} and autoregression-based filtering \citep{Hamilton2018}.  In the case of \acr,  it is coupled with elements of functional regression \citep{morris2015functional}.  

First, the problem yielding \textbf{\acc} shares various similarities with regularized forecast combination schemes such as that of  \cite{diebold2019machine}.  Namely,  we are looking to build an optimal weighted average  incorporating regularization inspired from what the default averaging should be in the specific application.   A key distinction between assemblage regression and combination schemes following from \cite{granger1984improved} is that,  as the name suggests,  we are putting parts of something together.  Forecast combination exercises, on the other hand, construct an optimal weighted average using noisy estimates \textit{of the same thing}.\footnote{An example, where forecasts are not the base material yet, the combination is built upon estimates of the same concept, is \cite{aruoba2012improving}, who combine GDP by revenues and GDP by expenses (two noisy estimates of GDP) to construct a more reliable GDP indicator.} \textcolor{black}{Following from} this distinction between supervised ensembling and assembling,   there are few design choices mirroring the different environments,   like the construction of the target and the whole regularization apparatus.  Nonetheless,  the connection with forecast combination regressions is useful to locate where \acc stands in the vast core inflation measures landscape.  Among other things,  \cite{granger1984improved}'s regression approach to forecast combinations constructs an optimal portfolio of forecasts taking into account their covariance structure,  which simpler "plug-in" approaches using  \cite{bates1969combination}'s formula do not.

In the unrealistic yet informative case where inflation components would be uncorrelated, reflecting on the least squares formula for weights offers some intuition.  \acc would downweight  components which do not covary significantly with headline at a pre-specified forecasting horizon, have high variance,  or both.   Clearly,  this matches the attributes of a component that should be excluded or downweighted in a core inflation building context: it is not persistent and highly volatile.   Moreover, if a component exhibits negative autocorrelation patterns that, on average,  have been self-absorbed within a time span corresponding to the forecast horizon,   then those transitory shocks will also be left out since they have hardly any predictive power \textit{at that horizon}.  Things inevitably get less trivial when components are heavily cross-correlated,  constraints are brought in,  and regularization is biting.  Still,  at a conceptual level,  focusing on maximal predictability is not a radical conceptual departure from approaches improving \cite{gordon1975alternative}'s popular suggestion by either excluding or reweighting items according to their overall volatility \citep{dow1993measuring,clark2001core,Acosta2018},  their cyclical volatility \citep{dolmas2009excluding}, their persistence \citep{BilkeStracca2007}, or even their sensitivity to the economic business cycle \citep{shapiro2017cyclical,ehrmann2018supercore,stock2020slack}.  In a way,  \acc encompasses most of these objectives,  but does so in a framework that keeps an eye on the prize (forecasting headline) and  accounting for the fact that we are aggregating correlated objects, which comes with interesting hedging possibilities if properly accounted for.   There have been rare instances in the literature where predictability for headline played a more direct role in informing the weighting in the components space.  However,  the proposed aggregation schemes overlooked the covariance structure \citep{RavazzoloVahey2009} or lacked,  among other things,   necessary nonnegativity restrictions \citep{gamber2019constructing}.

Core inflation estimation can be framed as a signal extraction problem and accordingly,  many variants of factor analysis have been suggested  \citep{cristadoro2005core,morana2007structural,khan2013common,StockWatson2016,BanburaBobeica2020}.  Compared to those methods,  especially when relying on Principal Component Analysis (PCA),   assemblage regression retains necessary shrinkage,  but makes sure, through supervision,  that the extracted signal is one we should care about.  Obviously,   by discarding some noise,  one expects the ensuing factors to incorporate relevant signals to forecast headline inflation.  However, it is not optimizing for it.  Factors are extracted as the linear combination of components that best explain the variation in all components.  The resulting product's correspondence with the "true" concept of core inflation inevitably comes from consistency under unverifiable assumptions embedded in the specification of the model.   \textcolor{black}{Therefore, the empirical evaluation is usually conducted ex-post through the backchannels of its implications for a proper core series (e.g.,  predictive power for headline and association with the strength of real activity). }
\acc and \acr surely feature their own  set of assumptions,  but they enjoy the distinct advantage that the optimization objective and the measurable notion of success perfectly coincide.  

The proposed method is also not entirely unrelated to approaches attempting to uncover trend inflation,  usually through compact state-space models  \citep{chan2016,eo2023understanding} or dynamic factor models  \citep{StockWatson2016}.   The link can be understood through the lenses of Hamilton filtering \citep{Hamilton2018}.  Indeed,  as an alternative to Hodrick-Prescott filtering (a smoothing splines problem),   Hamilton's suggestion is to use the residuals of direct autoregressive forecasting equation (at a user-specified horizon, a tuning parameter) as the detrended series.  \textcolor{black}{Assemblage regression in the component space can be seen as a constrained multivariate-to-univariate Hamilton filter, where instead of regressing the target on its own lags, we regress the targets on its lagged subcomponents. Furthermore, the target is the average path between the forecast date and the forecast horizon.} The usage of the average path,  typical in forecasting studies and particularly natural for inflation \citep{MDTM},   leads it to take an implicit mean over many "horizon tuning parameters",  a strategy that has been documented to be successful in output gap extraction applications \citep{quast2020reliable}.  Last but not least,  one should note that unlike \cite{Hamilton2018} and the ensuing literature,  the interest here lies in the analysis of the  extracted "trend" rather than deviations from it.  


\textbf{\acr} is our proposal for a maximally forward-looking core inflation based on \textit{temporary exclusion}.  It comes with its own distinct strand of literature \citep{bryan1991median,BryanCecchetti1993,bryan1997efficient,dolmas2005trimmed}.  Traditionally,  this is achieved by systematically ranking individual price changes from their lowest to highest values at each point in time and excluding the upper and lower tail of the distribution (either symmetrically or asymmetrically),  with median inflation \citep{bryan1991median} being the extreme case of keeping only the midpoint rank.  Trimming methods have received limited attention in the forecast combination literature \citep{wang2023forecast}, with most approaches zooming in on \textit{permanently} eliminating the least accurate contributors \citep{diebold2019machine,WangEtAl2022} or trying a few cutting points configurations for a trimmed mean estimator \citep{stock2004combination}.    

\acr features greater flexibility.  It assigns optimized weights to each rank rather than solely optimizing trimming points.   In terms of mechanics,   \acr is a constrained functional regression  \citep{morris2015functional},  where predictors (i.e.,  order statistics series)  are a function of the empirical distribution of price growth rates.  Applications of functional regression analysis and related methods in macroeconometrics are still scarce.  Recent ones include \cite{meeks2023heterogeneous} using functional PCA to summarize the distribution of inflation expectations distribution,  and \cite{chaudhuri2016forecasting} who study the  autoregressive properties of different parts of the inflation distribution.  As it was the case for the above discussion of PCA in components space,  the advantage of \acr over functional PCA and related methods is (i) supervision,  and (ii) constraints on the kinds of features we want to be extracted.  

Finally,  there have been numerous papers forecasting inflation using machine learning methods (see  \cite{medeiros2019},  \cite{HNN},  and references therein).  Recently,  there has been some attention in coupling those with components-level data \citep{barkan2023forecasting,boaretto2023forecasting,joseph2024forecasting}.  We differ from this strand of literature by our objective,  to create a core inflation measure with desirable properties.  This leads us to consider constrained linear forecasts based on components, as opposed to any functional form channeling in any dataset.

\vskip 0.15cm
{\sc \noindent \textbf{Empirical Results}.} As main applications,  we consider US and euro area (EA) inflation.  We also study the Canadian case more compactly in the appendix. In all instances,  we find that Albacore succeeds in being highly predictive of future headline inflation at various horizons.   It outperforms benchmark models in terms of predictive accuracy and yields low bias and variance compared to the official headline inflation rate. In terms of forecasting results, \acr is clearly our best performing model. Its gains mount to sizable margins, especially when taking a medium- to long-term perspective.   A look at the assembled ranks reveals a highly asymmetric trimming,  removing entirely the lower part of the monthly price growth distribution and putting the emphasis on the \textcolor{black}{60$^{\text{th}}$ to 75$^{\text{th}}$ percentile (depending on the application)}.   Unlike existing trimmed mean measures,  it keeps a non-trivial portion of the upper tail by assigning it a moderate weight. This proactive usage of the skewness in the monthly price growth rates distribution helps in staying alert to accumulating inflationary pressures,  like those that were occurring in early 2021.   
\acc's improvements are rather concentrated within the evaluation sample spanning the months following the Covid-19 pandemic.  Similarly to its supervised trimming counterpart, outperformance is more substantial for higher-order forecasts (i.e., 6 and 12 months ahead).  The optimized weighting scheme results in low weight assigned to highly volatile components,  such as energy and food,  and high weight on components related to the services sector, health, and housing. 

For the run-up of inflation, it is well known that developments in goods prices played an outsized role in the US. Accordingly, we find high contributions of goods inflation to the aggregate of \acr.
The officially reported trimmed mean, on the other hand, puts almost no weight on goods components, instead it is driven by shelter,  a notoriously lagging component \citep{bils2004sticky,cotton2023forecasting}. Given that this pattern manifests around inflation turning points, it can prove useful to downweight contributions from shelter using an asymmetric trimming scheme,  as is done by \acr.  Alternatively,  one can assign a downsized weight to problematic subcomponents (such as tenant rent) and compensate by upweighting others,  an opportunity which is leveraged by \acc, thanks to its high-dimensional regression setup.  This is less radical than excluding the shelter component entirely,  as in US supercore inflation \citep[i.e., nonhousing core services, see ][]{powell2022,leduc2023will}.  In the EA, sustained energy and food price shocks played a larger role than in the US, which makes the official core inflation rate a lagging indicator.    \acc and \acr avoid this predicament via positive contributions from food goods.  Regarding the early indication of the disinflationary phase,   both Albacore attribute it to easing goods price pressures for the US and the EA.


We explore a suite of extensions.   First,  we consider building core measures specialized for either upside or downside inflation risks by changing the squared error loss in assemblage regression to a quantile loss.  These can be equivalently interpreted as the optimal core inflation measure for an analyst facing asymmetric forecast error costs.    Interestingly,  we find more harmony between traditional core inflation measures and ours when customizing the latter for disinflation risk. \textcolor{black}{Thereby,  gains from our proposed approach are highest in the upper tail, especially so for our quantile extension of \acr post-Covid.}
Second,  we consider forward-looking yearly time-aggregation of headline inflation for the US.  We find that the autoregression in rank space decisively outperforms traditional autoregressions for this task.  It does so by eliminating large negative realizations of the month-over-month growth rate within the lookback window, and including positive ones with a moderate weight.  This suggests that,  in such simple models,  it is a more effective strategy to pay \textit{attention} to the position of an observation in the recent realized distribution rather than the exact location in time,  as is typically the case in standard autoregresssions.
Third,  we consider assembling \textit{headline} inflation from euro area member states to forecast the aggregate.  Albeit these do not necessarily qualify as a traditionally defined core inflation,  our results show that a downweighted average of the 4-5 highest state-level inflation rates at a given point in time offers substantial forecasting accuracy gains. 

\vskip 0.15cm
{\sc \noindent \textbf{Outline}.} We proceed as follows.   In Section \ref{sec:assreg},  we introduce the Assemblage Regression.  Section \ref{sec:emp} presents our empirical application to core inflation.  Section \ref{sec:exc} focuses on applications of \acr to other inflation-related aggregation problems. Section \ref{sec:concl} concludes and proposes various avenues for future research.  Appendix \ref{app} contains additional material, like our more condensed Canadian application.

\section{Assemblage Regression} \label{sec:assreg}

In this section,  we introduce the Assemblage Regression,  which,  in a nutshell,  is a generalized nonnegative ridge regression where the dependent variable is future headline inflation and the regressors are either inflation subcomponents or transformations of them.  Nonnegativeness comes from the desire of the output being a weighted average and "generalized" comes from the use of altered regularization schemes.  There will be two declinations of our ideas.   The first and most conceptually straightforward,  \acc,  aggregates and weights \textcolor{black}{inflation} components directly.  In essence,  it searches the space of basket weights for which consumers' today inflation  is most indicative of the average consumer inflation in the near-future.  The second,  marginally more sophisticated one  performs optimal trimming of inflation components (i.e., temporary exclusion/weighting).  \acr is the resulting measure -- it runs the assemblage regression in the rank space of the components data.  

\vskip 0.2em

{\sc \noindent \textbf{Ridge Regression Primer}.} A ridge regression solves what is more generally known as a penalized linear regression problem \citep{ESL}.  It is a traditional least squares problem plus a penalty on coefficients so to regularize the outcome of minimization.  Ridge regression coefficients are  obtained via
$$\hat{\boldsymbol{\beta}}_\text{Ridge} = \argmin_{\boldsymbol{\beta}} \sum_{t=1}^{T-1} ( y_{t+1} - \boldsymbol{\beta}'\boldsymbol{X}_{t})^2 + \lambda ||\boldsymbol{\beta}||_2  $$ 
where $ || .  ||_2$ is the $l_2$ norm.  The latter is equivalent to $\sum_{k=1}^{K} \beta_k^2$ in summation notation,  where $K$ \textcolor{black}{denotes the number of components}. The penalty term provides regularization by bringing in the mix the a priori that each coefficient should contribute to the fit, but  modestly.   In other words, it is \textit{shrinking} coefficients towards 0.  With a suitable $\lambda >0$,  a ridge regression curbs overfitting that plagues OLS (i.e.,  $\lambda=0$) when $K$ is large relative to $T$ and the signal-to-noise ratio is low.  \textcolor{black}{Everything good comes at a cost, and here,  the cost is bias.  $\hat{\boldsymbol{\beta}}_\text{Ridge}$  blends information from the data and our prior knowledge on $\boldsymbol{\beta}$.   Thus,  the specification of the penalty must be chosen wisely,  as must its strength.} $\lambda$ is typically tuned via cross-validation,  which is,  in essence,  a pseudo-out-of-sample evaluation metric.  The amount of restrictiveness, which will impact both forecasts and the coefficients (and thus the interpretation of the model),  is chosen so to maximize the model's predictive accuracy on unseen data.  Changing the $l_2$ norm for $l_1$ makes it the equally well-known LASSO problem.  In fact,  there is a constellation of possible penalty terms one can choose from depending on the application,  and we shall leverage that on the way to our main model \citep{hastie2015statistical}.   Lastly,  penalized regression problems are not scale-invariant, so either the predictors have to be scaled to exhibit the same variance,  or the penalty should be adjusted accordingly.

\vskip 0.2em

{\sc \noindent \textbf{Supervised Weighting}.}  We now provide the details of our first assemblage regression.  The permanent exclusion version,  or supervised weighting of basket components, is obtained via

\begin{equation}\label{eq:acc}
\hat{\boldsymbol{w}}_c = \argmin_{\boldsymbol{w}} \sum_{t=1}^{T-h} ( \pi_{t+1:t+h} - \boldsymbol{w}'\boldsymbol{\Pi}_{t})^2 + \lambda ||\boldsymbol{w}-\boldsymbol{w}_{\text{\tiny headline}}||_2 \quad  \text{st} \enskip \boldsymbol{w} \geq 0,   \enskip \boldsymbol{w}'\iota=1 
\end{equation}
 where $h$ is the forecasting horizon,  $T$ is the last training observation,  $\pi_{t+1:t+h}$ is average headline inflation between $t+1$ and $t+h$, 
and $ \boldsymbol{\Pi}_{t}$ is a matrix of component time series at a user-specified level of aggregation.  $\lambda ||\boldsymbol{w}-\boldsymbol{w}_{\text{\tiny headline}}||_2$ is the penalty term,  and it shrinks the solution towards $\boldsymbol{w}_{\text{\tiny headline}}$,  which are headline inflation weights for that level. \textcolor{black}{We constrain weights to be nonnegative ($\boldsymbol{w} \geq 0$) and sum to 1 ($\boldsymbol{w}'\iota=1$).} \textcolor{black}{$ \boldsymbol{\Pi}_{t}$ are expressed as price growth rates and in our applications,  it is the 3-months-over-3-months growth rate.  This allows for minimal time-smoothing while retaining timeliness \citep{BanburaBobeica2020},  and does not prevent for the measure to rightfully qualify as core inflation \textit{now}. }

\acc is defined as $ \pi^*_{\text{comps},  t} = \hat{\boldsymbol{w}}_c '\boldsymbol{\Pi}_{t}$ and is,  as such, a constrained forecast -- both in terms of the information set (only inflation rates) and how it is synthesized.  It is the linear aggregation,  or weighted average,  of components that is most closely related to future headline inflation as captured by $\pi_{t+1:t+h}$.  The adaptiveness is tied to the objective,  and it is natural to expect core measures to vary in composition along $h$ and depend on the loss function itself (see Sections \ref{sec:wcurves} and \ref{sec:quantile}, respectively).  Core measures that investment bankers, monitoring market reactions to inflation number releases, should care about can be different from those of interest to central bankers dealing with the long lags of monetary policy. Depending on the application, the user may choose the forecasting horizons $h$ for the target and the aggregation level of components.  While we consider various numbers of $h$ in our empirical applications to showcase the versatility of the approach,  the main analysis will focus on $h=12$ months which is widely regarded as the most relevant time frame for the use of core inflation measures in monetary policy decision-making. Further details are given in Section \ref{sec:emp}.

Shrinking to $\boldsymbol{w}_{\text{\tiny headline}}$ is intuitive as it represents the official statistical aggregation of the components and implies under $\lambda \rightarrow \infty$ a random walk forecast (if the moving average length of the target matches that of regressors).\footnote{Note that in this context,  opting for the traditional ridge penalty $\lambda ||\boldsymbol{w}||_2$ would imply shrinking to equal weights.  {The other two constraints (i.e., nonnegative weights which sum to 1) prevent the fit from collapsing to $\boldsymbol{0}$ and thus, shrink what is left to an identical value.} Nonetheless,  this {regularization scheme} may suffer from the fact that certain subcomponents of an upper level are simply more granular and numerous,  and would push the model to favor items that are highly numerous rather than those with higher consumption weights. } We resist the temptation of shrinking to core inflation (ex.  food and energy) weights, which surely implies  a less volatile "null" model than what we consider, and perhaps better empirical results.  This is motivated from the desire to see,  as a basic check, whether  \acc can discover this everlasting empirical wisdom without assuming half of the discovery first.  It is also plausible that some forward-looking core measures at shorter horizons could significantly differ from the typical core measures, which have historically been focused on medium- to long-term horizons.  \textcolor{black}{Taking on a different perspective, the assemblage regression can be viewed as an algorithm searching the space of basket weights for which consumers' inflation today is most indicative of the average consumer inflation tomorrow.  Although, the resulting assemblage does not explicitly fulfill any requirement to match a representative agent per se, it would be surprising if those two were vastly unrelated. Yet, if that were to prove insufficient according to pseudo-out-of-sample performance, the penalty  $\lambda ||\boldsymbol{w}-\boldsymbol{w}_{\text{\tiny headline}}||^2$ shrinks coefficients to the official weights (instead the usual 0).}

Lastly,  one needs to think about the strength of that regularization,  as captured by $\lambda$.  This can be set automatically,  but it needs to be done the right way.  The target time series is persistent,  particularly starting from $h>6$,  and so are some components. Thereby,  choosing $\lambda$ with standard cross-validation  designed for independent data will deliver an overconfident assessment  of the generalization error and a downward biased $\lambda$.   As recommended in \cite{GCFK} and others,  we opt for a non-overlapping blocks approach.  We divide the training sample in 10 contiguous segments and use those in 10 fold cross-validation.  The length of blocks thus depends on that of the training sample (e.g.,  an estimation sample of 20 years implies blocks of two years).  While longer blocks could be desirable for $h=12$ and above,  one must remember that too {few} blocks and lacking randomization likely would outweigh the benefits.

\vskip 0.2em

{\sc \noindent \textbf{Supervised Trimming}.} The second,  supervised trimmed inflation,  needs further thought in order to be cast within the assemblage regression apparatus described above.  After all,  trimming implies components jump in and out of the index every month,  implying a kind of time variation  in their weights based on how each component growth rate ranks compared to others in a given month.  Surely,  going straight at it in such a fashion would push us away from the "keep it sophisticatedly simple" principle  \citep{diebold1998elements}.  The opposite direction,  i.e., that of simplicity at the expense of generality and flexibility,  would bring things closer to \cite{bryan1997efficient}'s trimmed mean PCE inflation.   There,  components are either out (with a weight of 0), or in (with a weight proportional to their basket weight versus that of other non-trimmed components at time $t$).  The size of trimmed bands on both tails can then be optimized over two parameters using some criterion.  The resulting series is a robust weighted average.  However,  the mean (or the median for that matter), may or may not  \textcolor{black}{align with} where our eyes should be to extract leading signals for future headline inflation.

Instead, one can consider, more generally, retrieving the summary statistics of the current realized price growth distribution that best fulfill a specific statistical objective.  Or equivalently,  components get weights as a function of their location in the empirical distribution at time $t$.  A closer examination of what such time variation implies reveals that, in fact,  we can run a very similar assemblage regression as described above,   but using the empirical order statistics of $\boldsymbol{\Pi}_{t}$ as regressors.  To see this,  we can simply walk from the components space to the "rank space",  by writing the formula for fitted values in summation notation
\begin{align*}
 \pi^*_{\text{ranks}, t}  &= \sum_{k=1}^K w_{k,t} \Pi_{k,t} \\
&= \sum_{k=1}^K \underbrace{\sum_{r=1}^K w_r  I \left(\text{rank}(\Pi_{k,t})=r \right)}_{w_{k,t}}\Pi_{k,t} \\[-0.75em]
 &= \sum_{r=1}^K w_r  \underbrace{ \sum_{k=1}^K I \left(\text{rank}(\Pi_{k,t})=r \right)\Pi_{k,t}}_{O_{r,t}} \\[-1em]
 &= \sum_{r=1}^K w_r   O_{r,t}
\end{align*}
where $O_{r,t}$ is the $r^{\text{th}}$ order statistic (i.e.,  the value attached to rank $r$) at time $t$.  If $K$ were to be equal to 100,  $O_{1:100,t}$ would be a set of poor man's percentiles.  The above derivation informs us that this kind of time variation  in the component space implies time-invariant coefficients in the rank space (and vice versa).  We will later use this duality to map back the implication of $w_r$ in the component space in Section \ref{sec:sts}. Thus, ease of implementation can be restored from running the model in rank space,  i.e., using  $\boldsymbol{O}_{t}$,  which is effectively just sorting the components at each $t$ and stacking them in a matrix.   Precisely,  $ O_{1:K,t}=\texttt{sort}(\Pi_{1:K,t}) \enskip \forall t$.  The switch to rank space-based weighting can also be formalized in our preferred matrix notation,  with $\boldsymbol{O}_{t}= \boldsymbol{A}_{t}\boldsymbol{\Pi}_{t}$ where $\boldsymbol{A}_{t}$ is a $T \times T$ allocation matrix where  entries are 0s and 1s following $ I \left(\text{rank}(\Pi_{k,t})=r \right) \enskip \forall (r,k) $.  Thus, we run
\begin{equation}\label{eq:acr}
\hat{\boldsymbol{w}}_r = \argmin_{\boldsymbol{w}} \sum_{t=1}^{T-h} ( \pi_{t+1:t+h} - \boldsymbol{w}'\boldsymbol{O}_{t})^2 + \lambda ||D\boldsymbol{w}||_2 \quad  \text{st} \enskip \boldsymbol{w} \geq 0 ,   \enskip \bar{\pi}_{t+1:t+h} = \bar{ \pi}^*_{\text{ranks}, t}  
\end{equation}
where $D$ is the difference operator and \acr is defined as $ \pi^*_{\text{ranks}, t} = \hat{{\boldsymbol{w}}}_r '\boldsymbol{O}_{t}$.   Rather than learning which subcomponents to include,  the problem will now be learning which ranks (and with which weight) to include or exclude.    $||D\boldsymbol{w}||_2 $ (or equivalently $\sum_{r=1}^K \left(w_r - w_{r-1}\right)^2 $ in summation notation)  is a fused ridge penalty \citep{hastie2015statistical}.  It favors a smooth weighting scheme,  and in the $\lambda \rightarrow \infty$ limit, pushes the model towards the sample mean solution where each rank gets a weight of $\sfrac{1}{K}$.\footnote{ Note that,  in the $\lambda \rightarrow \infty$ limit,  solutions are equivalent in components and rank space ( $ \iota ' \boldsymbol{O}_t=  \iota ' \boldsymbol{A}_{t}\boldsymbol{\Pi}_{t} = \iota ' \boldsymbol{\Pi}_t$). }  In the other limit ($\lambda=0$),  the solution will be sparse for reasons we will come back to below.   

The equality constraint has been changed from $\boldsymbol{w}'\iota=1 $ to $\bar{\pi}_{t+1:t+h} = \bar{ \pi}^*_{ranks, t}$.  In the rank space,  $\boldsymbol{w}'\iota=1 $ would imply that the average of all weights is $\sfrac{1}{K}$,  which does not enforce symmetric trimming,  but rather would impose equivalent masses above and below $\sfrac{1}{K}$,  and thus blocking highly asymmetric outcomes which we will find to be successful in nearly all our experiments.  The $\bar{\pi}_{t+1:t+h} =  \bar{ \pi}^*_{\text{ranks}, t} $ constraint brings back discipline by forcing residuals to have mean 0.  They could deviate marginally because the model has no intercept,  and to minimize the sum of squared  residuals,  some bias could be traded for variance reduction.  We block this possibility by imposing the constraint that \acr has the same long-run mean as headline inflation (over the training sample).


{This stands in contrast to most trimming approaches, which opt for a preselected band within which ranks get assigned their official (components space) weights, and those outside of it have a weight of 0.} For instance,  the FRB Cleveland trimmed mean CPI sets to 0 the first and last 16\% of ranks,  and then a weighted average of the center band is reported.  In a similar vein,  the FRB Dallas' Trimmed Mean PCE trims out 24\% from the lower tail and 31\% from the upper tail \citep{dolmas2005trimmed}. \textcolor{black}{Then, there is the} Cleveland Fed Median CPI,  which keeps only one rank,  the middle one.  Those trimming approaches can be seen as knife-edge cases of the above,  where $\hat{\boldsymbol{w}}$ is chosen based on heuristics.  It is noteworthy that the fused penalty could be changed to $\lambda ||D\boldsymbol{w}||_1 $ (fused lasso) if one wanted to favor a sharp (boxy shaped) trim.  We do not opt for such a possibility given that our framework needs not be limited to such shapes.   Having this in mind,  it is important to note how our proposition differs from what is commonly done with trimmed measures.  It does not reweight  included components according to some rule based on their headline inflation weights, nor does it enforce that weights sum to 1 at every point in time.

\vskip 0.2em

{\sc \noindent \textbf{Additional Methodological Remarks}.} Note that there is no intercept included in our regressions,  which has a few implications.  First,  the fitted values are mechanically pushed (but not forced unless otherwise specified) towards having the same unconditional mean as $\pi_{t+1:t+h}$ through $w_{0} = \bar{\pi}_{t+1:t+h} -  \bar{ \pi}^*_{t} = 0$,  therefore making the fitted values already in headline inflation units by construction.  Second,  it implies a latent random walk hypothesis,  which is in synchronicity with the idea of a core inflation measure being the summary of inflation conditions right now that is most indicative of future developments in headline inflation.  

Additionally,  excluding the intercept $w_{0}$ is necessary to properly identify the elements of $\boldsymbol{w}$ in Equations \eqref{eq:acc} and  \eqref{eq:acr} at  longer horizons,  like $h=24$.  As $h$ grows,  the unconstrained regression solution puts a very small weight on current inflation conditions -- and a much larger one on the intercept.  This implies a very limited passthrough of information from $ \pi_{t+1:t+h}$ to $\boldsymbol{w}$,  which job is to effectively bundle the ensemble of components/ranks.  \textcolor{black}{ Thus,  with little variance from the supervisor $ \pi_{t+1:t+h}$ being canalized into the learning of $\boldsymbol{w}$,  its elements can only be estimated imprecisely,  and,  going towards the limit of an "intercept only" model,  are identified by the prior (i.e.,  the penalties).}   Evidently,  this problem does not occur when $\boldsymbol{w}$ is predetermined using heuristic rules,  like for most traditional core measures.   Interestingly,  we will find that excluding the intercept also improves the benchmark regressions in most cases and time periods,  but not nearly as much as for Albacores.   

The combination of various constraints has implications of its own.   In both regressions,  the exclusion of the intercept -- implying $w_{0} = \bar{\pi}_{t+1:t+h} -  \bar{ \pi}^*_{t} = \bar{\pi}_{t} -  \bar{ \pi}^*_{t} \approx 0$ --  combined with the nonnegativity constraint implies adaptive $l_1$ regularization. \textcolor{black}{To see this, note that in an attempt to mitigate variance, $\bar{ \pi}^*_{t}$ may fall marginally below $\bar{\pi}_{t}$, unless equality is forced as in \acr, and ergo classifies as an implicit \textit{soft} constraint. In other words,}  we have  $\bar{ \pi}^*_{t} \leq \bar{\pi}_{t}$,  which can,  by definition,  be rewritten as  \textcolor{black}{$\sum_{k=1}^K w_k \bar{ \pi}_{k,t} \leq \bar{\pi}_{t}$}.  Bringing in the nonnegativity constraint and denoting $\gamma_{k}  \equiv  \frac{\bar{ \pi}_{k,t}}{\bar{\pi}_{t}}$, we get
\begin{equation}\label{adalasso}
 \sum_k^K \gamma_{k}  |w_k| \leq 1
\end{equation}
which is the constrained optimization form of an adaptive LASSO constraint with a "budget" of 1 -- technically,  a tuning parameter. In the case of \acr, Equation \eqref{adalasso} holds with equality.  \textcolor{black}{In that of \acc sum-to-one constraint,  the implications are Equation \eqref{adalasso} holding with equality with $ \gamma_{k} = 1 \phantom{.} \forall  k$ ,  a plain LASSO constraint in addition to Equation \eqref{adalasso} itself.} Thereby,  unless other constraints are brought in,  the resulting model will likely be sparse.  Is this desirable? It depends.  In the high-dimensional \acc application,  this is undesirable given that we expect the forward-looking core measure to be dense,  i.e.,   based on more than a handful of highly disaggregated components.  At higher levels of aggregation,  sparsity becomes reasonable again, but may lead to a measure that is far from mapping into any realistic consumer,  which may impede the successful communication of the index.   In rank space,  sparsity can make sense,  because order statistics themselves are already dense combinations of underlying components through the rule underlying the allocation matrix.  A well-known sparse core inflation measure in rank space is \textcolor{black}{the weighted median inflation}.  If $\lambda=0$ is the outcome of cross-validation,  both  \acc and  \acr can embrace sparsity by letting the cocktail of constraints act as the single source of ($l_1$) regularization.  We will see that most often,  dense solutions are favored. 

\vskip 0.2em

{\sc \noindent \textbf{Implementation Details}.}  We use the \texttt{CVXR} package in R,  which provides fast solutions for linear convex programming problems.  Simplified versions of the above could be implemented via the well-known \texttt{glmnet} package.  However,  necessary equality constraints (such as $ \boldsymbol{w}'\iota=1$ in \acc) are beyond their functionalities,  and one cannot simply rearrange regressors as in OLS.  We provide the R package \texttt{assemblage},  which automatically implements the above given user-provided target and components.  Running estimations implying fewer than 20 components takes about 0.14 seconds on a Macbook Pro (with M2 chip), and a full-patch cross-validation of $\lambda$ takes 5.53 seconds (with a $\lambda$ grid of length 20 and 10 folds).  Larger models with above 200 components take 0.26  seconds for a single run and 7.91 seconds for cross-validation.   These numbers assume running cross-validation in parallel (coded as an option in the \texttt{assemblage} package) on 11 cores.  Using a single core implies the tuning of $\lambda$ now takes 23.21 seconds for the small model ($K=20$) and 39.17 seconds for the bigger one ($K=200$).

\section{Core Inflation Revisited} \label{sec:emp}

We construct Albacore for the US, the EA\textcolor{black}{, and Canada} with monthly price indices at different levels of disaggregation. \textcolor{black}{Results and implementation details for the latter are relegated to Appendix \ref{sec:canresults}.} For the US we focus on the price index for Personal Consumption Expenditure (PCE), which is taken from the Bureau of Economic Analysis (BEA). We base our core inflation measure on level 2, 3, and 6 including 15, 50, and 215 subcomponents, respectively. For the euro area, we use the Harmonized Index of Consumer Prices (HICP) from Eurostat amd  choose the two-, three-, and four-digit COICOP level, which comprises a number of 12, 39, and 92 subindices.\footnote{We exclude item CP0735 (Combined passenger transport) from level 4 in the EA due to the item's heavy irregularities during the Covid-19 pandemic.}   All disaggregated series are seasonally adjusted\footnote{Note that for the EA and Canada (unlike the US) there is no seasonally adjusted data published by statistical institutions. We provide details on the seasonal adjustment for both applications in Appendix \ref{sec:data}.} and expressed as 3-months-over-3-months changes,  which balances timeliness (as opposed to 12 months trailing averages) and noise reduction (versus the month-over-month growth rate).  Rank positions are calculated using month-over-month growth rates (as done for trimmed mean inflation), and then the order statistics time series are smoothed using the 3 months moving average. 

Constructing Albacore for each level involves predicting the average path of headline inflation for 1, 3, 6, 12, and 24 months ahead ($h \in \{1,3,6,12,24\}$).  Albacore based upon $h \in \{6,12,24\}$,  indicating medium-term developments,  is of interest for policymakers.  An index targeting shorter horizons ($h \in \{1,3\}$) can be useful to hedge funds performing reallocations based on anticipation of central banks rate decisions,  especially in times where there is sizable monetary policy uncertainty.

We choose two out-of-sample test sets spanning periods before the Covid-19 crisis (i.e., 2010m1 to 2019m12) and the post-Covid period (i.e., 2020m1 to 2023m12) and evaluate the point forecasting performance with root mean squared errors (RMSEs).  These are two vastly different regimes, with the first evaluation sample being characterized by stable low inflation and the second by unstable high inflation.  We base the estimation on a rolling window of 20 years,  allowing for mild structural changes in the composition of the maximally forward-looking inflation series. Note that for the EA this choice inevitably results in an expanding window since our disaggregated data set starts in 2003.

We compare the forecasting accuracy to a set of benchmark models.  For each country, this set includes the officially reported core inflation series (i.e., PCE/HICP excluding energy and food, henceforth, core PCE/HICPX), trimmed mean inflation (i.e., FRB Dallas' Trimmed Mean PCE for the US, 30\% trimmed mean inflation for the EA) as well as established model-based concepts of underlying inflation measures.  Given that the series address the problem from various angles,  the information they convey may differ. Hence, instead of focusing on a single measure of core inflation we enhance our benchmarks by combining them and, in this way, perform an ex-ante weighting of the best currently available measures \citep[in the spirit of ][]{Cogley2002}. 

Similar to our main setup, we include the different series in a nonnegative regression.  Our first benchmark (and numéraire for all RMSEs in our tables) is comprised of headline, core as well as trimmed mean inflation including an intercept ($\boldsymbol{X}_t^{\text{bm}}$), which allows the model to include the long-run average.  
The same set of variables without the intercept ($\boldsymbol{X}_t^{\text{bm}}$, $(w_0=0)$) is used as another competitor.  The former is more akin to a standard forecasting regression, while the latter is more in tune with traditional usage of core inflation measures and the inherent random walk hypothesis.  Accordingly,  we expect $\boldsymbol{X}_t^{\text{bm}}$  to be a tough benchmark during the stable inflation out-of-sample period,  and $\boldsymbol{X}_t^{\text{bm}}$, $(w_0=0)$ to be more competitive when facing the rapidly evolving inflation conditions of our second test sample.  We also consider a more comprehensive set of benchmarks by combining  7 publicly available inflation measures for the US,   and 8 for the EA ($\boldsymbol{X}_t^{\text{bm+}}$).\footnote{For the US we use PCE services other than housing, FRB Dallas' Trimmed Mean PCE, FRB Cleveland Median CPI, FRB Cleveland 16\% Trimmed Mean CPI, the Atlanta Fed Sticky CPI, core PCE (ex. food and energy) and PCE headline. The set for the EA includes HICP, HICP excluding food and energy, HICP excluding energy, HICP excluding energy and unprocessed food, Supercore, PCCI, PCCI excluding energy and the 30\% trimmed mean inflation.}  Again, we estimate it with and without an intercept ($\boldsymbol{X}_t^{\text{bm+}},  (w_0=0)$).

\subsection{Results for the United States} \label{sec:usresults}

\begin{figure}[t!]
  \caption{\normalsize{Albacore for the US}} \label{fig:albacore_us_h12}
  \begin{center}
    
    \begin{subfigure}[t]{\textwidth}
      \vspace*{-0.7cm}
      \centering
      \includegraphics[width=0.98\textwidth, trim = 0mm -10mm 0mm 0mm, clip]{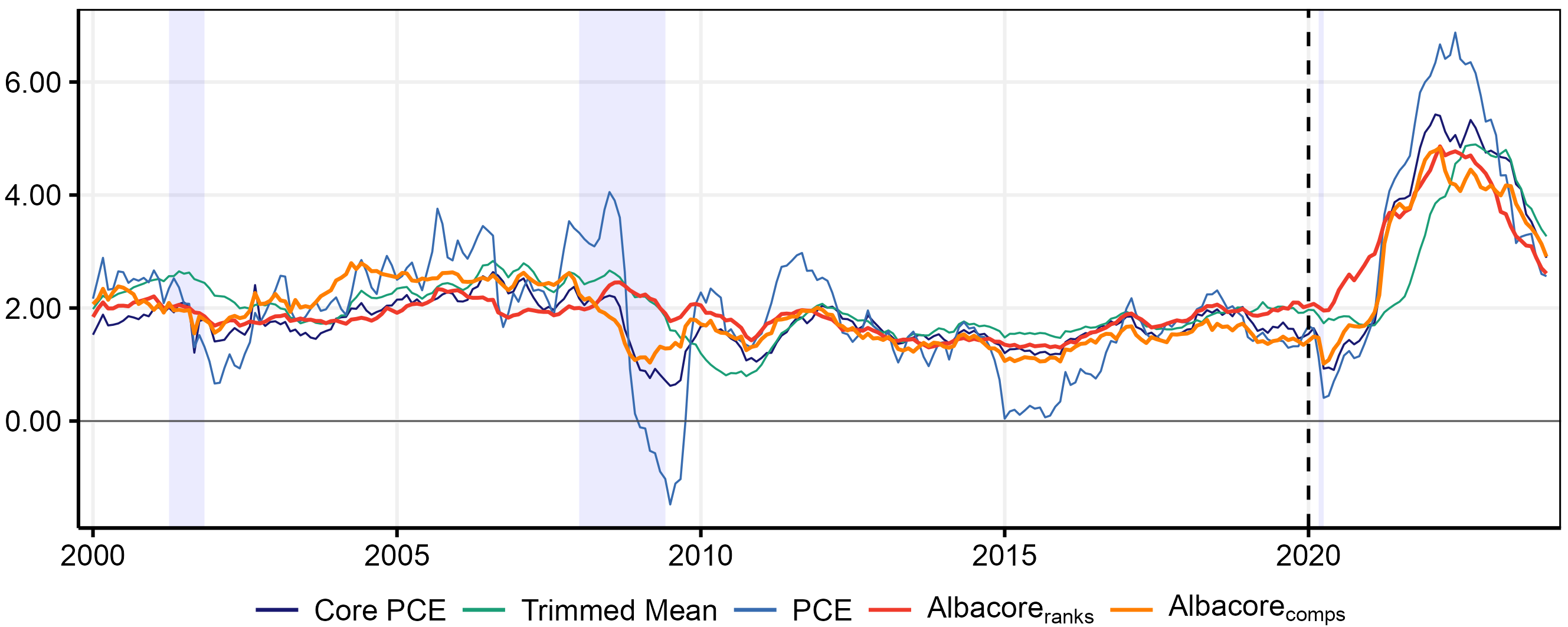}
    \end{subfigure}%

    \begin{subfigure}[t]{0.5\textwidth}
      \vspace*{-0.05cm}
      \centering
      \includegraphics[width=\textwidth, trim = -4mm -20mm 0mm 0mm, clip]{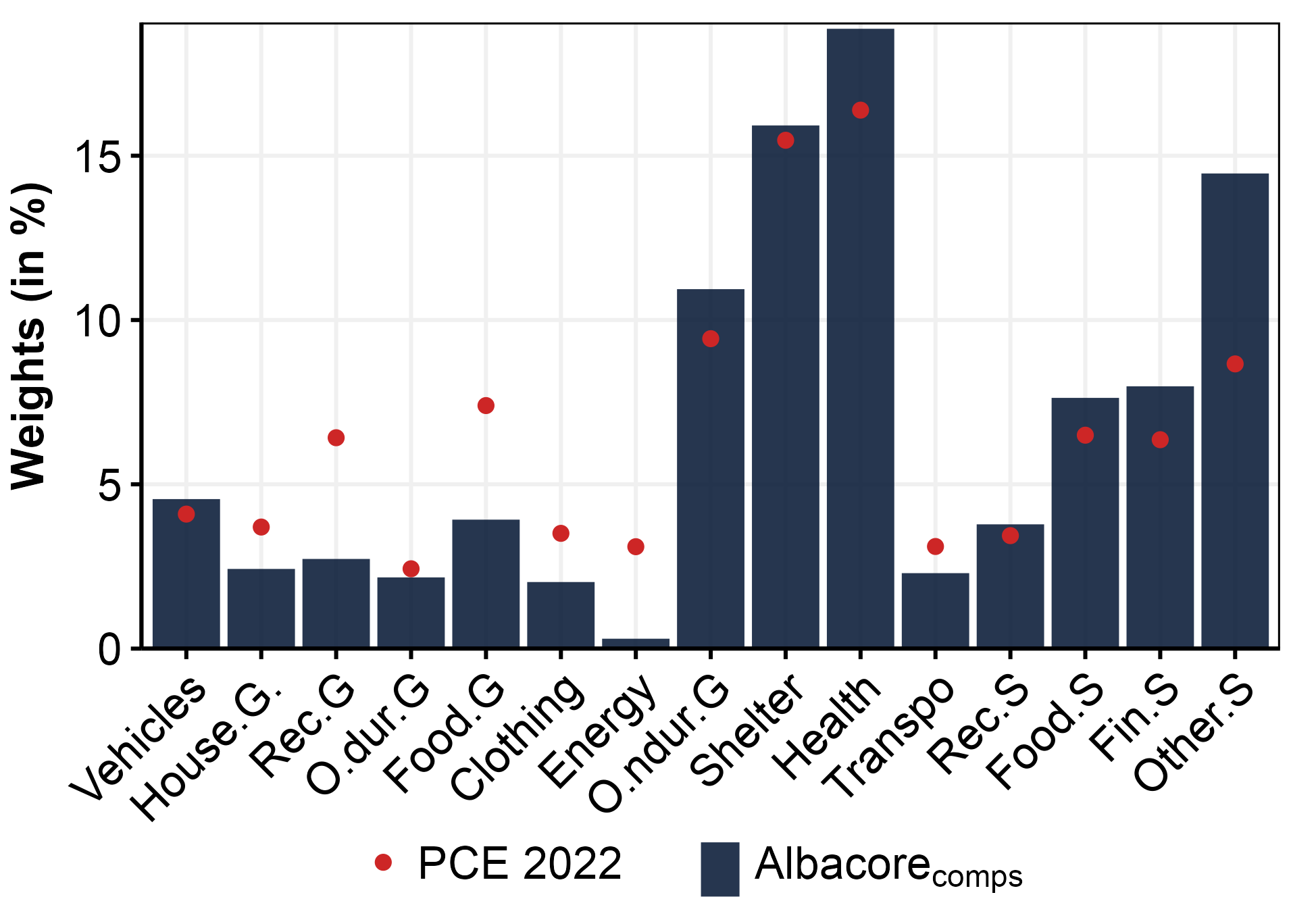}
      \vspace*{-0.6cm}
      \caption{Comparison of Component Weights}
    \end{subfigure}%
    \begin{subfigure}[t]{0.5\textwidth}
      \vspace*{-0.45cm}
      \centering
      \hspace*{0.07cm}
      \includegraphics[width=\textwidth, trim = 0mm -33mm 0mm -3mm, clip]{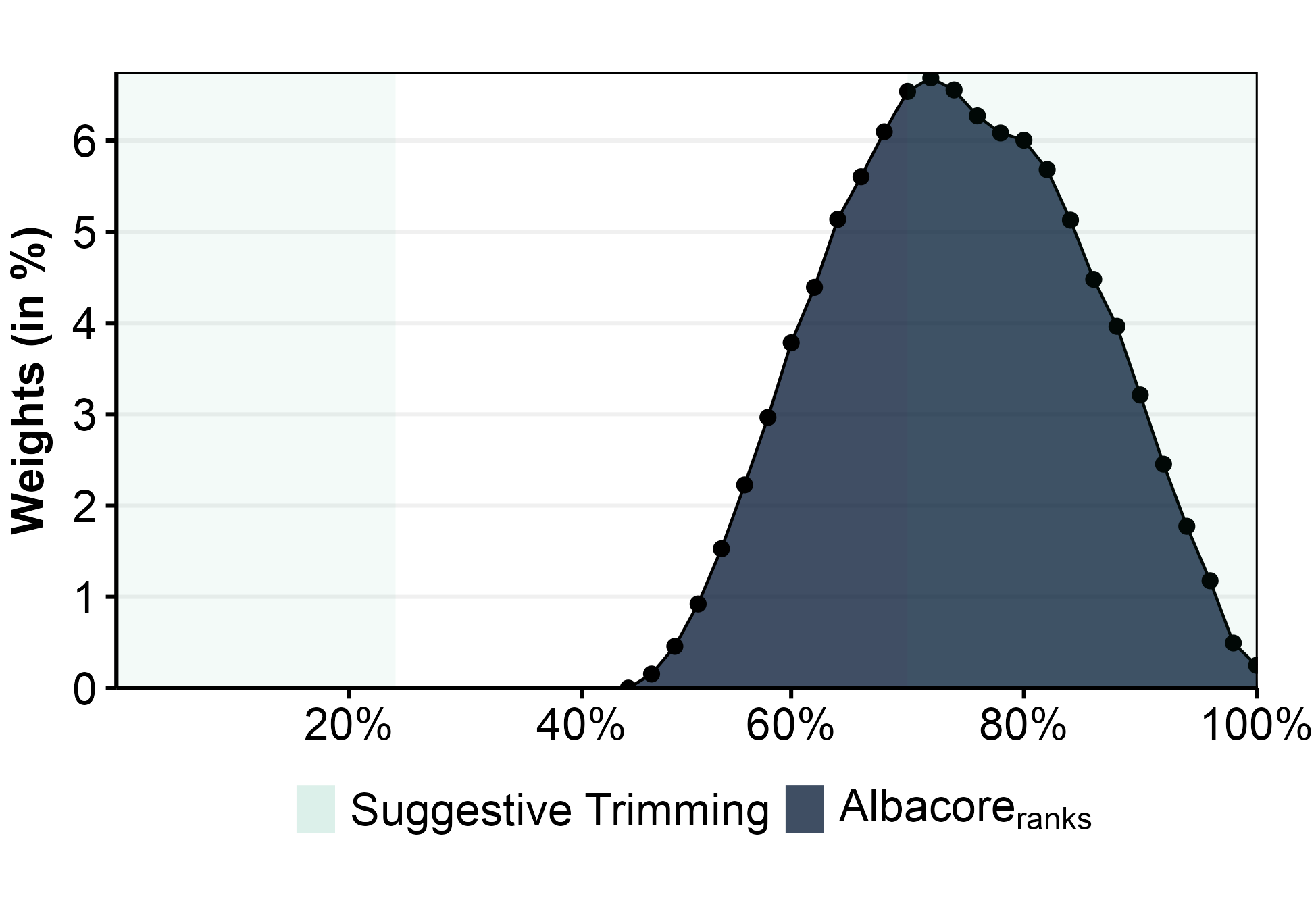}
      \vspace*{-0.7cm}
      \caption{Comparison of Rank Weights}
    \end{subfigure}
    
  \end{center}

    \begin{threeparttable}
    \centering
    \begin{minipage}{\textwidth}
    \vspace*{-0.5cm}
      \begin{tablenotes}[para,flushleft]
    \setlength{\lineskip}{0.2ex}
    \notsotiny 
  {\textit{Notes}: The \textbf{upper panel} shows Albacore (level 6 for Albacore$_\text{comps}$ and level 3 for Albacore$_\text{ranks}$) and benchmarks in yearly percentage changes. Up to 2020m1 we plot the in-sample results of Albacore and show out-of-sample predictions thereafter, indicated by the dashed line. The \textbf{lower left panel} presents the weights in percent for Albacore$_\text{comps}$. To ensure an adequate representation, they are aggregated to the lowest level of disaggregation (level 2) by averaging over the components' weights in each category. The red dots give the official PCE weights in 2022 determined by expenditure shares. Details on the labeling can be found in Appendix \ref{sec:data}. In the \textbf{lower right panel} we compare the resulting weights for Albacore$_\text{ranks}$ with the FRB Dallas' Trimmed Mean PCE. The green shaded area marks the excluded share of monthly price changes as is suggested by the officially reported trimmed mean before reweighting (24\% of the lower tail and 31\% of the upper tail). We show model results trained on data from 2001m1 to 2019m12.}
    \end{tablenotes}
  \end{minipage}
  \end{threeparttable}
\end{figure}

\begin{figure}[t]
        \caption{\normalsize{Albacore Decomposition for the United States}} \label{fig:albacore_us_decomp}
        \vspace*{-0.2cm}
        \centering
        \includegraphics[width=\textwidth, trim = 0mm 0mm 0mm 0mm, clip]{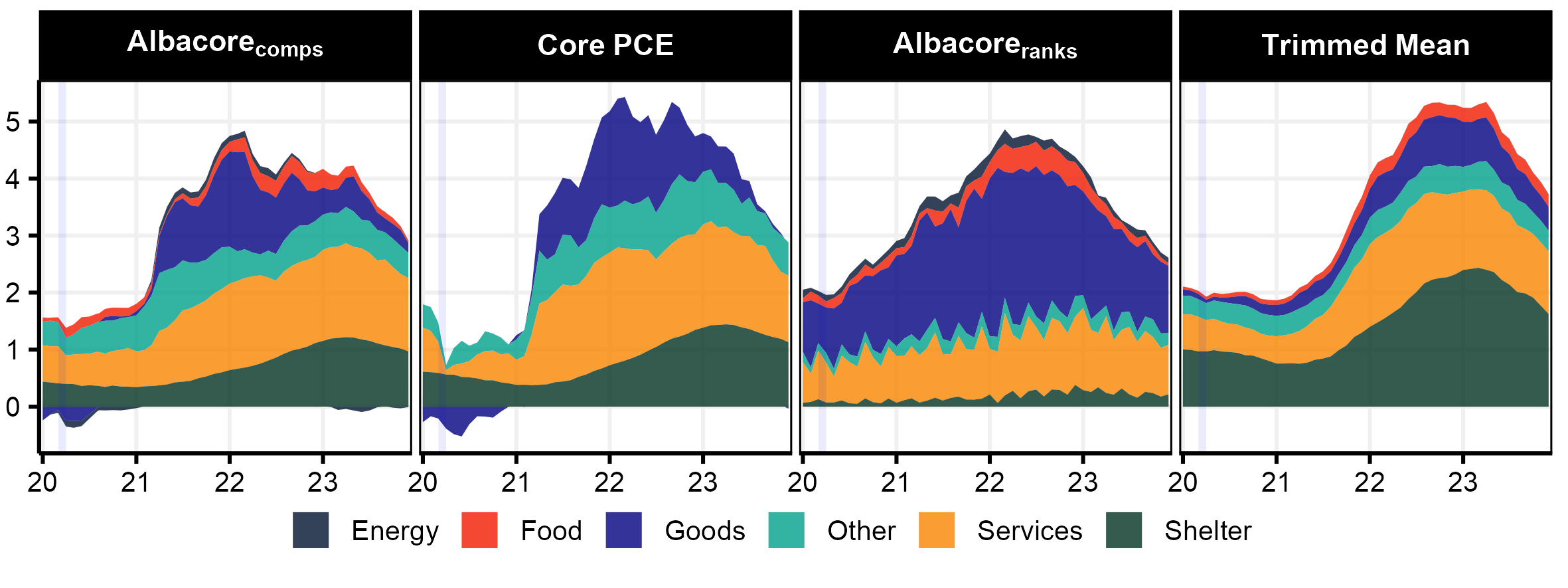}

  \begin{minipage}{\textwidth}
  \vspace*{-0.5cm}
    \begin{tablenotes}[para,flushleft]
      \setlength{\lineskip}{0.2ex}
      \notsotiny 
      {\textit{Notes}: We decompose each inflation series into a common lower level of six main aggregates. For Albacore$_\text{comps}$ ($h=12$ and level 6) and core PCE this is done by aggregating subcomponents to each group, i.e., energy, food, goods, services, shelter, and other, where other comprises transportation and health. Details can be found in Appendix \ref{sec:data}. Albacore$_\text{ranks}$ ($h=12$ and level 3) and the trimmed mean require a conversion to component space as discussed in Section \ref{sec:sts}.}
    \end{tablenotes}
  \end{minipage}
\end{figure}

First,  Table \ref{tab:results_us_pop3} contains the out-of-sample forecasting exercise results.  Next,  Figure \ref{fig:albacore_us_h12} shows the resulting time series as well as the weights assigned to ranks and components.   Finally,  in Figure \ref{fig:albacore_us_decomp},  we zoom onto the post-2020 period  and decompose Albacore and two classical core measures into the main aggregates.  This provides an understanding of their different assessments for both the surge and the slowdown.

\begin{table}[t!]
\vspace{0.3cm}
  \footnotesize
  \centering
  \begin{threeparttable}
    \caption{\normalsize {Forecasting Performance of Albacore for the US} \label{tab:results_us_pop3}
      \vspace{-0.3cm}}
    \setlength{\tabcolsep}{0.55em} 
    \setlength\extrarowheight{2.9pt}
    \begin{tabular}{l| rrrrrrrrr | rrrrrrrrr } 
      \toprule \toprule
      \addlinespace[2pt]
      &  \multicolumn{9}{c}{2010m1-2019m12} &  \multicolumn{9}{c}{2020m1-2023m12}  \\
      \cmidrule(lr){2-10} \cmidrule(lr){11-19}
      \multicolumn{1}{r|}{$h \rightarrow$}&  \multicolumn{1}{c}{$1$} & &  \multicolumn{1}{c}{$3$} & & \multicolumn{1}{c}{$6$} & & \multicolumn{1}{c}{$12$} & & \multicolumn{1}{c}{$24$}&  \multicolumn{1}{c}{$1$} & &  \multicolumn{1}{c}{$3$} & & \multicolumn{1}{c}{$6$} & & \multicolumn{1}{c}{$12$} & & \multicolumn{1}{c}{$24$}  \\
      \midrule
      \addlinespace[5pt] 
      \rowcolor{gray!15} 
      \multicolumn{6}{l}{Level 2 ($K = 15$)}   & & &&&&& & & & & & & \cellcolor{gray!15}  \\ \addlinespace[2pt]
      Albacore$_\text{\tiny comps}$ & 1.02 &  & 1.01 &  & 1.08 &  & 1.13 &  & 1.12 & 0.90 &  & 0.86 &  & 0.70 &  & 0.63 &  & 0.88 \\
      Albacore$_\text{\tiny ranks}$ & \textbf{1.01} &  & \textbf{0.97} &  & \textbf{0.99} &  & \textbf{0.98} &  & \textbf{0.87} & \textbf{0.84} &  & \textbf{\color{ForestGreen}0.74} &  & \textbf{\color{ForestGreen}0.57} &  & \textbf{0.59} &  &\textbf{ 0.69} \\ 
      \midrule 
      \rowcolor{gray!15} 
      \multicolumn{6}{l}{Level 3 ($K = 50$)}  & & &&&&& & & & & & & \cellcolor{gray!15}  \\ \addlinespace[2pt]
      Albacore$_\text{\tiny comps}$ & 0.99 &  & 1.02 &  & 1.10&  & 1.14 &  & 1.06 &0.87 &  & 0.84 &  & 0.80 &  & 0.75 &  & 1.01 \\  
      Albacore$_\text{\tiny ranks}$ & \textbf{\color{ForestGreen}0.97} &  & \textbf{0.92} &  & \textbf{0.93} &  & \textbf{0.87} &  & \textbf{\color{ForestGreen}0.73} & \textbf{\color{ForestGreen}0.81} &  & \textbf{0.76} &  &\textbf{0.61} &  & \textbf{\color{ForestGreen}0.56} &  & \textbf{\color{ForestGreen}0.63} \\ 
      \midrule 
      \rowcolor{gray!15}
      \multicolumn{6}{l}{Level 6 ($K = 215$)}  & & &&&&& & & & & & & \cellcolor{gray!15}  \\ \addlinespace[2pt]
      Albacore$_\text{\tiny comps}$ & 1.06 &  & 1.00 &  & 1.10 &  & 1.19 &  & 1.16 & 0.92 &  & 0.88 &  & 0.84 &  & 0.71 &  & 0.99 \\
      Albacore$_\text{\tiny ranks}$ & \textbf{0.98}  &  & \textbf{\color{ForestGreen}0.91} &  & \textbf{\color{ForestGreen}0.88} &  & \textbf{\color{ForestGreen}0.84} &  & \textbf{0.77}  & \textbf{0.86}  &  & \textbf{0.82}  &  & \textbf{0.70}  &  & \textbf{0.62}  &  & \textbf{0.67} \\ 
      \midrule 
      \rowcolor{gray!15}
      ${\mathbf{Benchmarks}}$  & & & & & & & &&&&& & & & & & & \cellcolor{gray!15}  \\ \addlinespace[2pt]
      $\boldsymbol{X}_t^{\text{bm}}$, $\phantom{..}$($w_0=0$) & 1.00 &  & 0.99 &  & 1.03 &  & 1.03 &  & 0.96 & 0.91 &  & 0.89 &  & 0.79 &  & 0.80 &  & 0.93 \\
      $\boldsymbol{X}_t^{\text{bm+}}$ & 1.01 &  & 1.04 &  & 1.04 &  & 1.04 &  & 1.02 & 1.37 &  & 1.48 &  & 1.18 &  & 1.14 &  & 1.09 \\ 
      $\boldsymbol{X}_t^{\text{bm+}}$, ($w_0=0$) & 1.02 &  & 1.04 &  & 1.07 &  & 1.07 &  & 0.97 & 1.10 &  & 1.15 &  & 0.86 &  & 0.87 &  & 0.96 \\
      
      \bottomrule \bottomrule
    \end{tabular}
   \begin{tablenotes}[para,flushleft]
    \scriptsize 
    \textit{Notes}: The table presents root mean square error (RMSE) relative to $\boldsymbol{X}_t^{\text{bm}} = [\text{PCE}_t \phantom{.}  \text{PCEcore}_t \phantom{.}  \text{PCEtrim}_t  ]$  with intercept. The remaining benchmarks are: $\boldsymbol{X}_t^{\text{bm}} = [\text{PCE}_t \phantom{.}  \text{PCEcore}_t \phantom{.}  \text{PCEtrim}_t  ]$ without an intercept (i.e., $w_0=0$), $\boldsymbol{X}_t^{\text{bm+}}$ with and without an intercept. Numbers in \textbf{bold} indicate the best model for each \textit{level} and each \textit{horizon} in each of the out-of-sample periods. Numbers highlighted in {\color{ForestGreen} green} show the best model per \textit{horizon} and out-of-sample period \textit{across levels}.
  \end{tablenotes}
  \end{threeparttable}
\end{table}

\vskip 0.2em
{\sc \noindent \textbf{Forecasting Performance}.} For the US data,  we find that \acr is the best performing model regardless of the forecasting horizon, level of disaggregation, and evaluation sample.   In most cases,  it outperforms benchmarks by an appreciable margin,  with the highest gains achieved for higher-order forecasts.   For the sample ending before the pandemic, \acr yields RMSEs well below all benchmarks with highest performance gains for medium-term predictions ($h \in \{12,24\}$) and higher numbers of assembled components ($K \in \{50,215\}$).  Lower-dimensional versions fare better post-2020,  and the middle-ground "level 3" option comes out as the most polyvalent.   \acc's outperformance is more local than that of its supervised trimming counterpart.  For the low inflation era,  it does not eclipse convex combinations of the usual benchmarks.  However,  its outperformance for the second test set is notable,  particularly for key horizons such as 6 and 12 months ahead.  

We also note from Table \ref{tab:results_us_pop3} that dispensing with the intercept is,  unsurprisingly so,  playing a non-trivial role in creasing performance for the second out-of-sample.   Indeed,  both the sparse and more comprehensive benchmarks deliver better results when excluding it.  \acr and \acc surely benefit from being part of the no-intercept family for the post-2020 evaluation period,  but they comfortably distance the benchmarks.  Of the $w_0=0$ cluster of models,   only \acr surpasses the performance of models featuring an intercept during the low/stable inflation era. 

Interestingly,  \acr also provides appreciable improvements for short-run forecasts ($h \in \{1,3\}$),  and does so for both test periods at nearly all levels.  This is notable given that short-run predictive accuracy is hard-earned -- beating headline itself for these targets is no small feat.   Most core inflation measures are designed with "monetary policy horizons" in mind,  and as such,  are suboptimal for such needs.    Thus,  \acr can also be of use to institutions interested in monitoring a single inflation series  that has an edge in foreseeing headline's next release ($h=1$) or its developments over the next 3 months.


\vskip 0.2em
{\sc \noindent \textbf{A Look at Time Series}.}  We focus on $h=12$ and the most granular level (i.e., level 6) for \acc and level 3 for \acr. This choice is based on \textcolor{black}{the importance of medium-term horizons for monetary policy decisions and forecasting results discussed above. Acknowledging the remarkable performance of lower levels of disaggregation, we present deeper insights on Albacore for level 2 in Appendix \ref{sec:sparse}.}  Figure \ref{fig:albacore_us_h12} presents Albacore estimated on data through 2019 by plotting the time series of the aggregate \textcolor{black}{and benchmarks in yearly percentage changes} (upper panel) and the weights of subcomponents/ranks in the lower panels.   \acr indicates inflation trends close to the 2\% target for the periods up to the Great Financial Crisis (GFC).  After showing slight upward pressures in 2009,  it falls below target in 2010 and remains low and stable up to 2020.  \acc deviates from both \acr and the target already in 2004.  It results in elevated inflation until 2008 before signaling downward pressures during the GFC.  From 2010 to 2019,  all core inflation measures move in accordance before diverging ahead of the Covid-19 pandemic.   Indeed,  permanent exclusion metrics (i.e., core PCE and \acc) follow headline downward in 2019 whereas the trimming-based approaches remain at the target. 

The subsequent periods, covering the pandemic-era inflation surge, demonstrate Albacore's remarkable forecasting performance out-of-sample.  We find that \acr shows upward pressures on inflation as early as mid-2020.  \acc, on the other hand, is not as timely for the initial surge but captures the turning point earlier (already after its peak in 2022m3).  \acr remains more persistent during 2022 but indicates a faster disinflationary process thereafter.  It lands at the lowest level (at 2.6\%) compared to the other measure at the end of our sample.

\vskip 0.2em
{\sc \noindent \textbf{Comparing Weights}.}  The lower panels of Figure \ref{fig:albacore_us_h12} reflect the importance of the different subcomponents (left panel) and ranks (right panel) for the aggregate.  For the left panel,  level 6 estimates have been re-aggregated back to level 2 for ease of communication.   In line with the official core inflation rate,  \acc excludes energy.  It assigns lower weight to food goods as well as clothing, housing, and recreational goods compared to the official headline rate.  Unlike core PCE,   which completely turns off food goods,  \acc shrinks it by half.  As we will see in Section \ref{sec:quantile},  only if we wish to specialize \acc for predicting low inflation risk will it be desirable to completely exclude food goods. 

Services, on the other hand,  get at least the weight they would get in the official series, with health and other services being even more important. Given that prices of services tend to be rather persistent and less prone to transitory shocks, they prove valuable in indicating the medium-term developments in inflation \citep{bils2004sticky}.  Moreover,  it is plausible that upweighting  these low-volatility components compensates for the inclusion of more volatile ones bearing leading signals,  like food goods.

The high-dimensional regression setup allows to investigate the importance of components at the disaggregated level, which brings to light some eye-catching elements. For better or worse, beer is the top-weighted component in food goods. In the group of vehicles, we find a significant role of the secondary market with increased weights for used cars and trucks prices. Personal computers, tablets, and television gain in importance when it comes to recreational goods.  So do medical insurances and prescription drugs for goods and services regarding health.  Downsizing happens for shoes and footwear, garments, gasoline, sporting equipment and air transportation.  At first sight these components do not have much in common. However, analyzing the persistence of each component's inflation rate reveals that upweighted items tend to yield higher persistence coefficients than downweighted ones.\footnote{We follow the literature on persistence weighted core inflation rates and use an AR(1) model to classify each component \citep{cutler2001core,BilkeStracca2007}.}

Looking at the weights of \acr reveals a highly asymmetric trim (see lower left panel in Figure \ref{fig:albacore_us_h12}). \textcolor{black}{Compared to the asymmetric trimming solution of the FRB Dallas' Trimmed Mean PCE (see green shaded area in Figure \ref{fig:albacore_us_h12}), \acr is much more aggressive in removing the lower part of the distribution.\footnote{Note that in commonly used trimming approaches subcomponents are reweighted with their official PCE weights, while \acr is not.} }  In Table \ref{tab:prop_us}  (in the appendix),   we see this helps in reducing both volatility and bias.   The absence of the sum-to-one constraint in Equation \eqref{eq:acr} is instrumental to this result.  By allowing rank weights to sum to any positive number,  \acr can leverage leading signals from a segment of the price growth distribution (here,  centered around the 75$^{\text{th}}$ percentile) which by construction,  grows in every period at a faster pace than 2\%,  and multiply them by a constant smaller than 1 to bring back the index to having the same long-run mean as headline inflation.  Note that this focus on the upper tail does not prevent \acr from flipping sign and entering deflation territory if need be.  This,  however,  would require for approximately 75\% of the level 6 components' prices to contract.

The benefits of highly asymmetric trimming can only be reaped if the shape of the price growth distribution is itself asymmetric and time-varying.  Otherwise,  there cannot be any comparative advantage versus simply monitoring (robust) location and scale.  The pandemic-era inflation episode affected the skewness of the short-term distribution of price changes, which shifted from negative to positive skewness.  As a results,  existing trimming approaches understated (early) trends \citep{rich2022corebias}, while \acr is particularly well equipped to catch those.  Moreover,  positive price changes are typically more persistent than negative ones.  As shown by the extensive literature on price setting behaviors of firms, prices tend to be adjusted faster when costs increase than when they decrease \citep[see, e.g., ][]{ball1994asymmetric,nakamura2008five,gautier2022new}. 
Thus,  \acr significantly upweights relative price changes that are more likely here to stay.


\vskip 0.2em
{\sc \noindent \textbf{Analysis of the Post-Covid Inflation Surge}.} We provide an in-depth analysis based on a decomposition of Albacore, the official core inflation rate and the trimmed mean (Figure \ref{fig:albacore_us_decomp}). We focus on the contributions of energy, food, goods, services, shelter, and others, where the latter is comprised of transportation and health. A detailed description of the included categories in each group is given in Appendix \ref{sec:data}.

In the US, a large bulk of the early acceleration in inflation can be attributed to goods prices. The pandemic-induced distortions in several key sectors combined with strong aggregate demand stemming from shifts in consumer spending,  accommodating monetary policy,   and fiscal stimulus led to high inflation and contributed to its exceptional nature \citep{ball2022understanding,guerrieri2022macroeconomic,blanchard2023caused,digiovanni2023quantifying}.
Clearly, \acr benefits from its focus on the upper half of the distribution. Contrary to \acc and core PCE, there is no downward pressure from goods inflation in 2020. Instead, \acr already spots signs of upward tendencies in mid-2020 that build up as more and more sectors face negative impacts from the mix of supply chain disruptions and sustained consumer spending. This stands in contrast to the trimmed mean,  which is lagging. Contributions from goods prices are muted until mid-2022 and remain comparably small throughout the sample. Instead, it attaches high weight to shelter; a notoriously lagging indicator \citep{bryan2010some,cotton2023forecasting}. \textcolor{black}{The stock of rental leases included in the index typically captures new and existing rentals, with the latter being sluggish in adjusting to market rents. Based on these grounds but without a priori excluding shelter components, as in US supercore inflation \citep[i.e., nonhousing core services, see ][]{powell2022,leduc2023will}, \acr benefits from the minimal contribution of shelter to the aggregate, on account of its highly asymmetric trimming scheme. }

Although, a substantial part of the pandemic-era inflation surge also reflects shocks in food and energy prices \citep{blanchard2023caused,gagliardone2023oil}, their contributions in all core inflation measures are small. Contrary to core PCE, neither \acc nor \acr exclude food or energy prices a priori, and yet, they both assign low weights to corresponding components, resulting in negligible contributions from energy and small, positive ones from food (given their exceptionally strong dynamics).

The deceleration of core inflation metrics in 2022 reflects fading supply chain disruption, accompanied by a moderation of goods inflation \citep{alcedo2022commerce,gagliardone2023oil}. All measures, except for the FRB Dallas' Trimmed Mean PCE, peak in 2022m3. Yet, \acc is the only measure indicating a clear turning point with goods inflation significantly decreasing thereafter.  Core PCE as well as \acr are persistent throughout 2022 before trending downward towards the end of the year. 
While goods inflation is decreasing in importance, pressures from the labor market become more dominant \citep{amiti2022pass,benigno2023pc,blanchard2023caused}. This is indicated by the rise of services inflation and its high persistence (since service sectors are typically labor-intensive), which is mainly observed for \acc and core PCE. Given that components related to services are getting higher weights in \acc, it proves to be a useful indicator for the second phase of the post-pandemic inflation era \citep{leduc2023will}.

\vskip 0.2em
{\sc \noindent \textbf{Robustness to Moving Average Choice}.} The choice of 3 months averages is motivated as being the maximal amount of time series averaging one can indulge in when building a measure of \textit{current} conditions in a forward-looking manner.   Nonetheless,  as a robustness check, we present results for different moving average transformations of the regressors (from month-over-month to year-over-year) in Tables \ref{tab:results_us_popALL1} and \ref{tab:results_us_popALL2} in the appendix.  As one would expect, \textcolor{black}{overall, we observe} that longer trailing averages perform better during calmer periods and worse during the post-2020 era.  However,  improvements  of year-over-year averages,  when applicable,  are rather quaint vis-à-vis the mechanically more timely 3 months average.   
\textcolor{black}{Interestingly, \acr yields  remarkably stable performance across transformations (from month-over-month to year-over-year) for both test sets,  as long as the level of disaggregation is deep enough ($K \geq 50$). So does \acc for all cases but the highly volatile month-over-month transformation for the pre-Covid sample.}
This suggests that when there is enough leeway for variance reduction  in the "cross-section" \textcolor{black}{of components},  additional time-smoothing has limited effects. We see this as desirable,  as averaging over long windows by necessity,  from a forward-looking perspective,  errs on the increased bias side of the time aggregation bias-variance trade-off. Variation in benchmarks performance is limited before 2020 and larger thereafter,  with a general preference for less timely averages.  All in all,  we find  \acc and \acr  in their original 3 months average setup to fare well overall,  even in the presence of a deluge of benchmarks and alternative specifications.

\subsection{Results for the Euro Area} \label{sec:earesults}

\begin{figure}[t!]
    \caption{\normalsize{Albacore for the Euro Area}} \label{fig:albacore_ea_h12}
  \begin{center}

    \begin{subfigure}[t]{\textwidth}
    \vspace*{-0.7cm}
        \centering
        \includegraphics[width=0.98\textwidth, trim = 0mm -10mm 0mm 0mm, clip]{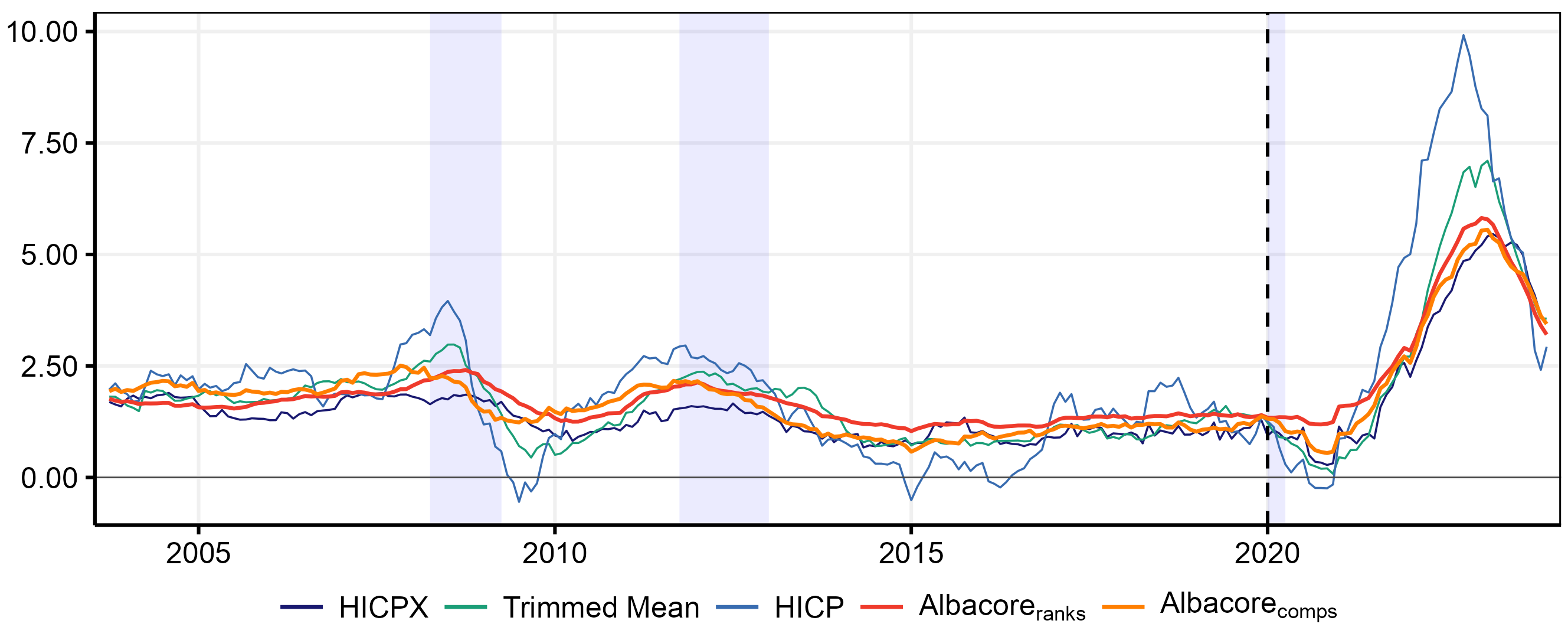}
    \end{subfigure}%

    \begin{subfigure}[t]{0.5\textwidth}
    \vspace*{-0.05cm}
        \centering
        \includegraphics[width=\textwidth, trim = -6mm -20mm 0mm 0mm, clip]{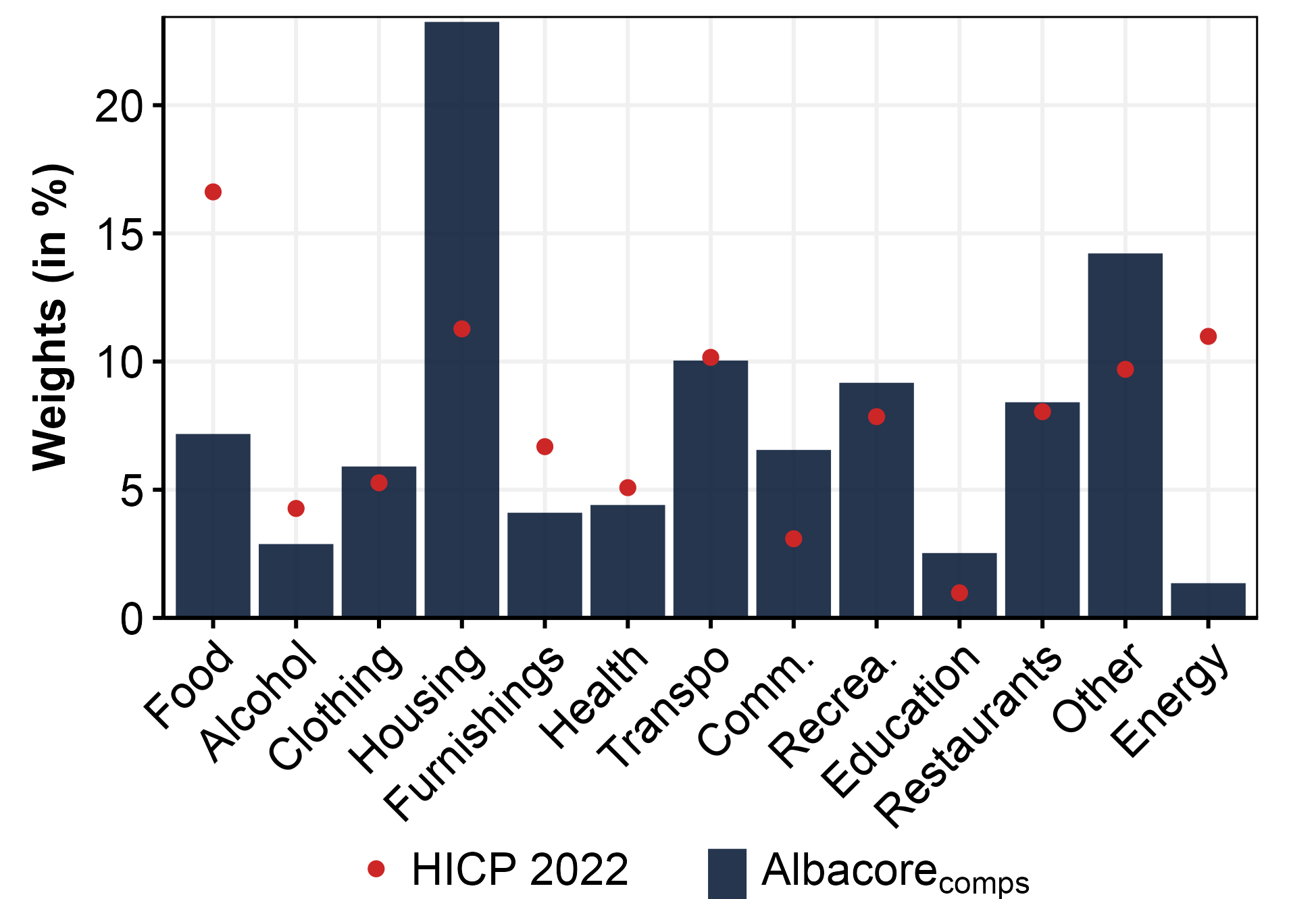}
        \vspace*{-0.7cm}
        \caption{Comparison of Component Weights}
    \end{subfigure}%
    \begin{subfigure}[t]{0.5\textwidth}
    \vspace*{-0.48cm}
        \centering
        \hspace*{0.06cm}
        \includegraphics[width=\textwidth, trim = 0mm -30mm 0mm 0mm, clip]{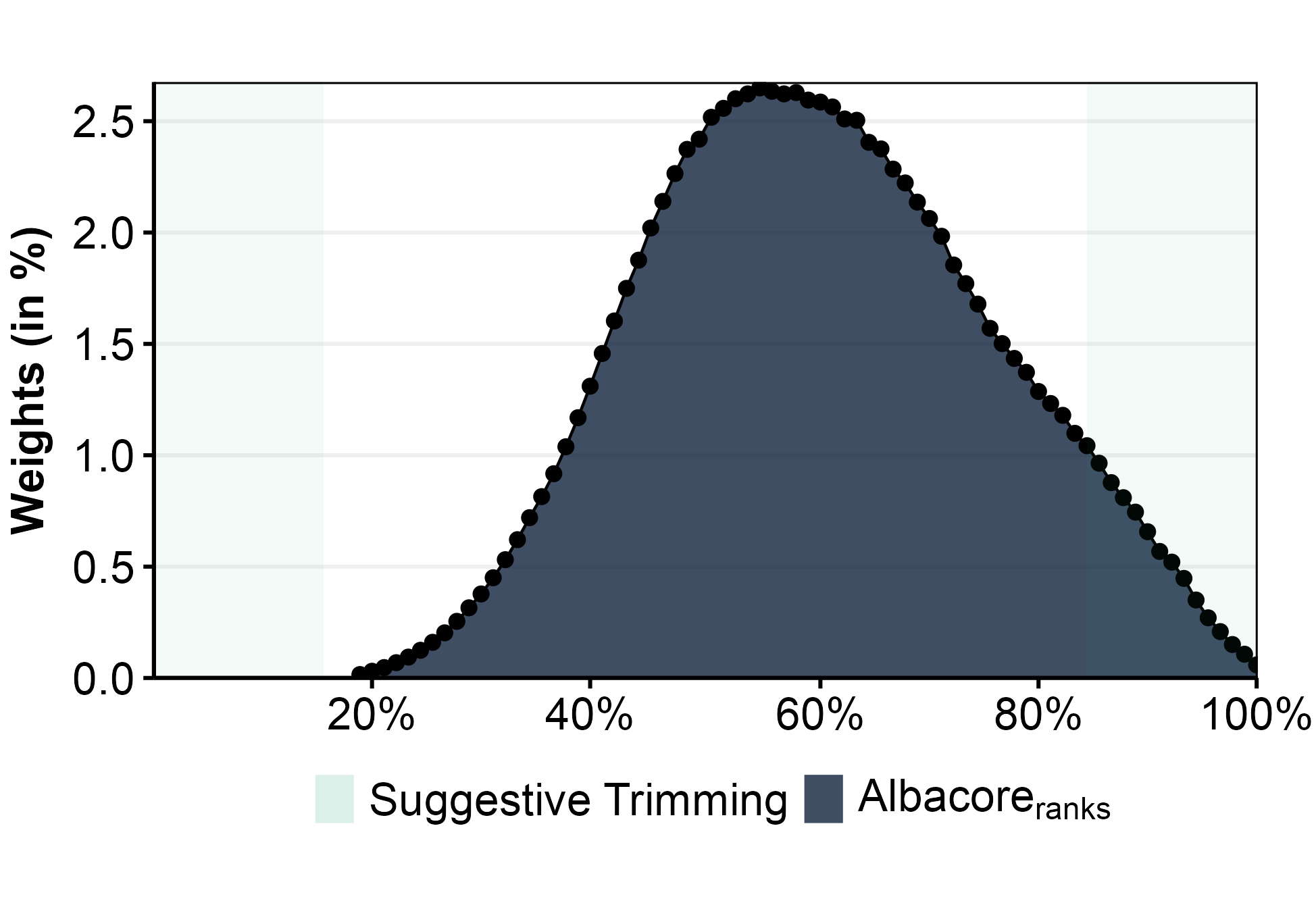}
        \vspace*{-0.67cm}
        \caption{Comparison of Rank Weights}
    \end{subfigure}

  \end{center}

    \begin{threeparttable}
    \centering
    \begin{minipage}{\textwidth}
    \vspace*{-0.6cm}
      \begin{tablenotes}[para,flushleft]
    \setlength{\lineskip}{0.2ex}
    \notsotiny 
  {\textit{Notes}: For more details we refer to Figure \ref{fig:albacore_us_h12}. Country-specific changes can be summarized as follows: For the EA we present Albacore in level 4. For the trimmed mean, we choose the officially reported 30\% trimmed mean inflation. The green shaded area marks the excluded share of monthly price changes as is suggested by the trimmed mean before reweighting (15\% of the lower tail and 15\% of the upper tail). Details on the labeling can be found in Appendix \ref{sec:data}.}
    \end{tablenotes}
  \end{minipage}
  \end{threeparttable}
\end{figure}

\begin{figure}[t!]
    \caption{\normalsize{Albacore Decomposition for the Euro Area}} \label{fig:albacore_ea_decomp}
    \vspace*{-0.2cm}
    \centering
    \includegraphics[width=\textwidth, trim = 0mm 0mm 0mm 0mm, clip]{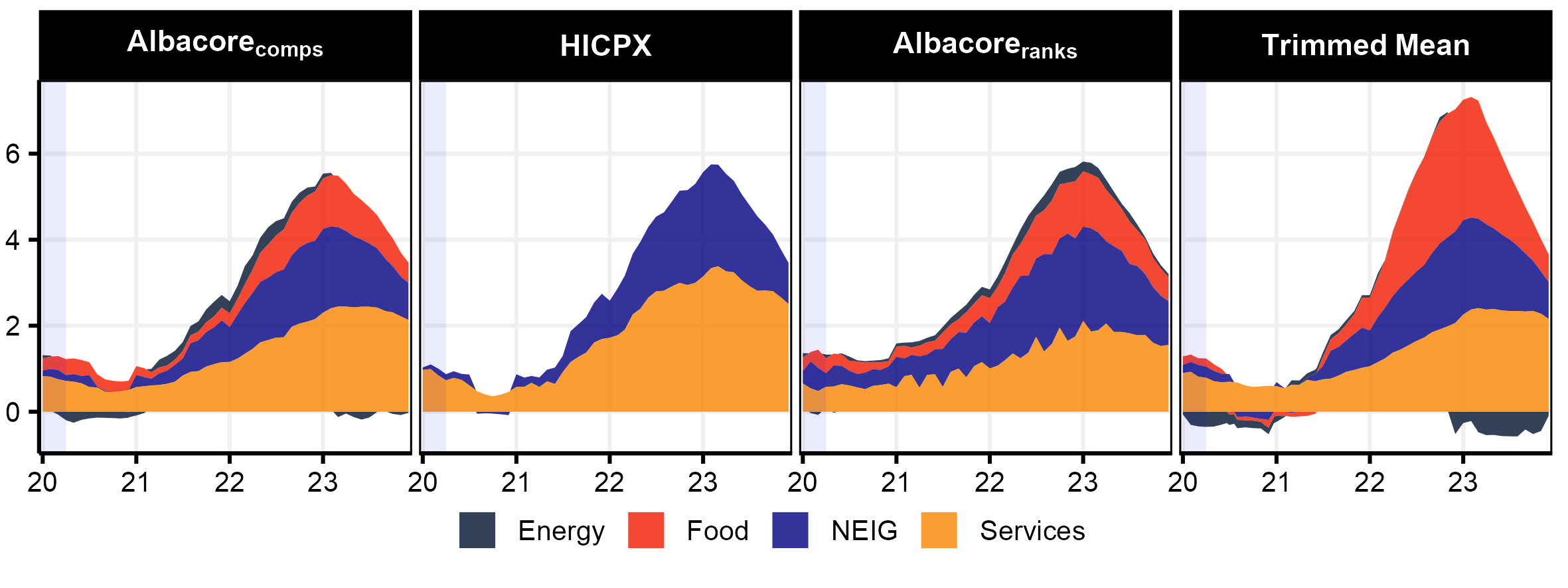}
  \begin{minipage}{\textwidth}
  \vspace*{-0.45cm}
    \begin{tablenotes}[para,flushleft]
      \setlength{\lineskip}{0.2ex}
      \notsotiny 
      {\textit{Notes}: We decompose each inflation series into a common lower level of four main aggregates. For Albacore$_\text{comps}$ ($h=12$ and level 4) and HICPX this is done by aggregating subcomponents to the four groups: energy, food, non-energy indusrial goods (NEIG), and services (as is suggested by Eurostat's ``Special Aggregates'' in COICOP classification, see Appendix \ref{sec:data}). Albacore$_\text{ranks}$ ($h=12$ and level 4) and the trimmed mean require a conversion to component space as discussed in Section \ref{sec:sts}.}
    \end{tablenotes}
  \end{minipage}
\end{figure}

In this section, we construct Albacore for the euro area. Again, we compare the model's forecasting performance with that of the most prominent core inflation series (Table \ref{tab:results_ea_pop3} in the appendix), provide shape and features of the proposed measure (Figure \ref{fig:albacore_ea_h12}) and delve into the recent surge (Figure \ref{fig:albacore_ea_decomp}). 

\vskip 0.2em
{\sc \noindent \textbf{Forecasting Performance}.}   Similar to our findings for the US, benchmarks are more difficult to beat in the short run. Table \ref{tab:results_ea_pop3} reveals that the convex combination of existing core inflation measures ($\boldsymbol{X}_t^{\text{bm+}}$) features competitive predictive power that either Albacores can match,  but struggle to surpass.  For medium-term forecasts,  we find that Albacore improves upon its competitors and yields the highest predictive accuracy for both evaluation samples.  However,  it does so by smaller margins than it did for the US,  which is not surprising given that a more significant part of EA headline inflation was attributable to, from the perspective of the model,  fundamentally unpredictable events. For short horizons, \acc tops the list with the lowest level of disaggregation (level 2). When it comes to predictive accuracy for medium-term horizons,  we find larger gains for \acr assembling a higher number of components (level 4).


\vskip 0.2em
{\sc \noindent \textbf{Comparing Weights}.}   In light of the results discussed so far, our in-depth analysis will be based on $h=12$,  the most granular level (i.e., level 4),  and the training sample ending in 2019m12.  We find that both \acc and \acr show a rather smooth path, moving close to the 2\% target.  In response to the two major crises in our sample, the GFC and the sovereign debt crisis, both series follow a downward trend with \acc indicating this drift earlier and moving below target for several years. \acr remains slightly higher (close to target) for the low inflation era preceding the Covid-19 crisis.  From 2020 onward,   \acr again showcase its quality as a leading indicator.  It does not give in to the deflationary tendencies at the onset of the pandemic and heeds high inflation warnings at an early stage.  Even though, \acc follows the common downward trend in 2020, it catches up quickly and both Albacore series plateau end-2022 (from 2022m12 to 2023m2), providing timely indication of the slowdown.

Similar to the US, \acc assigns low weight to energy and only half the weight to food and alcoholic beverages compared to the official HICP aggregate (see lower left panel in Figure \ref{fig:albacore_ea_h12}). Components getting higher weights are found in housing (in particular, rents, services and goods related to the dwellings), other services, and communication, which all feature  persistent behavior. \acr shows an asymmetric trim tilted to the upside of the distribution (see lower right panel in Figure \ref{fig:albacore_ea_h12}). This prevents \acr from decreasing in the aftermath of the initial Covid-19 shock and signals upward pressures before all other core inflation measures do. 

\vskip 0.2em
{\sc \noindent \textbf{Analysis of the Post-Covid Inflation Surge}.} We focus on Albacore versus HICPX and the 30\% trimmed mean and provide the contributions of energy, food, non-energy industrial goods (NEIG), and services to the aggregate in the decomposition exercise.\footnote{These groups are based on the ``Special Aggregates'' as suggested by Eurostat in the COICOP classification. Details are given in Appendix \ref{sec:data}.} Commodity prices, most importantly food and energy, accounted for the majority of the post-pandemic inflation acceleration in the EA. 
The standard argument that large swings in food and energy prices are mainly due to transitory shocks and tend to be quickly reversed, did not apply for the exceptional periods in 2021/2022.  Since many existing core inflation measures ignore commodity price shocks by construction, they captured very little of the  sizable and persistent increase in prices that ensued.  For both \acc and \acr we find positive contributions of food components throughout the sample (see Figure \ref{fig:albacore_ea_decomp}), which aligns with, e.g.,  \citet{peersman2022foodinflation}, highlighting the importance of global food commodity price shocks for euro area inflation dynamics in the medium term.

However,  focusing solely on energy and food would be too narrow \citep{nickel2022inflation,ascari2023euro,banbura2023drives}. Soaring goods inflation as a response to constrained supply and excess demand, put sizable upward pressures on inflation. \acc captures the increase via high weights on housing (including heavily affected durable goods) and transport goods, while \acr accounts for the upward pressure due to its focus on the middle-upper part of the price distribution.  The downward trend indicated by Albacore from 2023m1 onwards forestalls that of HICPX and the trimmed mean.  It is based on the easing of energy price pressures as well as goods inflation.  Services inflation, on the other hand, remains strong for all measures under study,  a reflection of  mounting domestic price pressures  \citep{banbura2023drives}.  For \acc, these persistent pressures mainly stem from rents, restaurants, and hotel services, as well as other services (e.g., insurance and financial services).

\subsection{From One Space to Another}\label{sec:sts}


A natural worry with trimming-based measures is that an item that we would deem important in component space,  shows up rarely in the trimmed  indicator.  This can happen if an important component separates from the pack (e.g.,  rent in Canada for 2022-2023),  potentially opening a rift between trimming-based (or median) inflation and headline.  We have seen, from earlier derivations,  that 
$$ \pi_{\text{ranks},t}^* =  \hat{\boldsymbol{w}}_r'\boldsymbol{O}_t  = \underbrace{{\hat{\boldsymbol{w}}_r' \boldsymbol{A}_t}}_{\boldsymbol{w}_{c,t}} \boldsymbol{\Pi}_{t}   \quad  \quad \text{and conversely}  \quad  \quad   \pi_{\text{comps},t}^* =  \hat{\boldsymbol{w}}_c'\boldsymbol{\Pi}_t  = \underbrace{{\hat{\boldsymbol{w}}_c' \boldsymbol{A}_t^{-1}}}_{\boldsymbol{w}_{r,t}} \boldsymbol{O}_{t}$$
and thereby, we can study \acr in components space and \acc in rank space -- as time-varying weights models.  


\begin{figure}[t!]
  \caption{\normalsize{Space-to-Space Results}} \label{fig:albacore_us_sts}
  
  \begin{center}
    
    \vspace*{-0.8cm}
    \begin{subfigure}[t]{0.5\textwidth}
      \centering
      \includegraphics[width=\textwidth, trim = 0mm -42mm 0mm 0mm, clip]{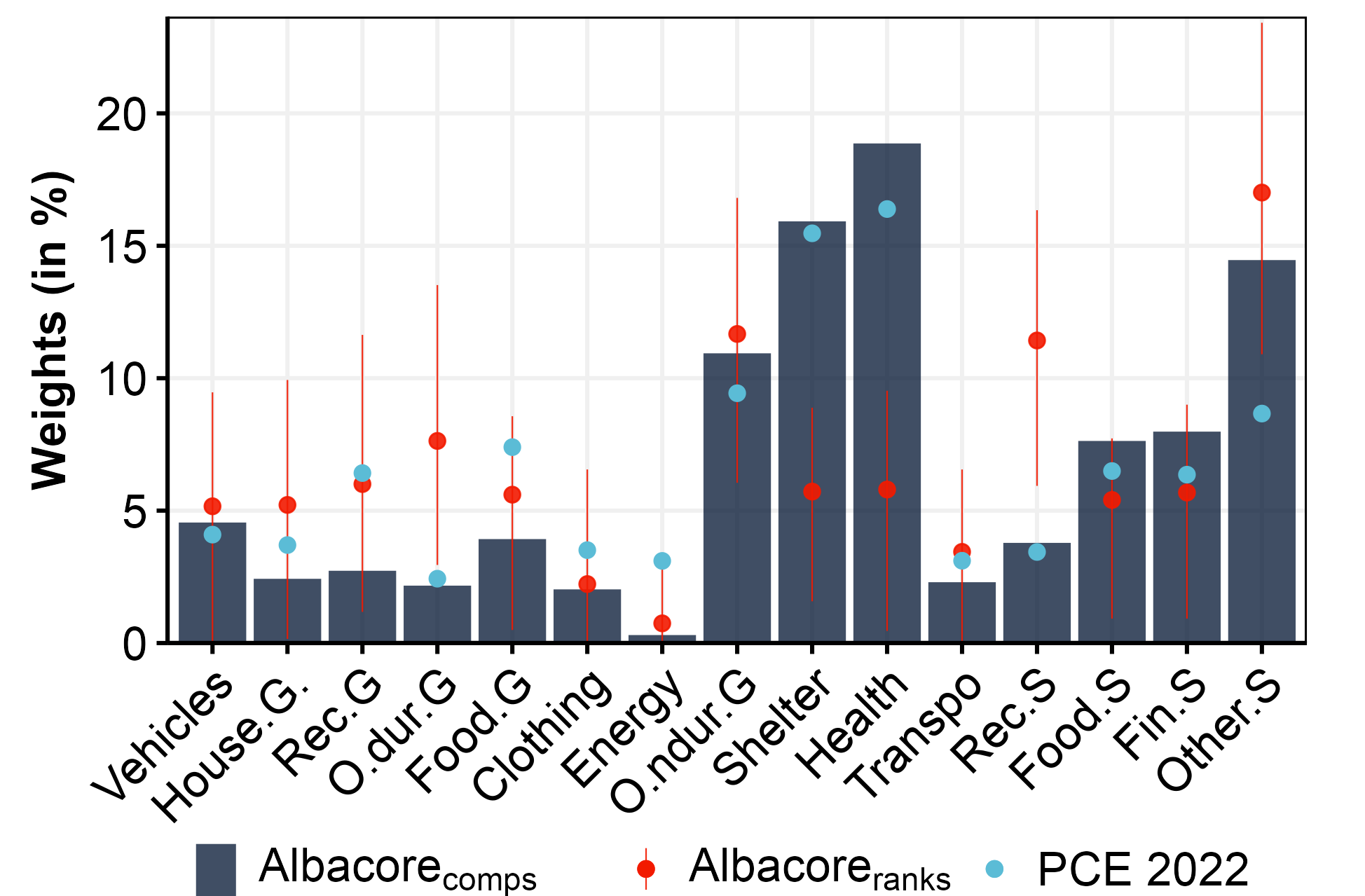}
      \vspace*{-1.1cm}
      \caption{\textbf{US} -- Comparison in Component Space}
    \end{subfigure}%
    \begin{subfigure}[t]{0.5\textwidth}
      \centering
      \hspace*{0.05cm}
      \includegraphics[width=\textwidth, trim = 0mm -25mm 0mm 0mm, clip]{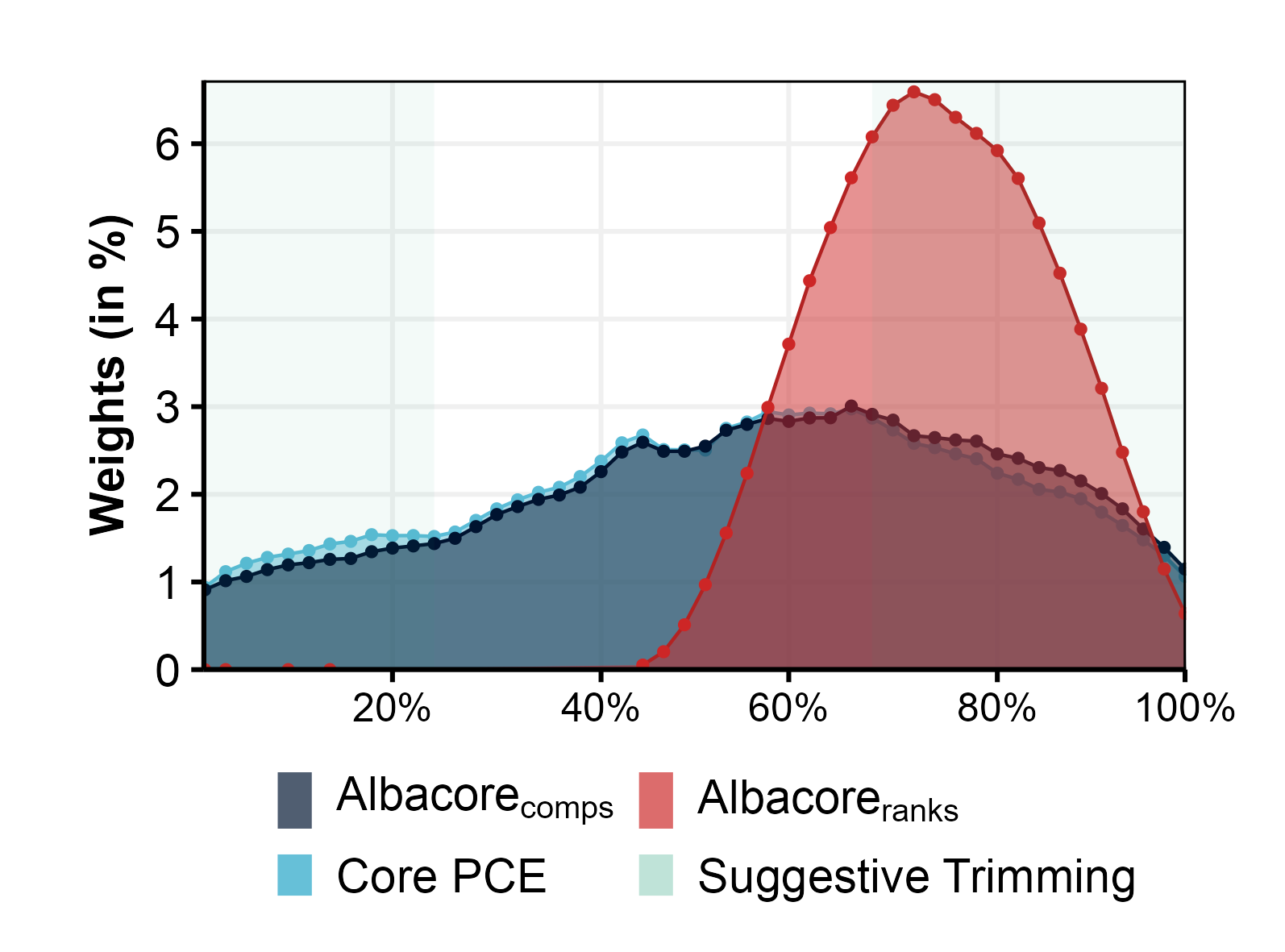}
      \vspace*{-1.1cm}
      \caption{\textbf{US} -- Comparison in Rank Space}
    \end{subfigure}
    
    \begin{subfigure}[t]{0.5\textwidth}
      \centering
      \hspace*{-0.1cm}
      \includegraphics[width=\textwidth, trim = 0mm -32mm 0mm 0mm, clip]{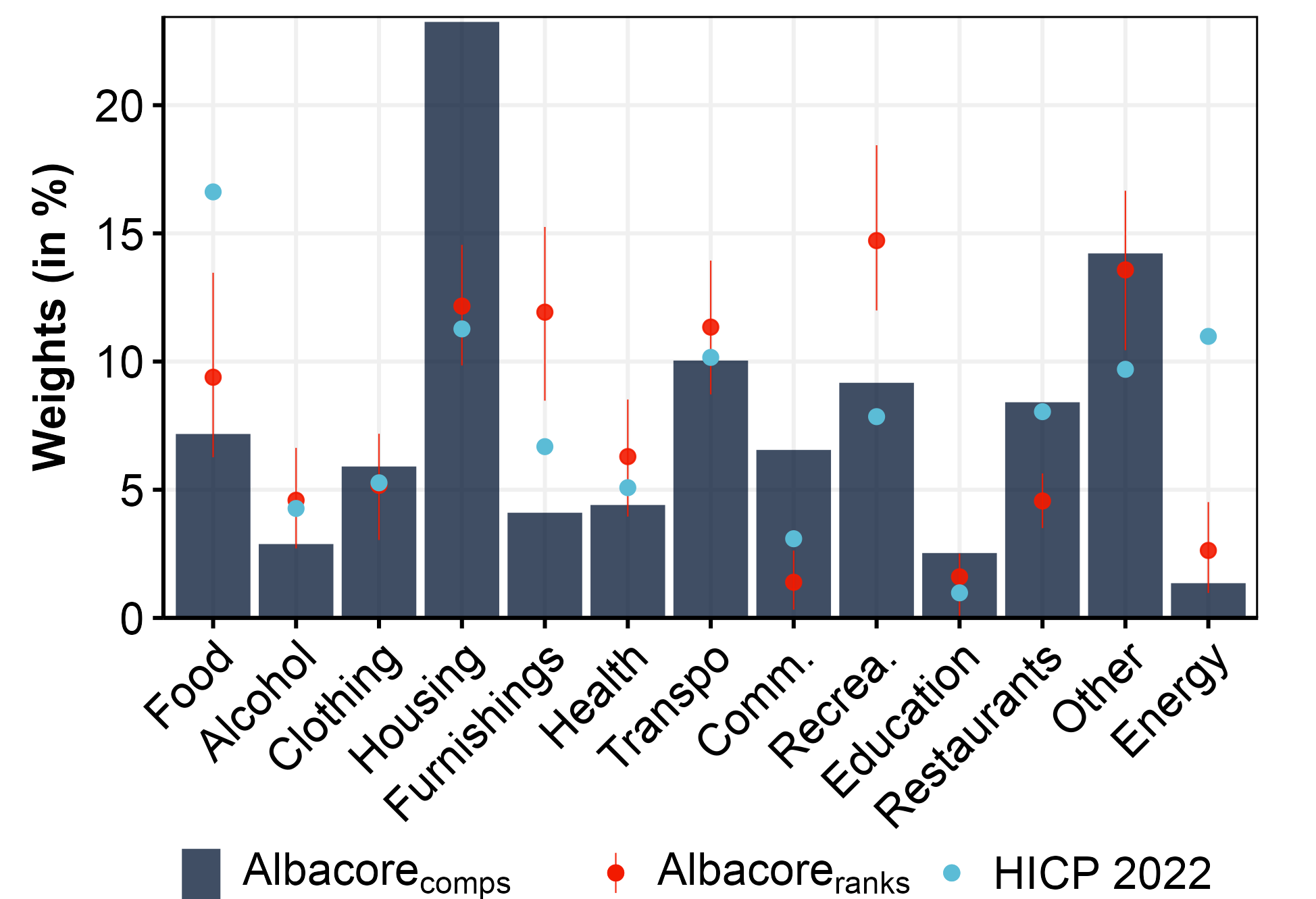}
      \vspace*{-0.9cm}
      \caption{\textbf{EA} -- Comparison in Component Space}
    \end{subfigure}%
    \begin{subfigure}[t]{0.5\textwidth}
      \centering
      \includegraphics[width=\textwidth, trim = 0mm -25mm 0mm 0mm, clip]{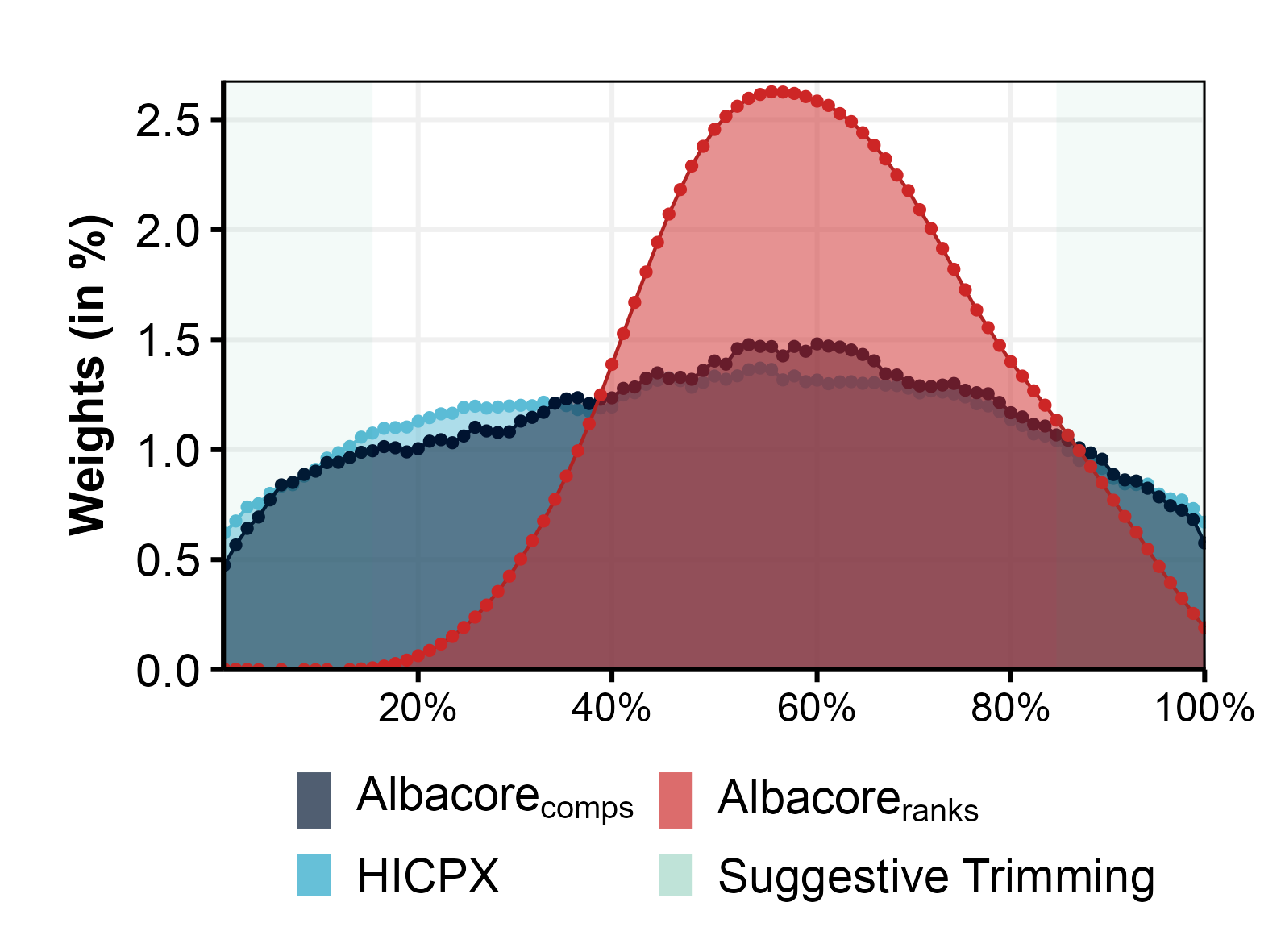}
      \vspace*{-0.9cm}
      \caption{\textbf{EA} -- Comparison in Rank Space}
    \end{subfigure}

  \end{center}

    \begin{threeparttable}
    \centering
    \begin{minipage}{\textwidth}
    \vspace*{-0.5cm}
      \begin{tablenotes}[para,flushleft]
    \setlength{\lineskip}{0.2ex}
    \notsotiny 
  {\textit{Notes}: All weights are based on models trained from 2001m1 to 2019m12. For the US, we present level 6 for Albacore$_\text{comps}$ and level 3 for Albacore$_\text{ranks}$. For the EA, both Albacore series are in level 4. In the left panel we aggregate the component weights to the lowest level of disaggregation (level 2) for each model. For Albacore$_\text{ranks}$ we show the average weight over time for each component with the red dots including the dispersion with the red, solid line. In the right panel we use a sliding average over two ranks to smooth the weights in rank space.}
    \end{tablenotes}
  \end{minipage}
  \end{threeparttable}
\end{figure}

\textcolor{black}{We report  results in  Figure \ref{fig:albacore_us_sts}.   In the left panel,   we conduct the comparison in components space.  This includes  \acc, the official PCE rate,  and \acr. For the latter,  we present $\boldsymbol{w}_{c,t}$ through its median (bullet) as well as the 16$^{\text{th}}$ and 84$^{\text{th}}$ quantiles (lines). Overall, we find that in the EA both Albacore versions are broadly consistent regarding weights assigned to subcomponents, while in the US the two models show a higher degree of disagreement. Most notably, \acr frequently excludes elements of health and shelter, likely due to their stickiness and their concentration in the middle of the distribution  \citep{bils2004sticky}. As pointed out in Section \ref{sec:usresults}, this supports forward-looking qualities of \acr since it devotes less attention to indicators with a rearview perspective.}

\textcolor{black}{Higher weights are assigned to recreation services and other durable goods. Both components exhibit high cyclical variation (where for the latter this is mainly attributable to jewelry and watches) and thereby provide early indication of domestic inflationary pressures  \citep{dolmas2009excluding,stock2020slack}. In the EA, we get similar results with recreational services and goods proving to frequently show up in the included part of the distribution. In line with findings in the US, their cyclical sensitivity is found to be strong and indicate price changes based on domestic business cycle movements \citep{frohling2011supercore}. In both regions energy gets trimmed out most of the time given its tendency to be located in the tails.}

The right panels of Figure \ref{fig:albacore_us_sts} present \acr, \acc, and the core PCE/HICPX rate in rank space.  This,  interestingly,  provides a view of the implied trimming of components space-based estimators.  As expected,  \acc and the official core inflation rate cover the whole distribution of price changes over time,  as they are not explicitly designed to remove a particular range of ranks.  Notably,  both solutions are  left-skewed,  with the extent of it being greater for the US.   Thus,  permanent exclusion choices from  \acc and core PCE  (like the absence of energy) implies  a more aggressive downweighing of the lower part of the price growth distribution -- and a subtle focus on the 60-70 percentiles region.  For the EA,  \acr accentuates the peak of the distribution at its original location,  whereas,  in the case of US,   \acr provides a concentrated mass located slightly to the left of the core PCE's highest weighted rank.  Therefore,  \acr seizes the opportunity to significantly amplify what components-space measures are doing only in a modest fashion.

\subsection{Weights Curves: A Look Across Forecasting Horizons}\label{sec:wcurves}

The composition of the maximally forward-looking core inflation measure will differ depending on the chosen forecasting horizon.  In this paper,  we have  focused on horizons typically of interest for the conduct of monetary policy.  But shorter (or longer) horizons are of great interest to many institutions.  In this subsection,  we investigate how \acc and \acr definitions adjust as we vary $h$.  As we can see in Tables \ref{tab:results_us_pop3} and \ref{tab:results_ea_pop3},  assemblage regression performs well at various forecasting horizons.

\begin{figure}[t!]
  \caption{\normalsize{Weights Curve for the US}} \label{fig:albacore_weights_curve}
  \begin{center}
    
    \begin{subfigure}[t]{0.5\textwidth}
      \vspace*{-0.75cm}
      \centering
      \includegraphics[width=0.98\textwidth, trim = 0mm 0mm 0mm 0mm, clip]{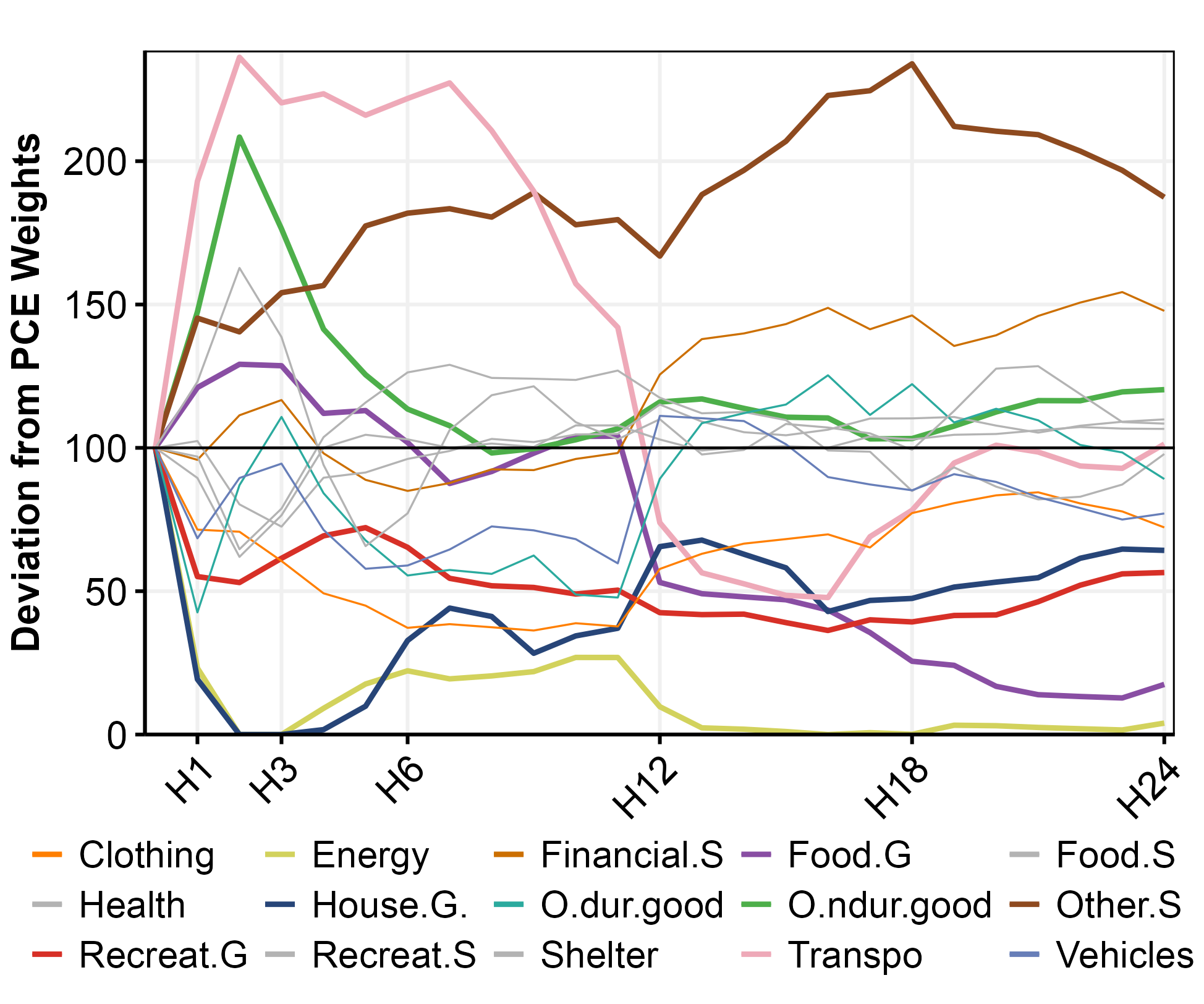}
      \vspace*{0.1cm}
      \caption{Albacore$_\text{comps}$}
    \end{subfigure}%
      \begin{subfigure}[t]{0.5\textwidth}
      \vspace*{-0.93cm}
      \centering
      \includegraphics[width=0.98\textwidth, trim = 0mm 0mm 0mm 0mm, clip]{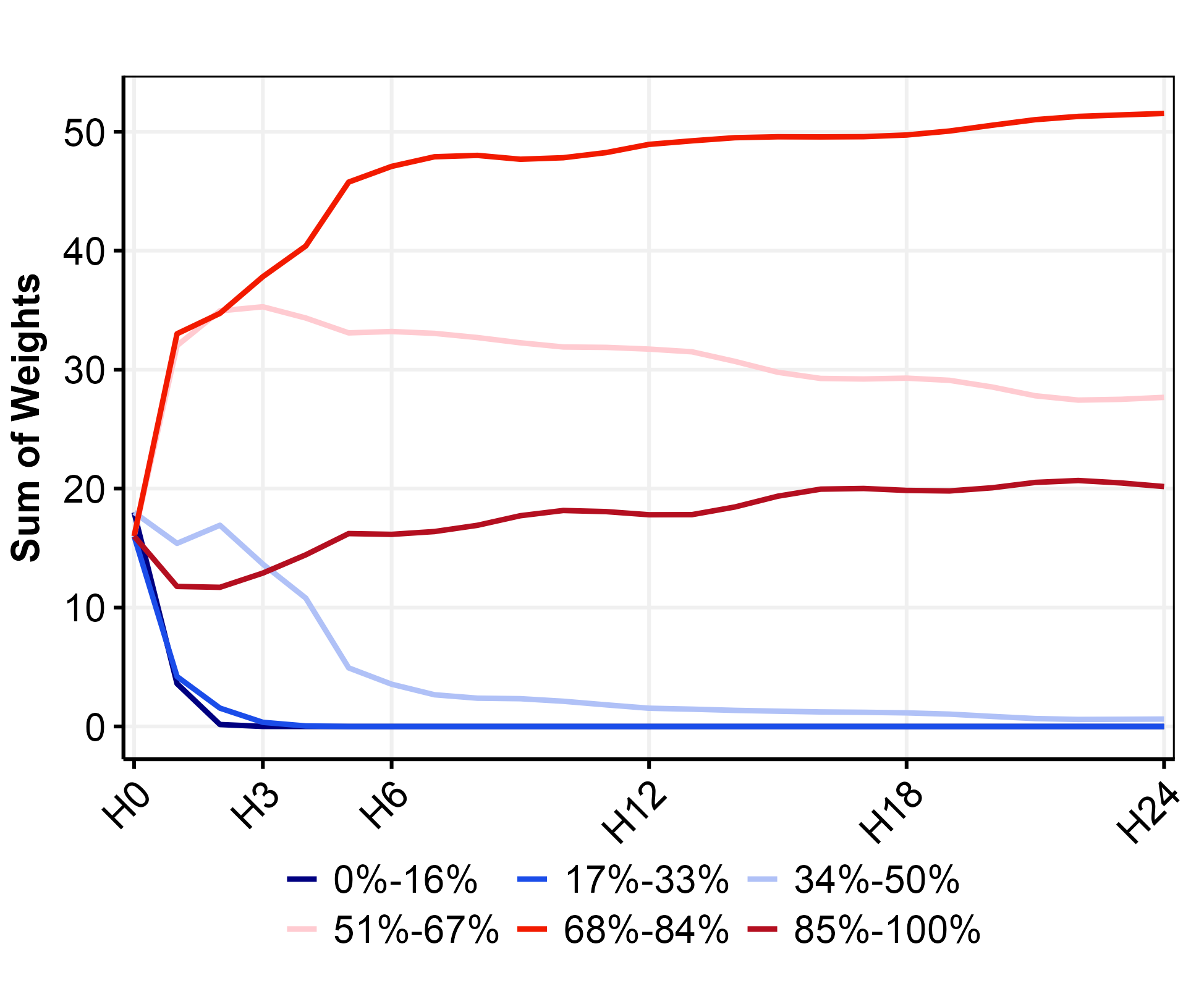}
      \vspace*{0.26cm}
      \caption{Albacore$_\text{ranks}$}
    \end{subfigure}%
    
    \begin{threeparttable}
    \centering
    \begin{minipage}{\textwidth}
    \vspace*{-0.1cm}
      \begin{tablenotes}[para,flushleft]
    \setlength{\lineskip}{0.2ex}
    \notsotiny 
  {\textit{Notes}: All weights are based on models trained from 2001m1 to 2019m12. We present level 6 for Albacore$_\text{comps}$ and level 3 for Albacore$_\text{ranks}$. 
  In the left panel, we show deviations of the component weights across varying forecasting horizons. A value of 100 indicates \textit{no} deviation, a value of 0 means a complete omission of the component, and a value of 200 indicates a doubling of the weight in Albacore$_\text{comps}$ relative to the official PCE weight.
  In the right panel, we show the aggregate weights for certain buckets of the rank distribution across various forecasting horizons.}
    \end{tablenotes}
  \end{minipage}
  \end{threeparttable}
    
  \end{center}

\end{figure}

Figure \ref{fig:albacore_weights_curve} reports weights of  \acc and \acr as $h$ increases from 1 to 24.  In the left panel,  we report \acc's weights (as a function of $h$) relative to those of headline PCE.   First,  let us focus on energy and food goods,  which are excluded from prototypical core measures.  We see a rapid decline in the importance of energy.  Energy components get close to 0 weight across all horizons, being entirely excluded for higher-order forecasts.  Food goods get initially upweighted at short horizons,   before decreasing steadily and converging towards 0 for $h>18$ months.  This suggests that food goods, despite their volatile behavior,  are at least indicative of short-run price trends.  

The services share of \acc is ever increasing in $h$.  This is visible for other services,  and also financial services which takes off after  $h=12$.   Some components experience the reverse pattern,  being more critical in the short term and becoming less significant over longer periods.  This symmetric lineup change is obviously inevitable by the sum-to-one constraint:  if some components come in,  others must go.  Among the latter,  we have other nondurable goods and transportation, which experience the largest relative increase for the first few forecasting horizons and then eventually decrease to the same level as the official rate (other nondurable goods) or below (transportation).
\textcolor{black}{The rationale for the weight curves of transportation and other nondurables can be drawn from their connection to commodity prices. Among PCE categories, energy prices mainly influence items related to fuel and transportation, while higher food commodity prices (e.g., crop) feed through to nondurables based on their high input shares \citep{hobijn2008commodity,gao2014oil}.\footnote{Note that crops are not only an important input for food goods but also for other nondurables since the latter include tobacco products, flowers, pots, and plants \citep{hobijn2008commodity}.} This rather fast passthrough of commodity prices makes transportation and nondurables better predictors for near-term developments rather than longer-term trends. 
Overall, we trace the slope of the weight curves to the components' characteristics with more persistent components being relevant for higher-order forecasts and faster-growing and/or more volatile ones contributing to the predictability in the short run.}

The right panel of Figure \ref{fig:albacore_weights_curve} presents the evolution of rank weights across horizons for \acr by grouping them into 6 "buckets" for visibility.  Results reveal that the lower percentiles are trimmed out early on,  echoing the exclusion of energy in the left panel.  The bucket covering the 34$^{th}$ to 50$^{th}$ percentile gets small positive weight for short horizons before approaching 0 from $h \geq 5$ onwards.  Highest weights throughout are assigned to the 68$^{th}$ to 84$^{th}$ percentiles followed by the 51$^{th}$ to 67$^{th}$ percentiles.  While the latter decreases in importance with increasing $h$, the former shows an upward sloping weight curve.  Similarly, the right tail of the distribution gets positive weight, which grows in $h$ and eventually converges to nearly twice its share from $h=1$.   This corroborates the view that its \textit{a-priori} exclusion leads to a sizable loss of information about longer-run inflation trends.

\subsection{Building Core Series Specialized for Inflation Tail Risk} \label{sec:quantile}

Overestimating and underestimating core inflation have different costs. Therefore,  it can be of interest to build core measures specialized  at foreshadowing either of the two key risks:  deflation and high inflation.  We do so by running quantile assemblage regression, which, as the name suggests,  now assembles components (or ranks) to minimize a quantile loss.\footnote{Following from a long tradition in financial risk management,  it is now widespread to forecast tail risks to macroeconomic variables.  \cite{adrian2019vulnerable} initiated a literature using quantile regression methods for forecasting tail risks to the real activity outlook \citep[see, e.g., ][]{giglio2016systemic,ferrara2022high,delle2023modeling,clark2024gar}. Of course,  some have also build analogous models for inflation,  which own palette of risks is of interest to central bankers,  investors,  governments, and others \citep[see, e.g., ][]{korobilis2017quantile,ghysels2018quantile,pfarrhofer2022modeling,lenza2023density}.} Focusing on rare events inevitably leads to a reduced effective number of training observations.  So that this number is not further reduced by significant overlapping of long averages for the target (e.g., $h \in \{12, 24\}$),  we now turn our attention to 3 and 6 steps ahead forecasts instead.  Moreover, we drop the 20 years rolling window in favor of an expanding window from 1990.  We predict the 15$^{th}$ quantile for the lower tail risks and the 85$^{th}$ quantile for the upper tail risks (i.e., $\tau \in \{0.15, 0.85\}$) for level 6 in case of \qacc and level 3 in case of \qacr. We modify Equations (\ref{eq:acc}) and (\ref{eq:acr}), respectively, to
\begin{align}\label{eq:qacc}
\hat{\boldsymbol{w}}_{c}^{\tau} =&  \argmin_{\boldsymbol{w}} \sum_{t=1}^{T-h} \rho_{\tau} ( \pi_{t+1:t+h} - \boldsymbol{w}'\boldsymbol{\Pi}_{t})^2 + \lambda ||\boldsymbol{w}-\boldsymbol{w}_{\text{\tiny headline}}||_2 \quad  \text{st} \enskip \boldsymbol{w} \geq 0,   \enskip \boldsymbol{w}'\iota=1 \times \frac{Q_{\tau}({\pi}_{t+1:t+h})}{\bar{\pi}_{t+1:t+h}} \\
\hat{\boldsymbol{w}}_{r}^{\tau} =& \argmin_{\boldsymbol{w}} \sum_{t=1}^{T-h} \rho_{\tau} ( \pi_{t+1:t+h} - \boldsymbol{w}'\boldsymbol{O}_{t})^2 + \lambda ||D\boldsymbol{w}||_2 \quad  \text{st} \enskip \boldsymbol{w} \geq 0 ,   \enskip   \bar{ \pi}^*_{\text{ranks}, t} = Q_{\tau}({\pi}_{t+1:t+h}) 
\end{align}
where the quantile loss function is given by $\rho_{\tau}(u) = (\tau - \mathbbm{1}\{u \leq 0 \}) u$ with $\mathbbm{1}\{u \leq 0 \}$ defining an indicator function, which returns the value 1 if $u \leq 0$ and 0 otherwise and $u$ denoting the error term \citep{gneiting2011making}.  {$Q_{\tau}\left(\cdot\right)$ calculates the empirical value of the unconditional quantile $\tau$.} Some modifications to our original equality constraints are necessary,  since they are not suitable for their quantile extensions.  Summing weights to 1 in case of \qacc and setting the long-run average of \qacr to that of headline inflation would prevent the algorithm from upweighting tail observations.  Rather,  we now want those constraints to hold up to a prefixed multiplicative constant $\left( \frac{Q_{\tau}({\pi}_{t+1:t+h})}{\bar{\pi}_{t+1:t+h}}\right)$, reflecting the difference in the long-run level of quantile $\tau$ versus the mean.    Thus, we modify the equality constraint for \qacc such that weights sum to our target's respective unconditional empirical quantile ($Q_{\tau}({\pi}_{t+1:t+h})$) relative to its sample mean.  For  \qacr,   this boils down to its training sample average being equal to that of the respective unconditional quantile of headline inflation $Q_{\tau}({\pi}_{t+1:t+h})$ instead of $\bar{\pi}_{t+1:t+h}$ as used in \acr.  

\begin{table}[t!]
  \footnotesize
  \centering
  \begin{threeparttable}
    \caption{\normalsize {Quantile Forecasting Performance of Albacore for the US} \label{tab:results_us_quant}
      \vspace{-0.3cm}}
    \setlength{\tabcolsep}{0.65em}
    \setlength\extrarowheight{2.5pt}
    \begin{tabular}{l| rrrlrrr|rrrlrrr}
      \toprule \toprule
      \addlinespace[2pt]
      &  \multicolumn{7}{c}{2010m1-2019m12}  & \multicolumn{7}{c}{2020m1-2023m12}  \\
      \cmidrule(lr){2-8} \cmidrule(lr){9-15}
      & \multicolumn{3}{c}{$h=3$} && \multicolumn{3}{c}{$h=6$}& \multicolumn{3}{c}{$h=3$} && \multicolumn{3}{c}{$h=6$}\\
      \cmidrule(lr){2-4} \cmidrule(lr){6-8} \cmidrule(lr){9-11} \cmidrule(lr){13-15}
      & \multicolumn{1}{c}{$\tau_{15}$} & \multicolumn{1}{c}{MSE}& \multicolumn{1}{c}{$\tau_{85}$} &&  \multicolumn{1}{c}{$\tau_{15}$} & \multicolumn{1}{c}{MSE}& \multicolumn{1}{c}{$\tau_{85}$} & \multicolumn{1}{c}{$\tau_{15}$} & \multicolumn{1}{c}{MSE}& \multicolumn{1}{c}{$\tau_{85}$} && \multicolumn{1}{c}{$\tau_{15}$} & \multicolumn{1}{c}{MSE}& \multicolumn{1}{c}{$\tau_{85}$}\\
      \midrule
      Qualbacore$_\text{\tiny comps}$ & 0.97 & 1.06 & 1.29 &  & 1.10 & 1.09 & 1.21 & 1.24 & 1.07 & {\color{ForestGreen}0.75} &  & 1.23 & 0.97 & {\color{ForestGreen}0.71} \\ 
  Qualbacore$_\text{\tiny ranks}$ & \textbf{\color{black} 0.94} & \textbf{\color{ForestGreen} 0.92} & \textbf{\color{black} 0.95 }&  & \textbf{\color{black} 0.94} & \textbf{\color{black} 0.93} & \textbf{\color{ForestGreen} 0.86} & \textbf{\color{black} 0.99 }& \textbf{\color{black} 0.80} & \textbf{\color{ForestGreen} 0.36} &  & \textbf{\color{black} 0.97} & \textbf{\color{black} 0.72} & \textbf{\color{ForestGreen} 0.40} \\ 
      \midrule 
      \rowcolor{gray!15}
      ${\mathbf{Benchmarks}}$  & & & & & & & &&&&& & &  \cellcolor{gray!15}  \\ \addlinespace[2pt]
      $\boldsymbol{X}_t^{\text{bm}}$, $\phantom{..}$($w_0=0$) & 0.96 & 0.99 & 1.12 &  & 0.97 & 1.01 & 1.11 & 1.04 & 0.96 & 0.81 &  & 1.00 & 0.94 & 0.90 \\
      $\boldsymbol{X}_t^{\text{bm+}}$ & 1.03 & 1.03 & 1.13 &  & 1.07 & 1.03 & 1.11 & 1.14 & 1.22 & 0.81 &  & 1.01 & 1.07 & 0.83 \\ 
      $\boldsymbol{X}_t^{\text{bm+}}$, ($w_0=0$) & 1.02 & 1.02 & 1.13 &  & 1.07 & 1.04 & 1.13 & 1.14 & 1.02 & 0.81 &  & 1.01 & 0.95 & 0.90 \\ 
      \bottomrule \bottomrule
    \end{tabular}
 \begin{tablenotes}[para,flushleft]
  \scriptsize 
  \textit{Notes}: The table presents root mean square error (RMSE) relative to $\boldsymbol{X}_t^{\text{bm}} = [\text{PCE}_t \phantom{.}  \text{PCEcore}_t \phantom{.}  \text{PCEtrim}_t  ]$  with intercept. The remaining benchmarks are: $\boldsymbol{X}_t^{\text{bm}} = [\text{PCE}_t \phantom{.}  \text{PCEcore}_t \phantom{.}  \text{PCEtrim}_t  ]$ without an intercept (i.e., $w_0=0$), $\boldsymbol{X}_t^{\text{bm+}}$ with and without an intercept. Numbers in \textbf{bold} indicate the best performing model for each pair of horizon and loss function.  Numbers in  {\color{ForestGreen} green} highlights the loss function for which there is the largest improvement with respect to the benchmark,  if applicable.   In the results in Section \ref{sec:usresults}, we present level 6 for Qualbacore$_\text{\tiny comps}$ (with $K = 215$) and level 3 for Qualbacore$_\text{\tiny ranks}$ (with $K = 50$).  Note that unlike previous results,  we consider an expanding window from 1990 to increase the inevitably scarce number of observations in the tails.
\end{tablenotes}
\end{threeparttable}
\end{table}

We start our evaluation of Qualbacore by contrasting its forecasting performance to the main set of benchmarks in Table \ref{tab:results_us_quant}. We present results for each quantile (i.e., $\tau_{15}$ and $\tau_{85}$) and include,  for reference, the  original loss function (i.e., MSE).  Aligned with our main findings in Section \ref{sec:usresults}, \qacr yields a remarkable performance regardless of the forecasting horizon, quantile, and evaluation sample.   Pre-Covid results reveal that gains achieved by \qacr are rather evenly distributed across the risk spectrum for $h=3$ and tilted to the upper tail of  $h=6$.   The post-2020 sample confirms that large improvements are to be found in the upper tail,  while those for deflation or low inflation risk are negligible.  We observe improvements by sizable margins for $\tau_{85}$, mounting to over 60\% reduction in quantile loss for both forecasting horizons. 

The predictive accuracy of \qacc is lower, mostly in line with the benchmarks for the first evaluation sample.  In contrast to \qacr, forecasting results of \qacc in this period (as well as that of the remaining benchmarks) tend to be inferior in the upper tail.  This finding reverses when analyzing our second evaluation sample, which encompasses the pandemic and post-pandemic periods. \qacc and benchmarks perform well for the 85$^{th}$ quantile but fall short in the lower quantile. \textcolor{black}{This underperformance can be localized to the disinflationary shock in mid-2020 driven by the Covid-19 induced energy price shock.} In general,  the tendency  for substantial outperformance to be localized in the upper tail is not entirely surprising: key disinflationary events of the last 30 years are almost always due to unpredictable (and often momentaneous) oil price shocks.


\begin{figure}[t!]
  \caption{\normalsize{Qualbacore for the US ($h=6$)}} \label{fig:albacore_us_quant_h6}
  
  \begin{center}
    
    \vspace{-0.7cm}
    \begin{subfigure}[t]{0.5\textwidth}
      \centering
      \includegraphics[width=\textwidth, trim = 0mm 5mm 0mm 30mm, clip]{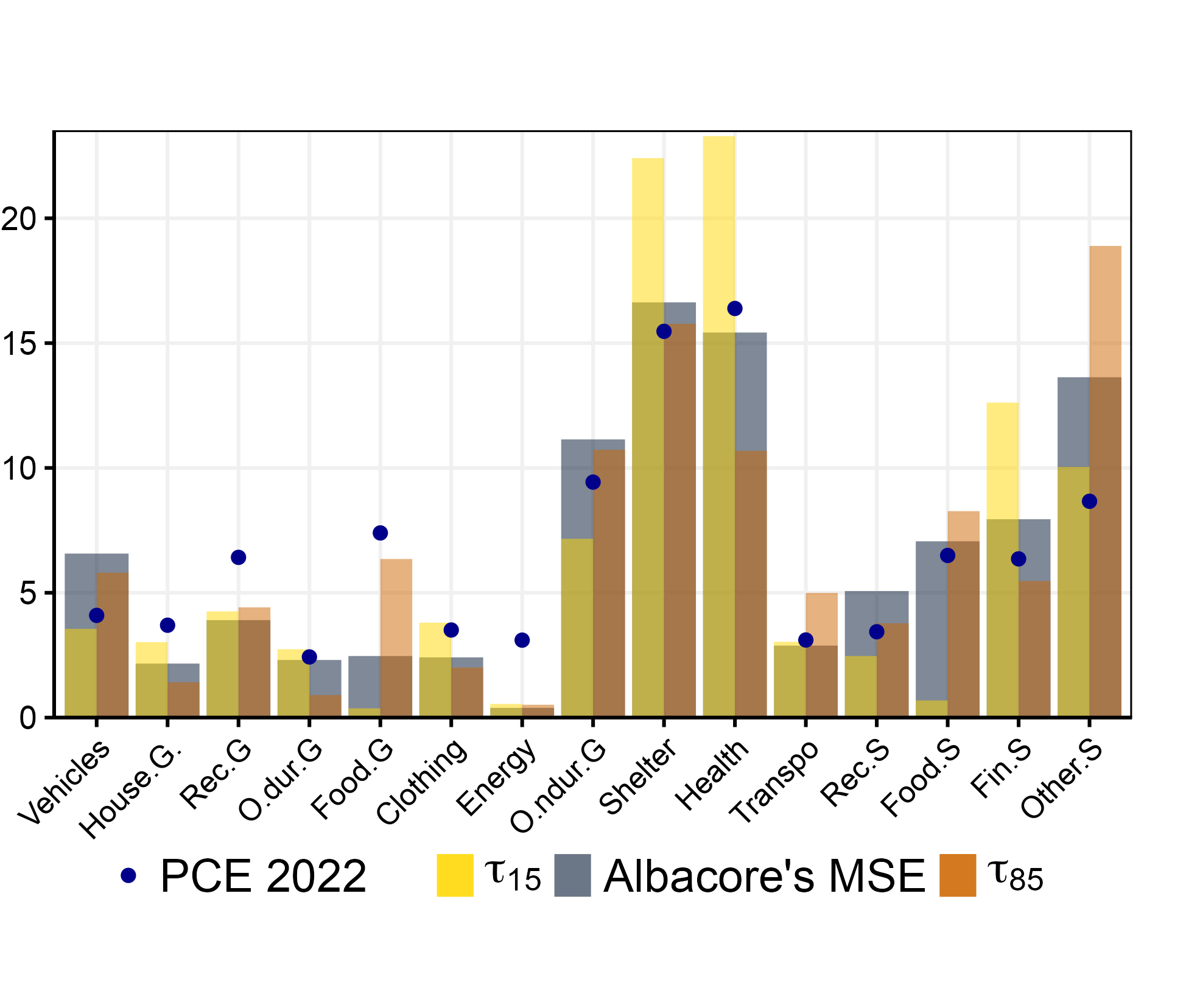}
      \vspace{-0.8cm}
      \caption{Comparison of Component Weights}
    \end{subfigure}%
    \begin{subfigure}[t]{0.515\textwidth}
      \centering
      \includegraphics[width=\textwidth, trim = 0mm 0mm 0mm 41mm, clip]{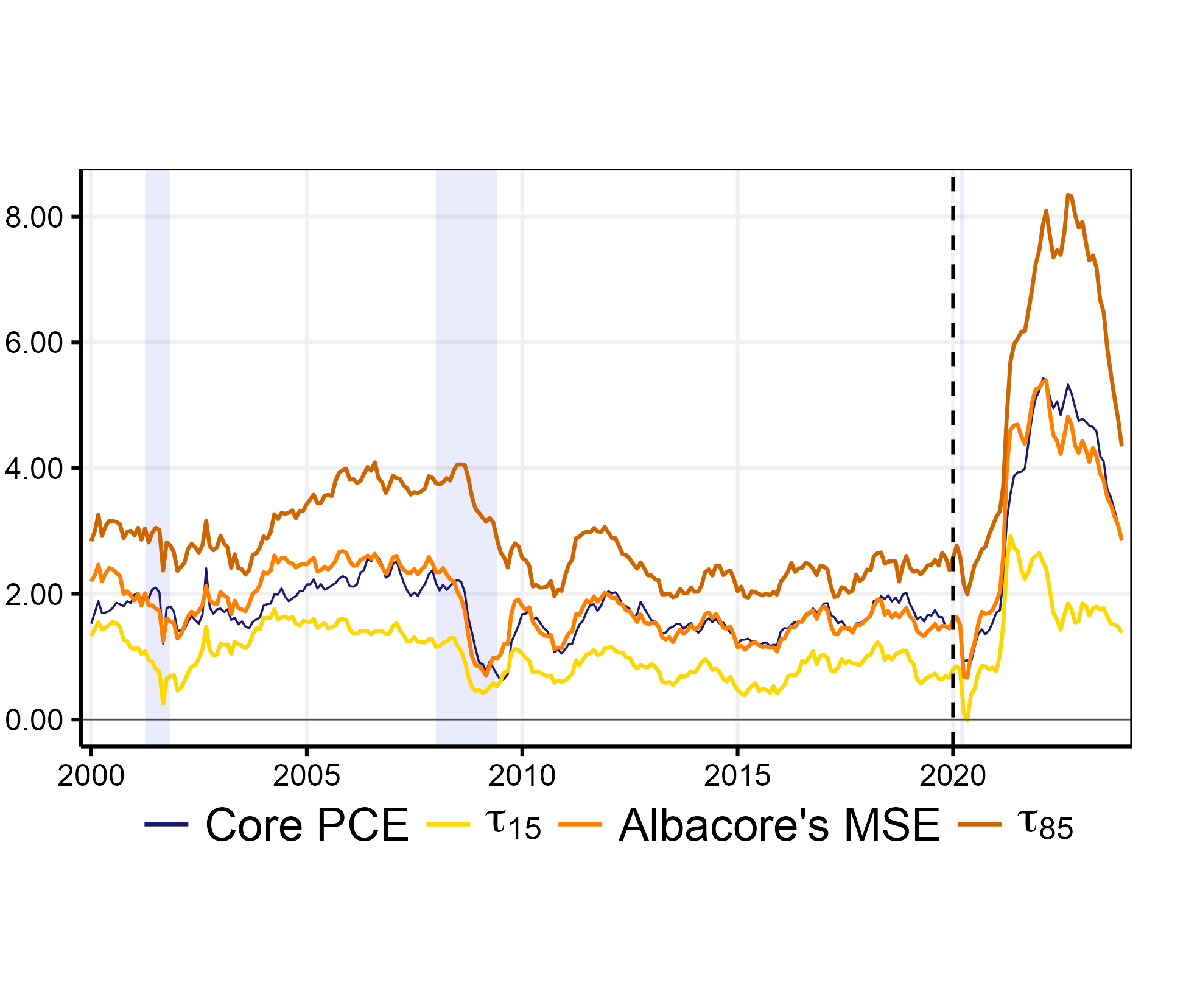}
      \vspace{-0.8cm}
      \caption{Time Series Comparison of \qacc}
    \end{subfigure}

    \vspace{0.1cm}
    \begin{subfigure}[t]{0.5\textwidth}
      \centering
      \includegraphics[width=\textwidth, trim = 0mm -2mm 0mm 40mm, clip]{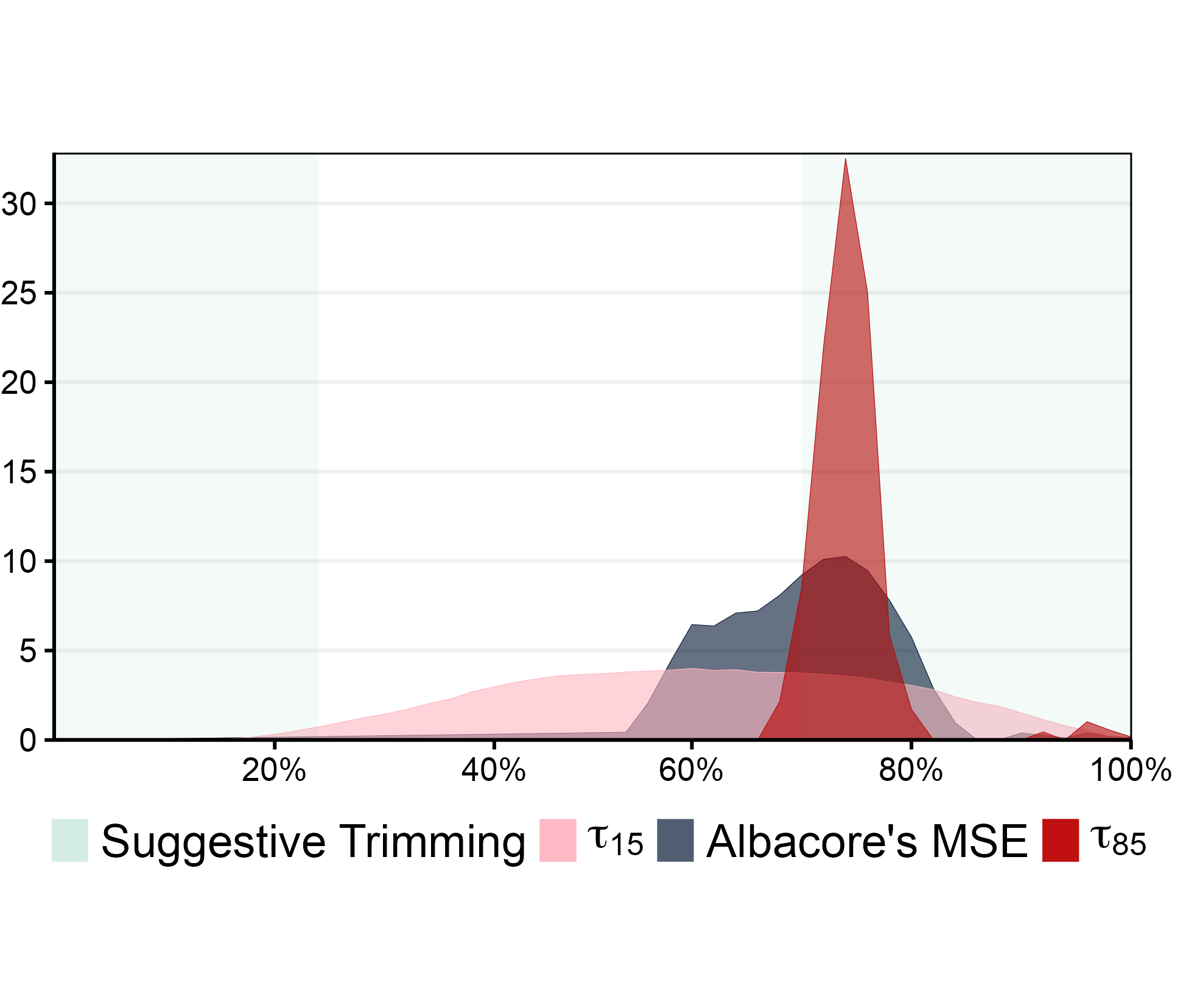}
      \vspace{-1.0cm}
      \caption{Comparison of Rank Weights}\label{fig:albacore_us_quant_h6c}
    \end{subfigure}%
    \begin{subfigure}[t]{0.515\textwidth}
      \centering
      \includegraphics[width=\textwidth, trim = 0mm 0mm 0mm 41mm, clip]{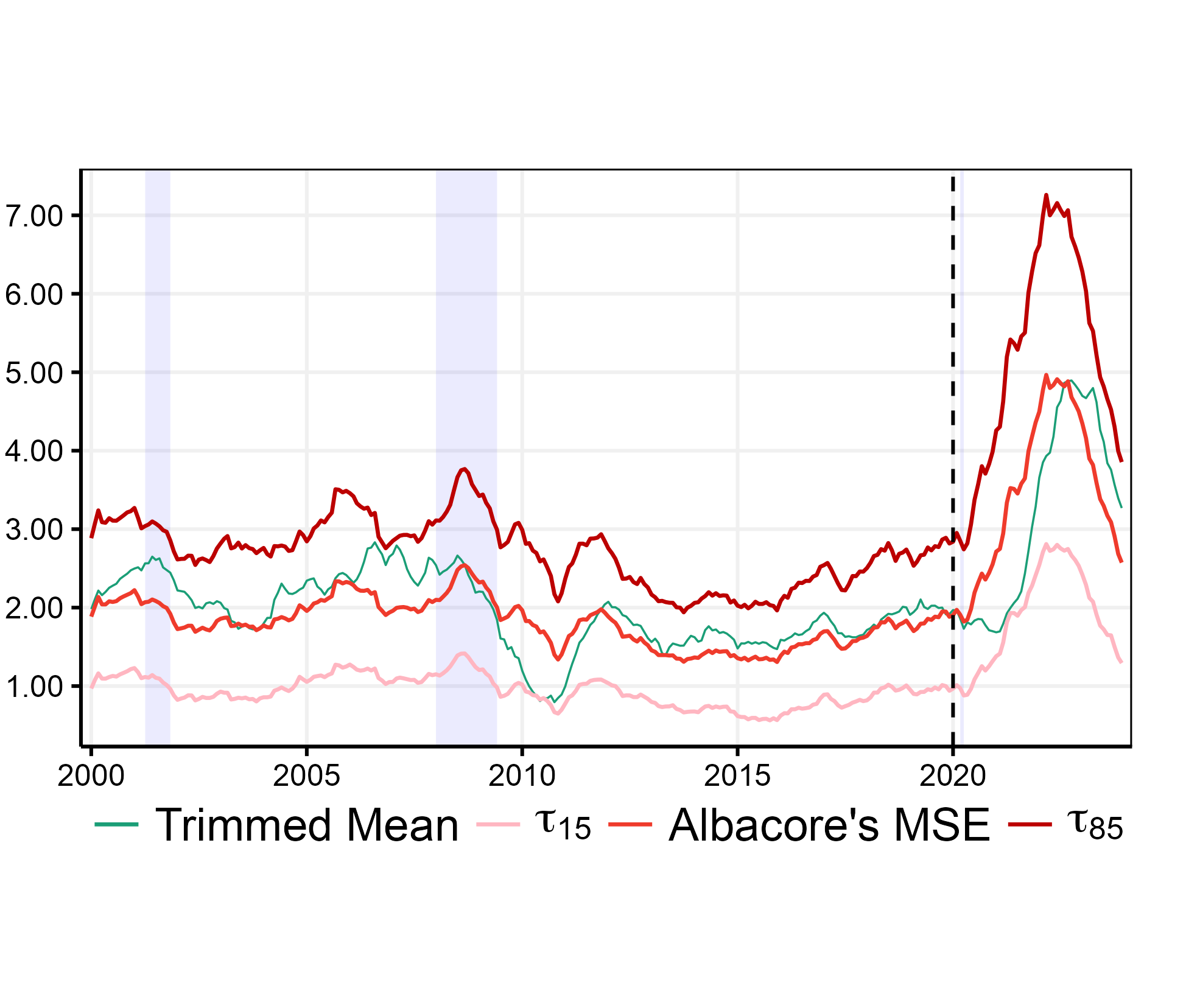}
      \vspace{-1.0cm}
      \caption{Time Series Comparison of \qacr}
    \end{subfigure}

  \end{center}
  
  \begin{threeparttable}
    \centering
    \begin{minipage}{\textwidth}
    \vspace*{-0.9cm}
      \begin{tablenotes}[para,flushleft]
    \setlength{\lineskip}{0.2ex}
    \notsotiny 
  {\textit{Notes}: All weights are based on models trained from 2001m1 to 2019m12. We present level 6 for Qualbacore$_\text{comps}$ and level 3 for Qualbacore$_\text{ranks}$. In the upper left panel we aggregate the component weights to the lowest level of disaggregation (level 2) for each quantile.}
    \end{tablenotes}
  \end{minipage}
  \end{threeparttable}

\end{figure}

To probe deeper, Figure \ref{fig:albacore_us_quant_h6} presents the resulting time series of Qualbacore (right panels) as well as the corresponding weights of assembled components and ranks (left panels) for the 6 steps ahead horizon. Details for the shorter horizon ($h=3$) are relegated to the appendix (Figure \ref{fig:albacore_us_quant_h3}). 
Studying the results from \qacc's upper tail shows that food goods and food services get high weights as well as components related to transportation and other services. Items assigned lower weights are mostly related to services and durable goods (including clothing, house goods, and other durable goods). For the lower tail, the opposite holds true. Shelter, health, and financial services, as well as durable goods enter the spotlight while food goods and food services get almost no weight. Energy, comprising the most volatile items, is excluded from both tails. \textcolor{black}{Given the usual wisdom, one might have expected a similar outcome for food components.  Yet, their assigned weights for the upper tail is nearly that of headline.  As shown in the literature, underestimation of future inflation is often linked to positive food price shocks \citep{bowles2007ecb,cecchetti2008commodity}. They tend to exhibit stronger second-round effects than energy price shocks and are more likely to propagate to headline eventually \citep{de2012commodity,de2016macroeconomic}.  Thus, it is less surprising that upweighting (with respect to core PCE) developments of food prices in the upper tail contributes to the predictive power of \qacc.} 

For \qacr, the resulting rank weights differ significantly when focusing on either the upper or lower quantile. $\tau_{85}$ yields a rather sparse solution centered at the upper part of the distribution of monthly price changes while the weights for $\tau_{15}$ are distributed densely covering most of the distribution except for the lowest part.  Hence, when specializing our supervised trimming measure to disinflation/deflation risk,  it is much more aligned with the specification of traditional trimmed mean inflation, as reflected by the shaded regions in  Figure \ref{fig:albacore_us_quant_h6c}.  An even clearer instance of this phenomenon can be visualized in Figure \ref{fig:albacore_us_quant_h3c},  where Qualbacore$_{\text{ranks}}(\tau_{15})$ is almost literally median inflation.  This leads us to conclude that traditional trimming-based core inflation measures,  with a sharp focus on the center of the price growth distribution,  are better equipped to foresee low inflation,  which was, from 1990 to 2020, seen as the main source of risk.  Our methods can capture this equally well,  but highlight that when balancing both types of risks within a single loss function (i.e.,  the MSE),  deviating from putting the accent on midpoint ranks is preferable.

\section{Assemblage Excursions}\label{sec:exc}

In this brief section,  we explore additional inflation-related applications of the assemblage regression,  with a particular focus on the rank space variant.  We deviate from aggregating inflation subcomponents.   First,  given the widespread use of year-over-year measures,  we consider aggregating lags into a supervised (trimmed) moving average of 12 months and compare it with \textcolor{black}{the usual headline rate in yearly growth rates} (Year-over-year,  YoY).  Second,  we devise a forward-looking combination of country-level headline inflation rates in the euro area.  

In both applications, we find that assembling in rank space is more successful than the more common components space version.  This suggests that,  when deciding on heterogeneous aggregation weights,  it can be preferable to draw more attention to the location of a realization in the overall distribution rather than its exact position in time or space. This finding is obviously, for now,  bounded to an environment where we are devising easily-interpretable forward-looking linear aggregations.  More work is needed to see if this holds as strongly elsewhere,  e.g.,   in more sophisticated data-rich nonparametric models or for other targets than inflation.  

\subsection{Rank Autoregression}\label{sec:lags} 

Year-over-year measures,  or equivalently,  12 months averages  of headline inflation are a common way to visually inspect price pressure metrics.  In this section,  we answer the following question: if one wishes to construct some average of the last 12 months of headline inflation that is most predictive of the next 6,  12,  and 24 months,  what would such an average look like?   Given the focus on time-smoothing (with the prefixed 12-month window) and the usage of only autoregressive information,  the outcome is conceptually  closer to  a trend inflation extraction, again,  by acknowledging the connection to \cite{Hamilton2018}'s filtering.

We deploy assemblage regression in components and rank space.  We focus on the US,  and the ``components'' are now 12 lags of month-over-month PCE growth rates.  \textcolor{black}{Of course,   in components space,  this is simply a constrained autoregression (AR).} The interest rather lies in the rank version,  which,  instead of permanently upweighting or downweighting lags,  performs temporary exclusion.  Its usage of past realizations of the time series differs from that of a standard autoregression. \textcolor{black}{The latter orders these lagged observations based on the distance in time between the observation and the target,  and allows for corresponding heterogeneous weights.}  Instead,  the rank version fixes a lookback window (here,  12 months) and  weights past realizations based on their location in the window's realized distribution. \textcolor{black}{We thereby shift} the focus away from timing with potentially sophisticated dynamics with lag-specific coefficients  to consider each lag as a draw and ask ourselves which average of such draws should we be paying \textit{attention} to.  The elements discussed in Section \ref{sec:sts} apply,  {implying} in this context that the order statistics-based autoregression is a time-varying autoregression in \textit{lags space}.

The usage of the term "realized" to describe month-to-month inflation rates here is deliberate,  as what we are building is a summary statistic from high(er)-frequency data designed to be used as a predictor for a lower frequency target.  Thus,  there is a natural connection to the construction of realized volatility (RV) from intra-day data in finance \citep{andersen2003modeling,mcaleer2008realized}.   A key methodological distinction is that here,  the nature of the statistic itself is \textit{estimated} via rank coefficients rather than being \textit{predefined} (and proven) to be a consistent estimator for a variance,  a quantile \citep{dimitriadis2022realized},  a mean,  or else. The theoretical environment backing RV is that true volatility is fundamentally latent and can be estimated consistently from high-frequency data under various assumptions.  Inevitably,  we are within the realm of unsupervised learning because the desired supervisor's "true volatility" is unobserved.\footnote{Nonetheless,  in the spirit of this paper,   it is thinkable that such measures could be defined in a "supervised" fashion by maximizing usefulness in predicting another \textit{observed} variable (or maximizing general economic value) with rank weights being shrunk to what would be implied by traditional RV.}  In our rank AR inflation context,  there is no quest for a particular summary statistic\footnote{For instance,  if inflation were hypothetically defined as a yearly process,  the intra-year squared observations could not be averaged directly to build a variance estimator because month-to-month observations are not uncorrelated \citep{mcaleer2008realized}.},  but rather,  a search of any that will fulfill the empirical task of being closely associated with future headline inflation, as measured by statistical agencies.  Validity comes from predictive power holding both in- and out-of-sample.

\begin{table}[h]
	
	\caption{Forward-Looking 12 Months Moving Average  Results  \vspace*{-.5cm}}\label{tab:lagpoly}
	 \hspace{-.373cm}
	\begin{minipage}{0.502\linewidth}
		\vspace*{0.1cm}
		\centering
		\includegraphics[width=1.0\textwidth, trim = 10mm 20mm 5mm 20mm, clip]{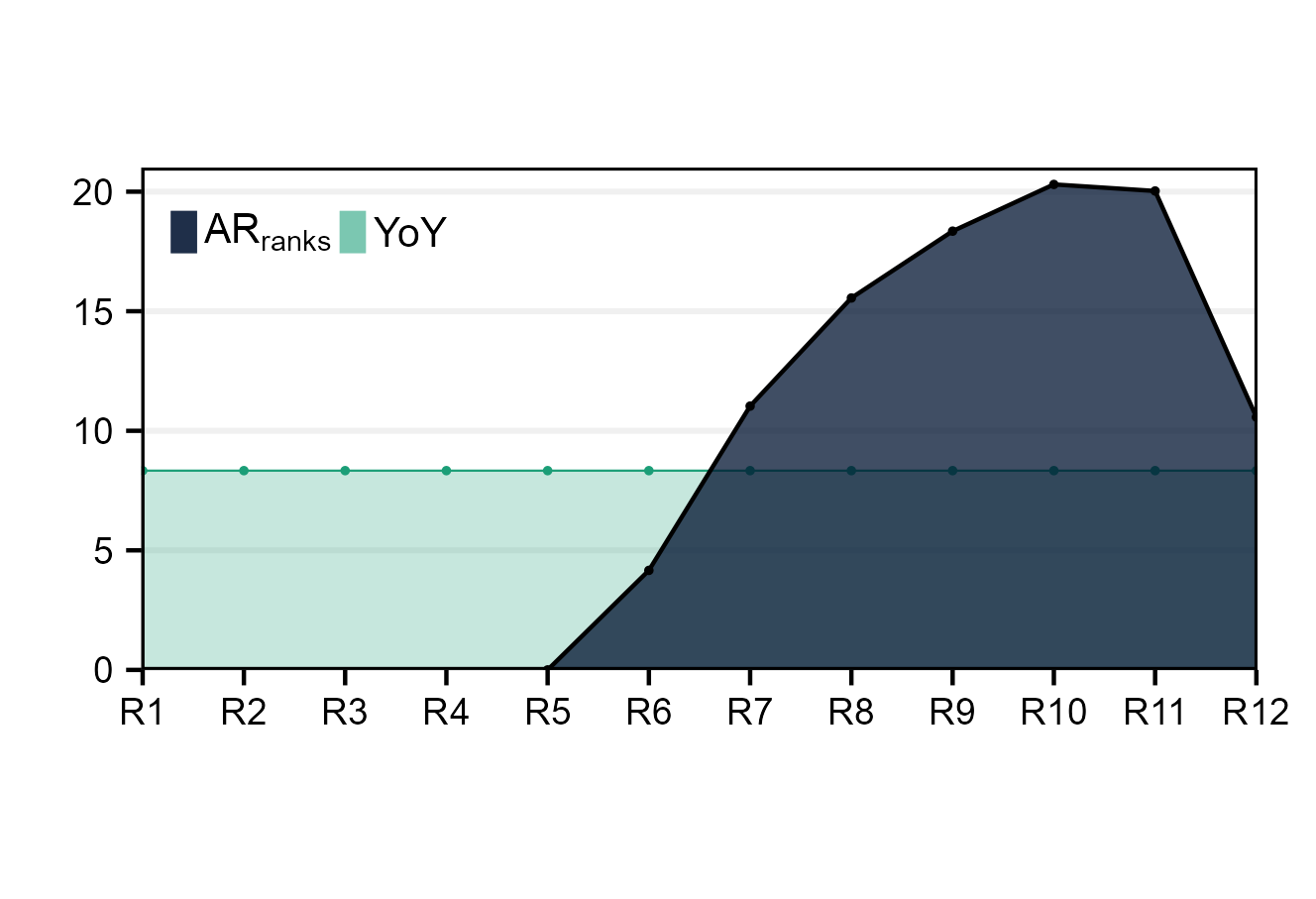}
			\end{minipage}  \hspace{-.115cm}
				\begin{minipage}{0.502\linewidth}
	\includegraphics[width=\textwidth, trim = 50mm 10mm 40mm 0mm, clip]{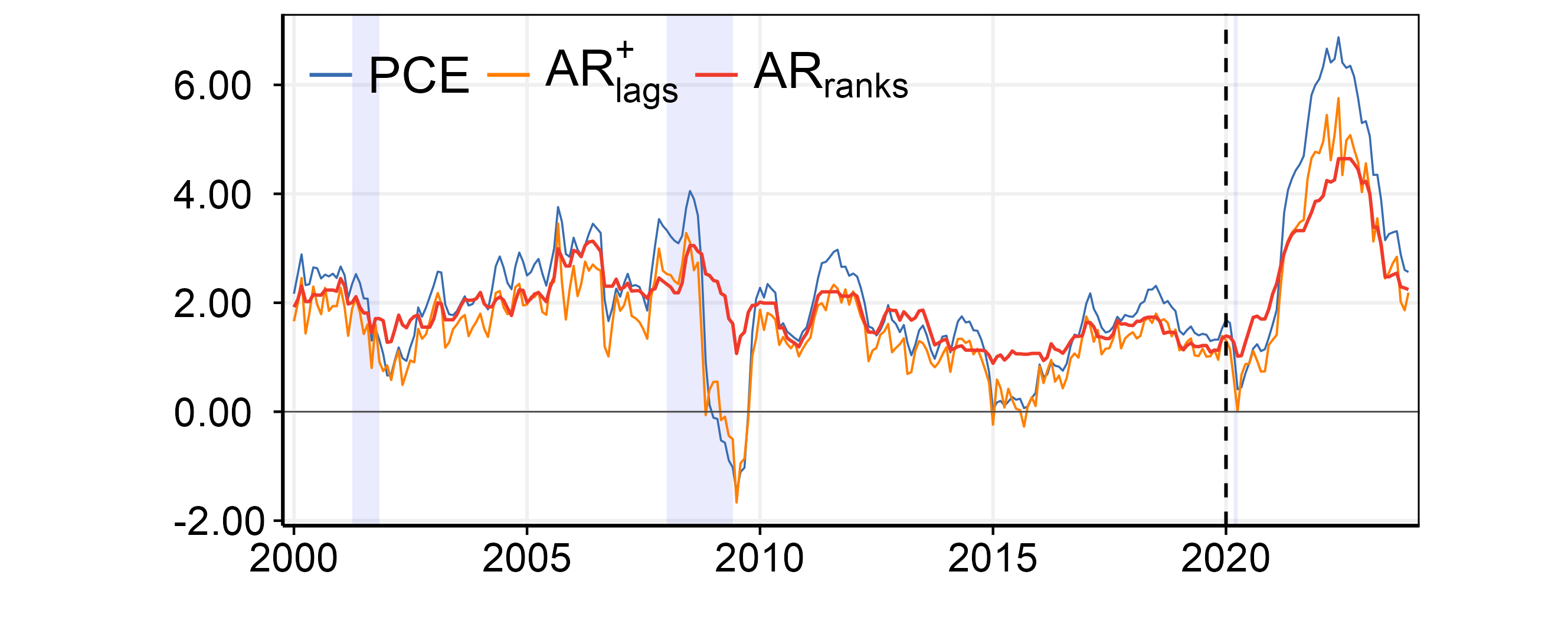}
	\end{minipage}    
\end{table}
\begin{table}[h]
 \footnotesize
	\vspace*{-0.5cm}
	\centering
	\rowcolors{2}{white}{gray!10}
	\fontsize{10.5}{10.5}\selectfont
	\begin{threeparttable}
		\centering
		\setlength{\tabcolsep}{0.61em} 
		\setlength\extrarowheight{2.7pt}
		\begin{tabular}{l| rrrrrrrrr | rrrrrrrrr } 
			\toprule \toprule
			\addlinespace[2pt]
			\rowcolor{white} 
			&  \multicolumn{9}{c}{2010m1-2019m12} &  \multicolumn{9}{c}{2020m1-2023m12}  \\
			\cmidrule(lr){2-10} \cmidrule(lr){11-19}
			\rowcolor{white} 
			\multicolumn{1}{r|}{$h \rightarrow$}&  \multicolumn{1}{c}{$1$} & &  \multicolumn{1}{c}{$3$} & & \multicolumn{1}{c}{$6$} & & \multicolumn{1}{c}{$12$} & & \multicolumn{1}{c}{$24$}&  \multicolumn{1}{c}{$1$} & &  \multicolumn{1}{c}{$3$} & & \multicolumn{1}{c}{$6$} & & \multicolumn{1}{c}{$12$} & & \multicolumn{1}{c}{$24$}  \\
			\midrule
			\rowcolor{white} 
			AR$_{\text{lags}}$  & 0.99 &  & 0.99 &  & 0.98 &  & 0.99 &  & {0.95 } & { 1.12} &  & 1.30 &  & 1.29 &  & 1.08 &  & 0.99 \\  
			\rowcolor{white} 
			AR$^{\text{+}}_{\text{lags}}$ & \textbf{0.95} &  & 1.00 &  & 0.99 &  & 0.98 &  &{0.99 } & \textbf{ 0.95} &  & \textbf{0.93} &  & 0.96 &  & 0.98 &  & 0.99 \\   
			\rowcolor{white} 
			AR$_{\text{ranks}}$ \phantom{oo00} & 0.99 &  &\textbf{0.96} &  & \textbf{0.93} &  & \textbf{0.85} &  & \textbf{0.75 } & { 0.98} &  & 0.95 &  & \textbf{0.93} &  & \textbf{0.87} &  & \textbf{0.94} \\   \midrule 
			\rowcolor{white} 
			YoY in AR(1) \  & 0.99 &  & 0.97 &  & 0.94 &  & 0.94 &  & {0.90 } & { 1.17} &  & 1.34 &  & 1.27 &  & 1.08 &  & 1.00 \\   
			
			\bottomrule \bottomrule
		\end{tabular}
		\begin{tablenotes}[para,flushleft]
			\scriptsize
						\textit{Notes}: The table presents RMSE relative to a random walk forecast using year-over-year (YoY) data. AR$_{\text{lags}}$ indicates an unconstrained AR(12), AR$^{\text{+}}_{\text{lags}}$ denotes an AR(12) with nonnegative coefficients and without an intercept and AR$_\text{ranks}$ is the corresponding AR(12) in rank space, all using monthly growth rates. Numbers highlighted in \textbf{bold} show the best model per \textit{horizon} and out-of-sample period. The upper right panel shows AR$_\text{ranks}$, AR$^{\text{+}}_{\text{lags}}$ and YoY PCE inflation.  We plot in-sample results up to 2020m1 and show out-of-sample ones thereafter. In the lower right panel,  we compare the corresponding weights for AR$_\text{ranks}$ with the YoY PCE benchmark (in \%).
		\end{tablenotes}
	\end{threeparttable}
\end{table}

Table \ref{tab:lagpoly} presents the performance of the ranked (AR$_{\text{ranks}}$) and constrained (AR$^{\text{+}}_{\text{lags}}$) autoregression compared to a simple AR$_{\text{lags}}$(12) with monthly growth rates, an AR(1), and a random walk, both using year-over-year data (the latter as the numéraire).  During periods of low and stable inflation,  putting the emphasis on persistence is an established strategy  \citep{clark2014evaluating,StockWatson2016}.  Accordingly,  AR$_{\text{ranks}}$ procures good performance during the pre-2020 sample by leveraging little information.   What is more striking is that we find decisive improvements from AR$_{\text{ranks}}$ over AR alternatives for medium- and long-run horizons -- and for both out-of-sample periods.   At horizon 3,  AR$_{\text{ranks}}$ and AR$^{\text{+}}_{\text{lags}}$ report similar performance,  with $\approx 5 \%$ improvements over the numéraire.  Only at $h=1$ does AR$^{\text{+}}_{\text{lags}}$ prevail marginally,  suggesting that the gains of paying attention to timing with lags (as apposed to distributional aspects with AR$_{\text{ranks}}$) are circumscribed to very short horizons.

The successful AR$_{\text{ranks}}$ uses a downweighted average of the 6 highest-valued lags, \textcolor{black}{i.e. those 6 out of the 12 lags that show the highest month-over-month increase} (see lower right panel in Table \ref{tab:lagpoly}).  The inclusion of several lags equips the measure with adequate smoothness, \textcolor{black}{while} the exclusion of lower realizations allegedly mitigates positive "base effects" following negative price shocks.  The resulting average puts a high bar for the index to show signs of lasting deflation,  as it would require that order statistics corresponding to ranks $R_5$ and above showcase negative growth, implying that a large number of (nearly) consecutive months witnesses deflation.  This is consistent with recent history: we find episodes of deflation (2009) or near deflation (2020) in YoY due to a handful of months showing abnormal negative shocks,  but those quickly reversed.  Depending on the precise event itself,  one may attribute this reversal to the very nature of these (oil) shocks or the expansionary monetary policy often triggered in the aftermath.     Whatever the cause might be, acknowledging that such shocks have little predictive power for headline inflation conditions in 6, 12, or 24 months,  AR$_{\text{ranks}}$ trims them out.  Disinflation is not as exceptional of an outcome for AR$_{\text{ranks}}$  as deflation is.  If the upper order statistics showcase lower growth than they usually do (by construction they are trending above the target),  then AR$_{\text{ranks}}$  will turn in a disinflation forecast, as is the case for 2014 to 2021. This asymmetric outcome is not a major surprise given our main results in Section \ref{sec:usresults}.  The prevalence of temporary negative shocks with no predictive power for future headline inflation can be identified among components at a fixed point in time,  or \textcolor{black}{alternatively},  by observing repeatedly an aggregate with a time-varying level of exposition to those \textcolor{black}{negative shocks}.

While AR$_{\text{ranks}}$ does well in our post-2020 sample and is clearly the best {among the models using this limited information set},  it is no match for those using components-level data (Table \ref{tab:results_ea_pop3}).   12 months proves too long of a lookback window and solely focusing on autoregressive information is insufficient,  motivating the usage of the more involved  \acr of Section \ref{sec:assreg}, which fares well in all regimes.  Thus,  we find AR$_{\text{ranks}}$ and \acr  to share similarities before 2020,  but less so thereafter.  Nonetheless,  given its decisive merits within the class of data-poor models,  the autoregressive inflation model in rank space is a handy tool to keep in one's arsenal.  It may be of  even greater interest for countries where component level data is less reliable or not available for an extended period.

\subsection{Geographic Assemblage} \label{sec:geo}

Although sharing the same currency, and thus a single monetary policy, inflation rates of EA member states are far from being homogenous.  As a result, the predefined weighted average of HICP rates forming the EA aggregate may or may not be the most timely indicator for the whole area $h$ months from now. A natural curiosity that arises thereof is whether an alternative aggregation of member states inflation rates forms a better leading indicator for HICP than HICP itself.  To answer this question, we introduce \gac, a geographic assemblage of \textit{Albacore}, with the two representatives \gacr and \gacc. Different from \acc, which assembled the subcomponents of HICP in the predictor matrix, $\boldsymbol{\Pi}_{t}$ is now composed of the headline inflation numbers of each of the $M$ EA member states: $\boldsymbol{\Pi}^G_{t} = \left[HICP^1, ..., HICP^M\right]$.  Similar to \acr, the order statistics matrix $\boldsymbol{O}^G_{t}$ for \gacr, is then just the ranked version of $\boldsymbol{\Pi}^G_{t}$.

The geographic versions of \acc and \acr are designed as follows: the weights of Albacore$^\text{G}_\text{\tiny comps} $ are shrunk towards the annual expenditure shares for each country.\footnote{Expenditure shares are the percentage of total household final monetary consumption expenditure based on national accounts data \citep[see ][]{EURegulation20201148}. Data is taken from Eurostat (\texttt{prc\_hicp\_cow}).} Weights are not constrained to sum to unity,  dropping the random walk hypothesis is hardly appropriate when only combining headlines.  Sum-to-one can be restored  by interpreting the resulting (lower) sum of weights as an AR(1) coefficient,  which, being smaller than 1, reduces variance.  The loosening of this constraint, which was introduced to discipline the model when assembling a large number of \textit{components}, prevents the solution from converging to HICP itself.   Furthermore, we demean each country rate and add back to it the EA average (i.e.,   a country fixed effect).    Variance reduction in \gacr is obtained through different channels,  and thus,  neither of these modifications  to our baseline assemblage regression proposition is needed.   It keeps the fused ridge penalty, while constraining the resulting conditional mean  to match the unconditional mean of our target.

\begin{table}[t!]
	\caption{Forward-Looking  Euro Area Aggregation  \vspace*{-.3cm}}\label{tab:results_ea_bonus}
	
	\hspace{.1cm}
	\begin{minipage}{0.37\linewidth}
		\includegraphics[width=1.122\textwidth, trim = 80mm 35mm 30mm -3mm, clip]{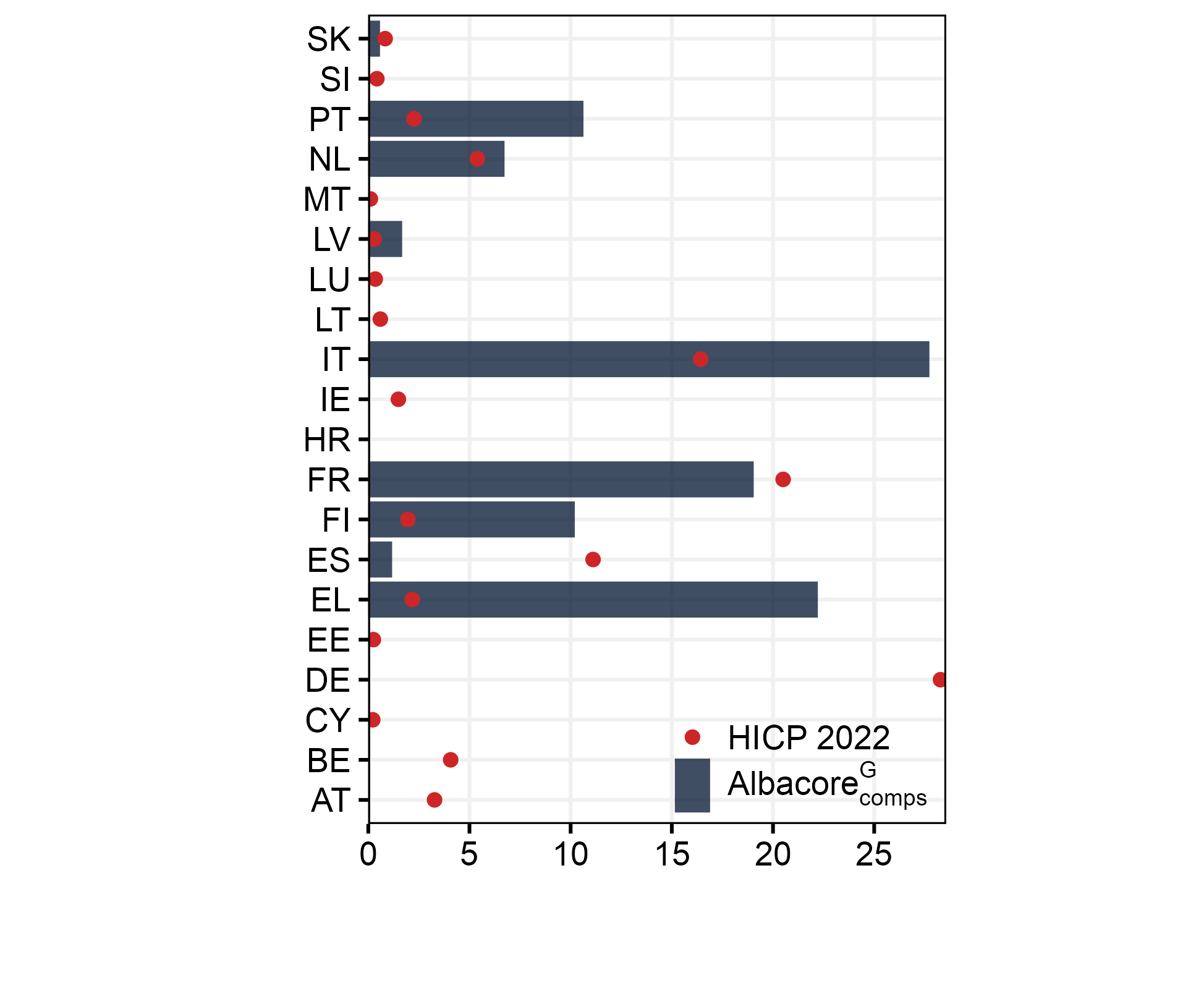}
	\end{minipage} \hspace{-.85cm} 
	\begin{minipage}{0.65\linewidth}
		\centering
			\includegraphics[width=.952\textwidth, trim = 18mm 92mm 0mm 76mm, clip]{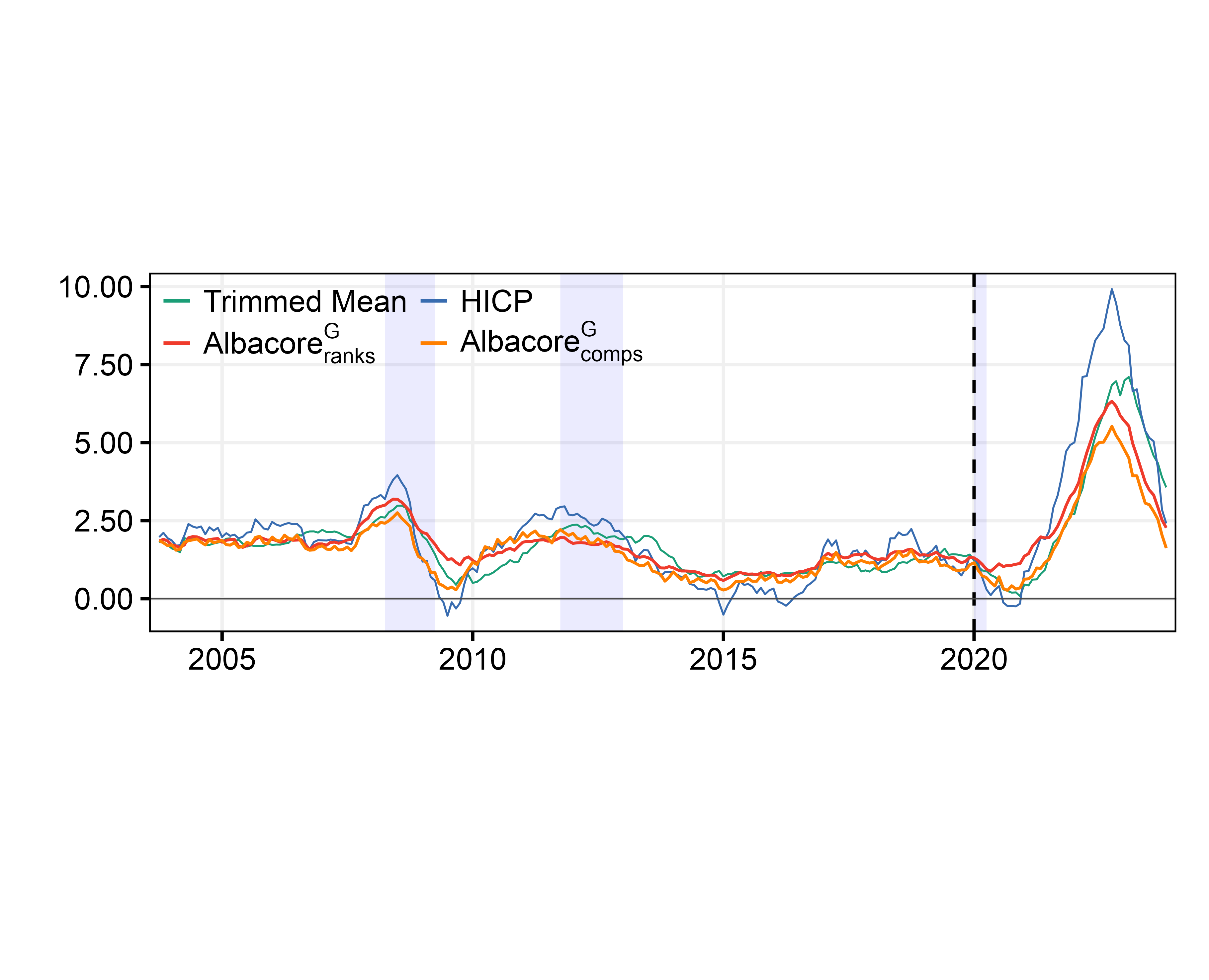}
		\includegraphics[width=.962\textwidth, trim = 5mm 80mm 0mm 80mm, clip]{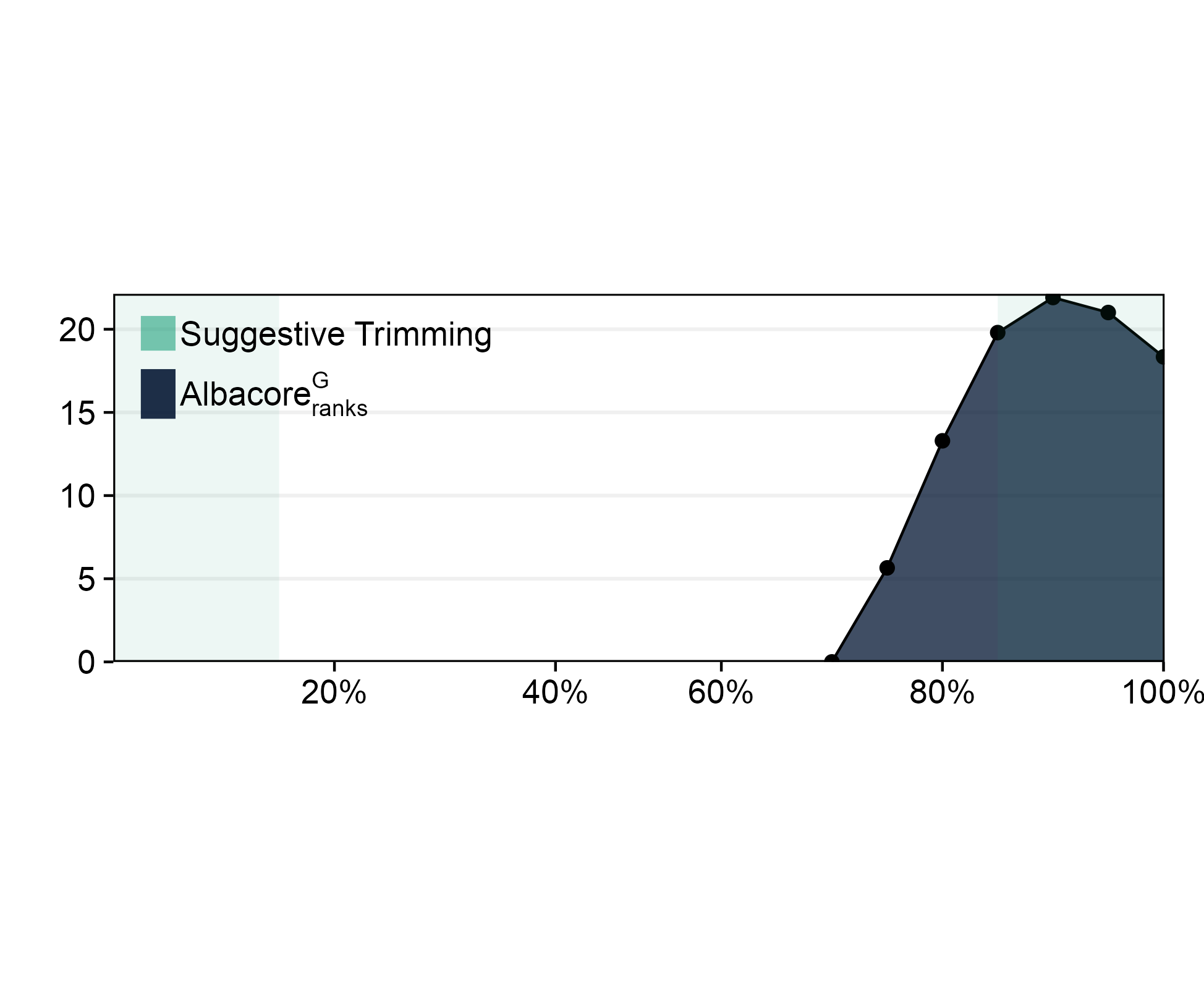}
	\end{minipage}    
\end{table}

\begin{table}[t!]\vspace{-.35cm}
	\footnotesize
	\centering
	\begin{threeparttable}
		\setlength{\tabcolsep}{0.53em} 
		\setlength\extrarowheight{1.5pt}
		\begin{tabular}{l| rrrrrrrrr | rrrrrrrrr } 
			\toprule \toprule
			\addlinespace[2pt]
			&  \multicolumn{9}{c}{2010m1-2019m12} &  \multicolumn{9}{c}{2020m1-2023m12}  \\
			\cmidrule(lr){2-10} \cmidrule(lr){11-19}
			\multicolumn{1}{r|}{$h \rightarrow$}&  \multicolumn{1}{c}{$1$} & &  \multicolumn{1}{c}{$3$} & & \multicolumn{1}{c}{$6$} & & \multicolumn{1}{c}{$12$} & & \multicolumn{1}{c}{$24$}&  \multicolumn{1}{c}{$1$} & &  \multicolumn{1}{c}{$3$} & & \multicolumn{1}{c}{$6$} & & \multicolumn{1}{c}{$12$} & & \multicolumn{1}{c}{$24$}  \\
			\midrule
			\addlinespace[5pt] 
			
			Albacore$^\text{G}_\text{\tiny comps} $ & 1.00 &  & 1.03 &  & 0.96 &  & 0.81 &  & 0.70 & 0.99 &  & 0.91 &  & 0.90 &  & 0.96 &  & 1.12 \\ \addlinespace[2pt]

			Albacore$^\text{G}_\text{\tiny ranks} $ & \textbf{0.97} &  & \textbf{0.93} &  & \textbf{0.86} &  & \textbf{0.74} &  & \textbf{0.66} & \textbf{0.94} &  & \textbf{0.84} &  & \textbf{0.80} &  & \textbf{0.87} &  &\textbf{0.93} \\ \addlinespace[2pt]

			\midrule 

			
			HICP-RW & 1.08 &  & 1.23 &  & 1.31 &  & 1.22 &  & 1.20 & 0.99 &  & 0.89 &  & 0.86 &  & 1.03 &  & 1.03 \\ 
			HICP-AR(1) & 1.00 &  & 1.02 &  & 1.01 &  & 1.01 &  & 1.06 & 1.06 &  & 1.07 &  & 1.05 &  & 1.02 &  & 0.98 \\ 
			Best Core Benchmark  & 0.98 &  & 0.95 &  & 0.93 &  & 0.87 &  & 0.75 & 0.95 &  & 0.89 &  & 0.92 &  & 0.97 &  & 0.99 \\

			\bottomrule \bottomrule
		\end{tabular}
		\begin{tablenotes}[para,flushleft]
			\scriptsize 
			\textit{Notes}: The table presents root mean square error (RMSE) relative to $\boldsymbol{X}_t^{\text{bm}} = [\text{PCE}_t \phantom{.}  \text{PCEcore}_t \phantom{.}  \text{PCEtrim}_t  ]$  with intercept. The remaining benchmarks are: HICP RW is a random walk model of EA headline inflation, HICP-AR(1) is the corresponding AR(1) model, and the last row shows the best performing benchmark considered in the main analysis (as in Table \ref{tab:results_ea_pop3}). Numbers in \textbf{bold} show the best model per horizon and out-of-sample period. The upper left panel shows Albacore$_\text{comps}^{G}$'s allocation of weights to countries (in \%) compared to official country weights provided by Eurostat. The plots on the right show Albacore$^{G}$ and benchmarks in yearly percentage changes (top) and the corresponding weights of Albacore$_\text{ranks}^{G}$ (in \%) compared to the 30\% trimmed mean (bottom). 
		\end{tablenotes}
	\end{threeparttable}
\end{table}

From Table \ref{tab:results_ea_bonus},  it is apparent that \gacr trumps all competitors in both subsamples and across all forecasting horizons,  whereas the component version is mostly aligned with the performance of the best core benchmarks.  We show corresponding time series and the weights for \gacc and \gacr in the panels above. Similar to Figure \ref{fig:albacore_ea_h12}, \gacr anticipates the broiling inflationary pressures at a much earlier stage than HICPX or the trimmed mean. Furthermore, it calls the peak and the following downward trend precisely.   \gacr showcases strong asymmetric trimming: the three lower quartiles are entirely discarded, and weight is given only to the very right tail of the distribution.  While it might appear like a radical outcome, we find reasonable arguments from both a statistical as well as an economic point of view. From a time series perspective, one should keep in mind that  we are trimming highly aggregated indices,  which feature much less dramatic movements in the tails than subcomponents-level series. Empirically, results suggest that higher inflation rates across countries are more likely to be contagious and generalize to the whole area.  This also means that low inflation will be expected in the 6, 12, or 24 months only if the highest inflation rates over the last 3 months are lower than they typically are. Alternatively,  this may be linked to tails reacting more quickly to shocks and an intensified spillover of inflation during turbulent times, as is found in the literature \citep{aharon2022infection,bouri2023global}.  Thus,  regardless of the size of EA member states, their expenditure intensity, or other country-specific characteristics, a forward-looking approach may want to pay attention to those with higher inflation rates at a given point in time.  In particular,  as a thumb rule,  a downweighted average of the 5 highest headline inflation rates among member states is found to be a good tracker of forthcoming EA-wide price pressures.

In terms of RMSE,  \gacc outperforms the official HICP for longer horizons,  but not the rank version discussed above.   It moves rather in lockstep with the other core measures (HICPX and trimmed mean) at the outset of the pandemic-induced recession, and is late to calling the mounting inflationary pressures. For the turning point, it is close to \gacr and resembles that of a compressed headline. This behavior can be traced back to the allocation of country weights, which deviates substantially from the official ones (see Table \ref{tab:results_ea_bonus}).  \gacc essentially overweights Italy, Finland, Portugal, and Greece, and keeps France and the Netherlands in the set with close to official weights. While Greece and the Netherlands experienced the post-pandemic inflationary pressures late, they were among the first member states to start the disinflationary process. France and Finland, on the other hand, contribute to the smoothness of the series given their comparably moderate increases. 
At first sight, it might seem surprising to see \gacc excluding Germany, which constitutes the biggest EA economy with most influential trade relations \citep{auer2019international,tiwari2019analysing}. However, when taking on a forward-looking perspective, inflation rates of smaller (or peripheral) countries may provide relevant signals since they are more susceptible to shocks. In our sample, Greece stands out as it was heavily affected by the GFC, triggering the subsequent sovereign debt crisis.  It experienced low (and negative) inflation before the whole area itself,  which likely contributes to its high weights.  This is, obviously,  quite contextual to the training sample including 2010s,  and highlights why \gacr, with its temporary inclusions and exclusions scheme, fares better overall than \gacc, which focuses on a fixed portfolio of countries.

\section{Concluding Remarks and Directions for Future Research} \label{sec:concl}

\noindent We introduced the assemblage regression to build a maximally forward-looking measure of core inflation.  While predictive power for future headline inflation is generally an afterthought for the construction of core inflation measures,  our  \acc and \acr are explicitly optimized to perform well with respect to this criterion.  {Both yield favorable forecasting results coupled with economic insights of high relevance.} The most striking {finding} is the asymmetric trimming obtained from \acr,  which is particularly pronounced in the US.  This allows \acr to perform well in quieter times and capture early signs of the post-pandemic surge months in advance of trimmed mean and other alternatives.  {In terms of supervised weighting of components, we find a special role assigned to food goods. It is not totally excluded from  \acc (as typically done in core PCE), and when it comes to upside inflation risks its overall weight even matches that of headline.  These results, and additional insights from our quantile regression extension,  suggest that the core inflation measures we should monitor to flag disinflation/deflation risks may differ from what is needed to detect the risks that came to materialize from 2021 onward. } 

There are numerous avenues for future research.   First,   and closest to the analysis conducted in this paper,  there are a few natural extensions to  Albacore that one may wish to investigate.  While we focused on aggregating components in a single space at a time,  one may think of combining components and rank selection using more flexible nonlinear optimization and a layered approach (e.g.,  a structured neural network). Alternatively,  one could consider a "reservoir" approach \citep{tanaka2019recent},  which would boil down to an extremely high-dimensional ridge regression.


Second,  assemblage schemes considered in \textcolor{black}{our work} are built to fulfill a set of simple statistical requirements.  One can depart from "forward-looking" \textcolor{black}{objective},  or,  at least, include a broader set of \textcolor{black}{objectives}.  One possibility is using \cite{MACE}'s algorithm  to construct a core inflation indicator that is maximally reactive to monetary policy and real activity.  Combining these supervised aggregation ideas with the forward-looking criterion could lead to {maximizing} the overall coherency in a  vector autoregressive setting (i.e.,  its likelihood).  In the latter case,  this would imply that the assembled series would be both highly reactive to key aggregates \textit{and} predictive of them.  This could involve assembling a wider basket of indicators -- like GDP,   which is seldomly exposed to sectoral shocks that are a distraction from the general macroeconomic trend.


Third, the econometric idea of running  assemblage regression in rank space to build a maximally predictive summary statistic can be exported beyond inflation applications. \textcolor{black}{For example, supervised trimmed forecast combinations might extend trimming methods used in the literature, which only involve choosing the endpoints in a similar vein to  trimmed mean inflation  (see \cite{wang2023forecast} for a survey).}  While traditional regressions in forecasters space upweight and downweight individuals based on their past skills,  running it in rank space allows to detect which section of the crowd's distribution has more wisdom (whatever its time-varying composition in terms of individuals is).  Regressions using order statistics may also be better equipped to deal with the inevitable entry and exit of forecasters from the panel.  Another, more operationally ambitious avenue,   is supervised temporal aggregation of intraday financial returns data to maximize the economic value derived from the resulting series.  Possible  outcomes, depending on the \textit{observable} target,   include some realized volatility series,  one its many refinements and extensions \citep{mcaleer2008realized},  a quantile,  or something else entirely.   

All in all, the supervised aggregation ideas developed in this paper are exportable to a variety of settings where the official aggregate series (or manual alterations of it) offers room for improvement. This is, essentially,  the foundational idea of deep learning.  Rather than manually building predictors from disaggregates, one can endogenously construct the features from the input through multiple layers which are jointly optimized along with the predictive model using them.  In an econometric modeling environment,  there are obvious constraints on such constructions, so that the extracted features make economic sense.   Nonetheless,  this paper has shown -- within the context of an elementary forecasting model -- that such cohabitation is possible.


 \pagebreak
 \clearpage
 \setlength\bibsep{5pt}
		
\bibliographystyle{apalike}

\setstretch{0.75}
\bibliography{PCE}

\clearpage

\appendix
\newcounter{saveeqn}
\setcounter{saveeqn}{\value{section}}
\renewcommand{\theequation}{\mbox{\Alph{saveeqn}.\arabic{equation}}} \setcounter{saveeqn}{1}
\setcounter{equation}{0}
	
\section{Appendix}\label{app}
\setstretch{1.25}

\subsection{Additional Results}\label{sec:addresults}


\begin{table}[h]
  \footnotesize
  \centering
  \begin{threeparttable}
  \caption{\normalsize {Forecasting Performance of Albacore for the Euro Area} \label{tab:results_ea_pop3}
    \vspace{-0.3cm}}
    \setlength{\tabcolsep}{0.55em} 
          \setlength\extrarowheight{2.9pt}
 \begin{tabular}{l| rrrrrrrrr | rrrrrrrrr } 
\toprule \toprule
\addlinespace[2pt]
&  \multicolumn{9}{c}{2010m1-2019m12} &  \multicolumn{9}{c}{2020m1-2023m12}  \\
\cmidrule(lr){2-10} \cmidrule(lr){11-19}
\multicolumn{1}{r|}{$h \rightarrow$}&  \multicolumn{1}{c}{$1$} & &  \multicolumn{1}{c}{$3$} & & \multicolumn{1}{c}{$6$} & & \multicolumn{1}{c}{$12$} & & \multicolumn{1}{c}{$24$}&  \multicolumn{1}{c}{$1$} & &  \multicolumn{1}{c}{$3$} & & \multicolumn{1}{c}{$6$} & & \multicolumn{1}{c}{$12$} & & \multicolumn{1}{c}{$24$}  \\
\midrule
\addlinespace[5pt] 
\rowcolor{gray!15} 
\multicolumn{6}{l}{Level 2 ($K = 12$)}   & & &&&&& & & & & & & \cellcolor{gray!15}  \\ \addlinespace[2pt]
Albacore$_\text{\tiny comps}$ & 1.00 &  & 0.98 &  & 0.94 &  & \textbf{\color{ForestGreen}0.84} &  & \textbf{\color{ForestGreen}0.75} & \textbf{0.96} &  & \textbf{\color{ForestGreen}0.90} &  & 0.96 &  & 0.99 &  & 1.02 \\ 
Albacore$_\text{\tiny ranks}$ & \textbf{\color{ForestGreen}0.98} &  & \textbf{\color{ForestGreen}0.95} &  & \textbf{\color{ForestGreen}0.90} &  & \textbf{\color{ForestGreen}0.84} &  & 0.78 & 1.02 &  & 0.97 &  & \textbf{0.95} &  & \textbf{0.94} &  & \textbf{0.95} \\ 
 \midrule 
 \rowcolor{gray!15} 
  \multicolumn{6}{l}{Level 3 ($K = 39$)}  & & &&&&& & & & & & & \cellcolor{gray!15}  \\ \addlinespace[2pt]
Albacore$_\text{\tiny comps}$ & \textbf{0.99} &  & \textbf{0.98} &  & 0.93 &  & 0.88 &  & \textbf{0.81} & 1.00 &  & \textbf{1.00} &  & 1.01 &  & 1.00 &  & 1.07 \\ 
Albacore$_\text{\tiny ranks}$ & 1.00 &  & 0.98 &  & \textbf{\color{ForestGreen}0.90} &  & \textbf{0.85} &  & 0.85 & 1.02 &  & \textbf{0.98} &  & \textbf{0.97} &  & \textbf{0.95} &  & \textbf{0.94} \\ 
 \midrule 
  \rowcolor{gray!15}
 \multicolumn{6}{l}{Level 4 ($K = 92$)}  & & &&&&& & & & & & & \cellcolor{gray!15}  \\ \addlinespace[2pt]
  Albacore$_\text{\tiny comps}$ & \textbf{1.00} &  & 1.03 &  & 0.96 &  & 1.00 &  & \textbf{0.85} & 0.99 &  & 0.96 &  & 1.06 &  & 0.97 &  & 1.04 \\ 
  Albacore$_\text{\tiny ranks}$ & 1.02 &  & \textbf{1.00} &  & \textbf{0.92} &  & \textbf{0.87} &  & 0.87 & \textbf{\color{ForestGreen}0.95} &  & \textbf{0.94} &  & \textbf{\color{ForestGreen}0.93} &  & \textbf{\color{ForestGreen}0.93} &  & \textbf{\color{ForestGreen}0.91} \\ 
\midrule 
  \rowcolor{gray!15}
 ${\mathbf{Benchmarks}}$  & & & & & & & &&&&& & & & & & & \cellcolor{gray!15}  \\ \addlinespace[2pt]
$\boldsymbol{X}_t^{\text{bm}}$, $\phantom{..}$($w_0=0$) & 1.00 &  & 0.98 &  & 0.95 &  & 0.90 &  & 0.78 & 0.97 &  & 0.96 &  & 0.98 &  & 1.02 &  & 1.09 \\ 
$\boldsymbol{X}_t^{\text{bm+}}$ & \textbf{\color{ForestGreen}0.98} &  & \textbf{\color{ForestGreen}0.95} &  & 0.92 &  & 0.98 &  & 1.00 & \textbf{\color{ForestGreen}0.95} &  & \textbf{\color{ForestGreen}0.90} &  & \textbf{\color{ForestGreen}0.93} &  & 0.97 &  & 1.00 \\ 
$\boldsymbol{X}_t^{\text{bm+}}$, ($w_0=0$) & 0.99 &  & 0.97 &  & 0.91 &  & 0.86 &  & 0.80 & 0.98 &  & 1.00 &  & 1.02 &  & 0.98 &  & 1.05 \\ 
\bottomrule \bottomrule
  \end{tabular}
  \begin{tablenotes}[para,flushleft]
  \scriptsize 
    \textit{Notes}: The table presents root mean square error (RMSE) relative to $\boldsymbol{X}_t^{\text{bm}} = [\text{HICP}_t \phantom{.}  \text{HICPX}_t \phantom{.}  \text{HICPtrim}_t  ]$  with intercept. The remaining benchmarks are: $\boldsymbol{X}_t^{\text{bm}} = [\text{HICP}_t \phantom{.}  \text{HICPX}_t \phantom{.}  \text{HICPtrim}_t  ]$ without an intercept (i.e., $w_0=0$), $\boldsymbol{X}_t^{\text{bm+}}$ with and without an intercept. Numbers in \textbf{bold} indicate the best model for each \textit{level} and each \textit{horizon} in each of the out-of-sample periods. Numbers highlighted in {\color{ForestGreen} green} show the best model per \textit{horizon} and out-of-sample period \textit{across levels}.
  \end{tablenotes}
\end{threeparttable}
\end{table}


\begin{figure}[h!]
  \caption{\normalsize{Qualbacore for the US ($h=3$)}} \label{fig:albacore_us_quant_h3}
  
  \begin{center}
    
    \vspace{-0.7cm}
    \begin{subfigure}[t]{0.5\textwidth}
      \centering
      \includegraphics[width=\textwidth, trim = 0mm 5mm 0mm 30mm, clip]{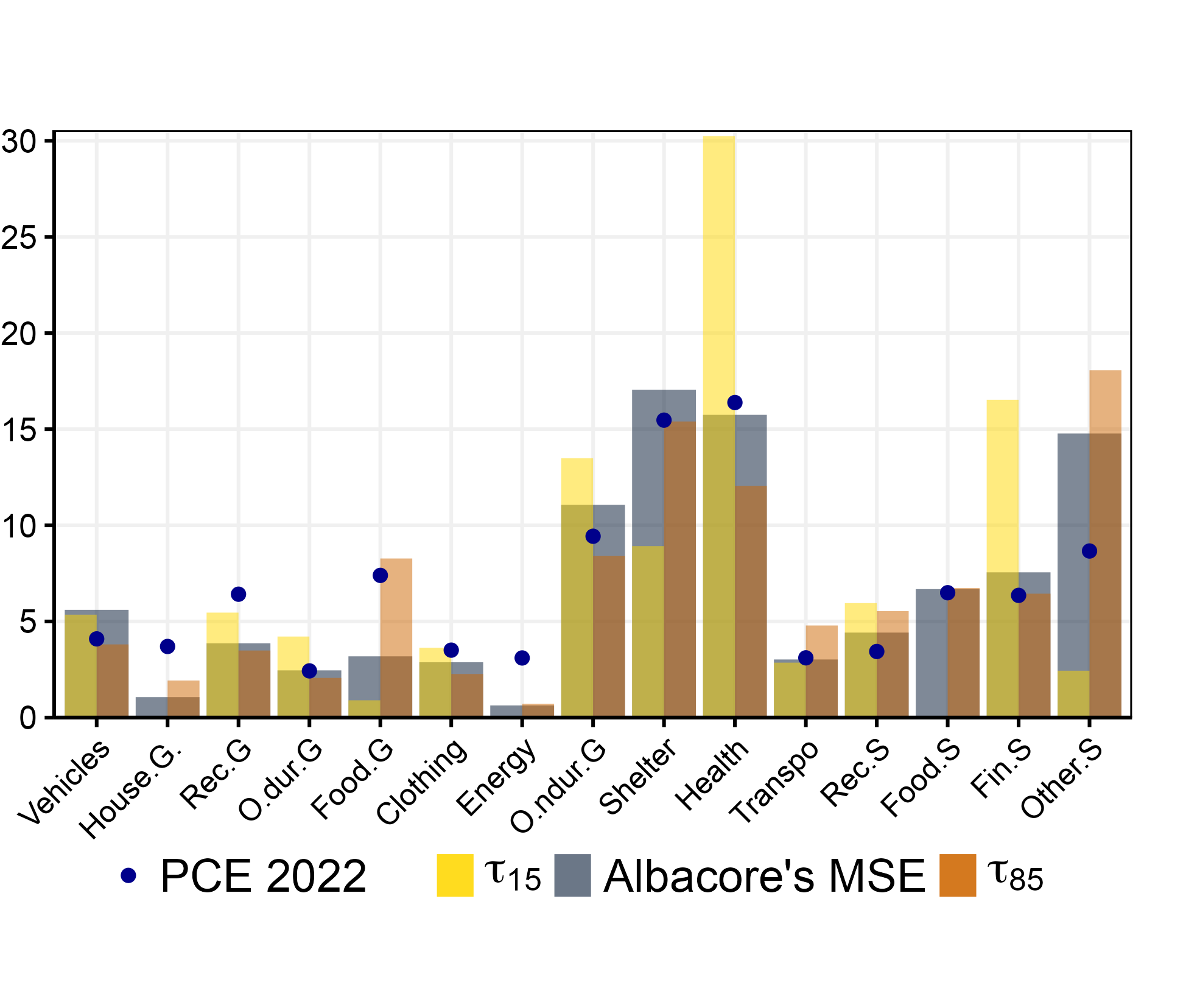}
      \vspace{-0.8cm}
      \caption{Comparison of Component Weights}
    \end{subfigure}%
    \begin{subfigure}[t]{0.515\textwidth}
      \centering
      \includegraphics[width=\textwidth, trim = 0mm 0mm 0mm 41mm, clip]{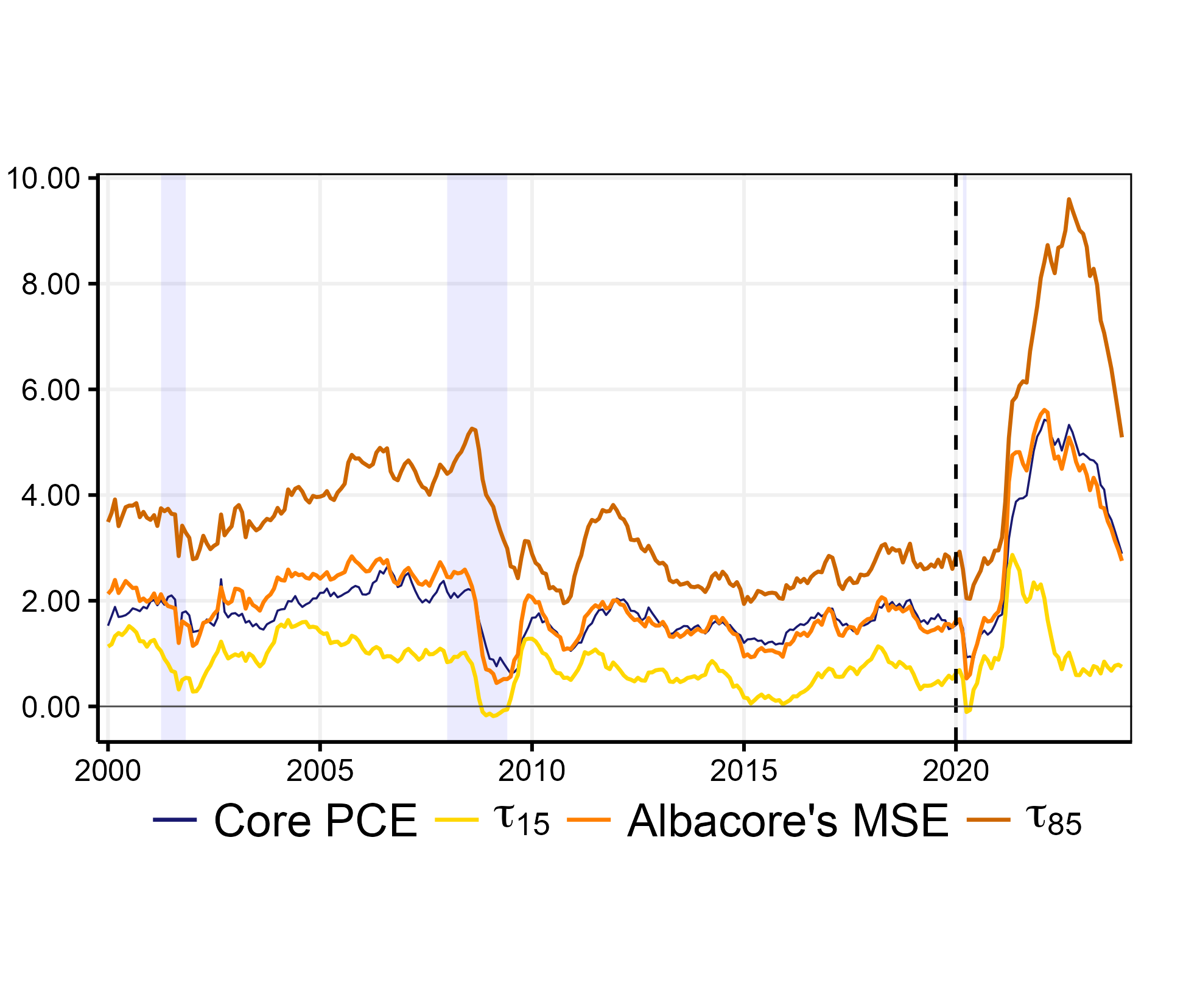}
      \vspace{-0.8cm}
      \caption{Time Series Comparison of Qualbacore$_\text{comps}$}
    \end{subfigure}

    \vspace{0.1cm}
    \begin{subfigure}[t]{0.5\textwidth}
      \centering
      \includegraphics[width=\textwidth, trim = 0mm -2mm 0mm 40mm, clip]{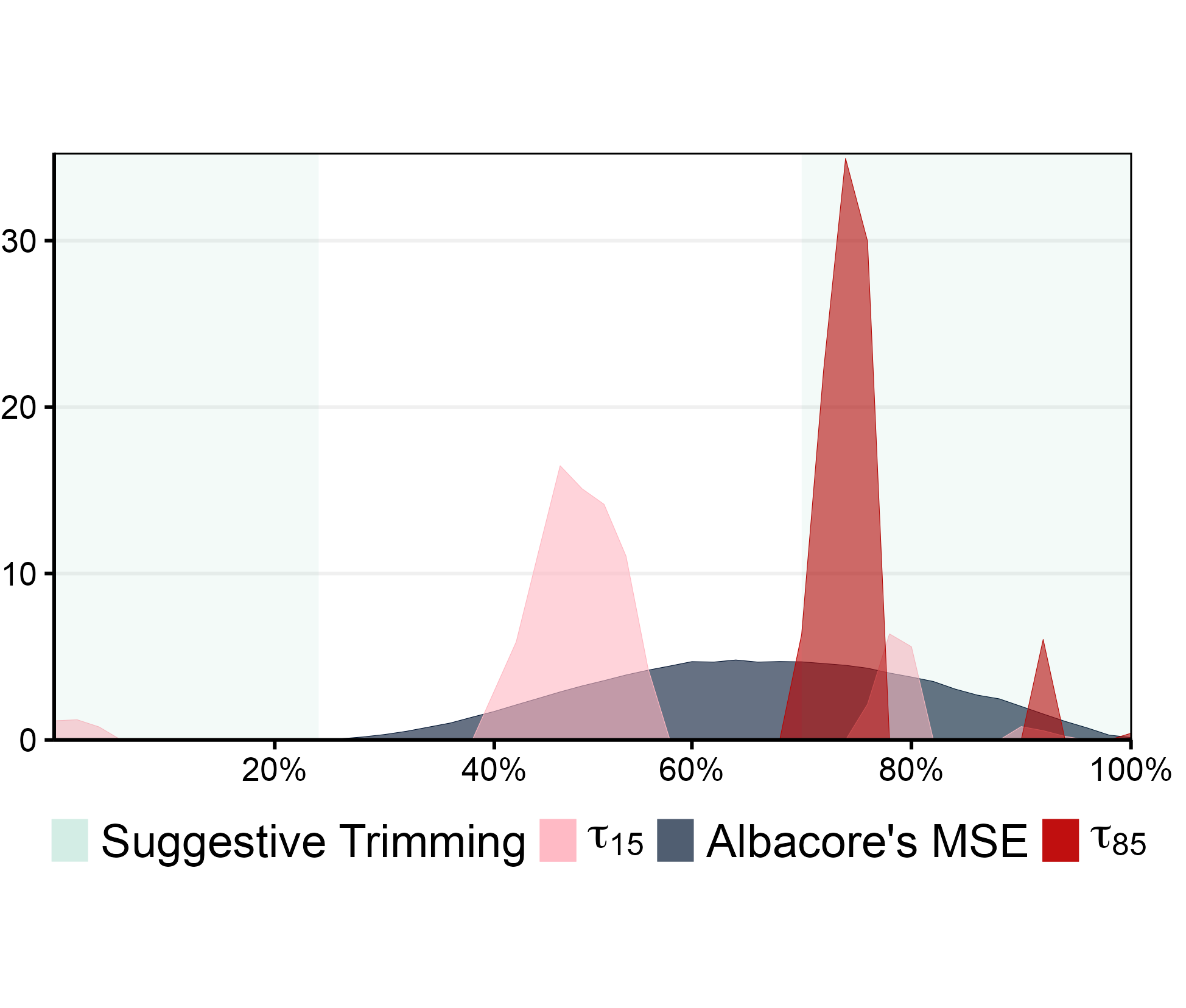}
      \vspace{-1.0cm}
      \caption{Comparison of Rank Weights}\label{fig:albacore_us_quant_h3c}
    \end{subfigure}%
    \begin{subfigure}[t]{0.515\textwidth}
      \centering
      \includegraphics[width=\textwidth, trim = 0mm 0mm 0mm 41mm, clip]{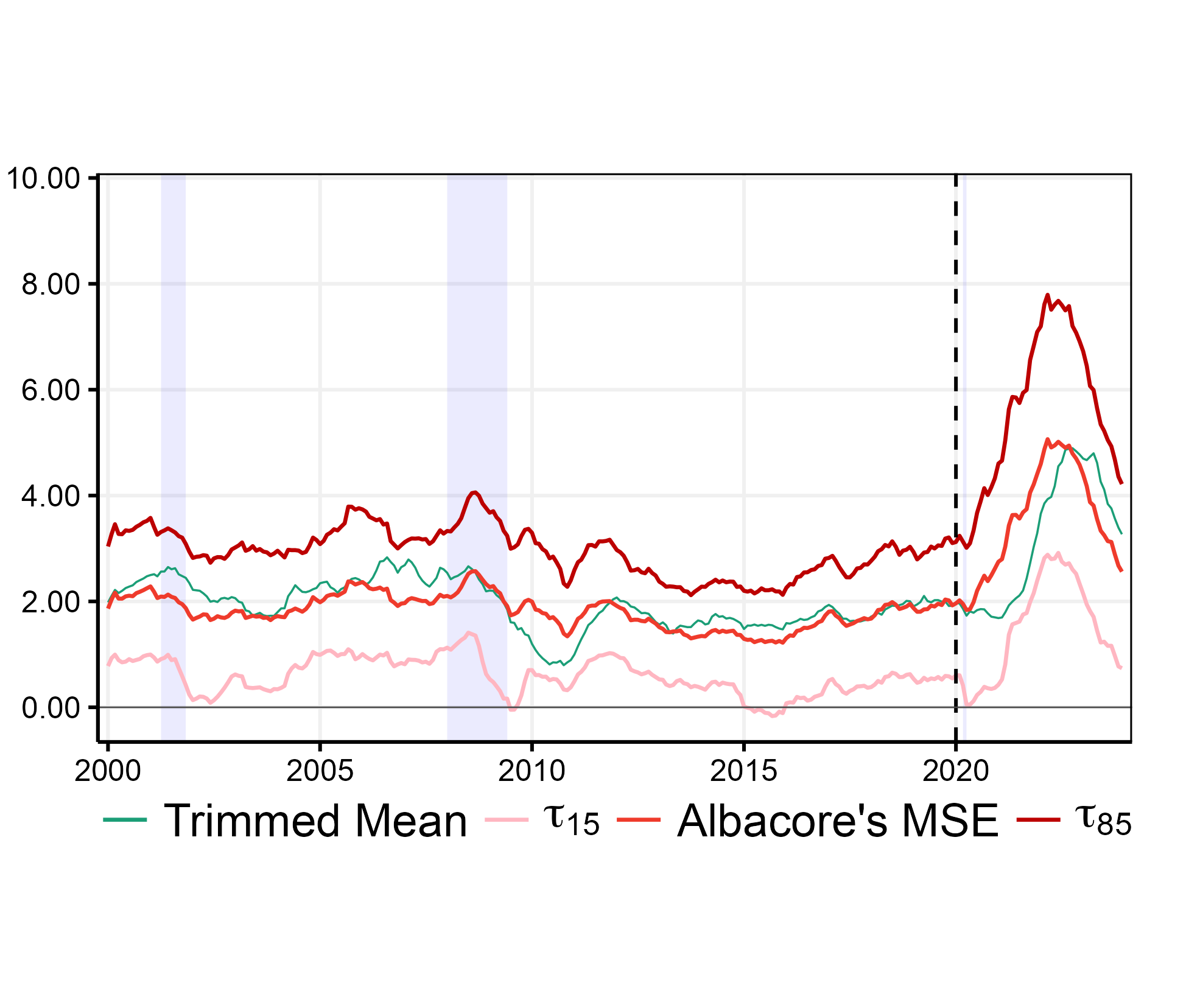}
      \vspace{-1.0cm}
      \caption{Time Series Comparison of Qualbacore$_\text{ranks}$}
    \end{subfigure}

  \end{center}

   \begin{threeparttable}
    \centering
    \begin{minipage}{\textwidth}
    \vspace*{-0.9cm}
      \begin{tablenotes}[para,flushleft]
    \setlength{\lineskip}{0.2ex}
    \notsotiny 
  {\textit{Notes}: All weights are based on models trained from 2001m1 to 2019m12. We present level 6 for Qualbacore$_\text{comps}$ and level 3 for Qualbacore$_\text{ranks}$. In the upper left panel we aggregate the component weights to the lowest level of disaggregation (level 2) for each quantile.}
    \end{tablenotes}
  \end{minipage}
  \end{threeparttable}

\end{figure}

\begin{table}[t!]
  \footnotesize
  \centering
  \begin{threeparttable}
    \caption{\normalsize  {Forecasting Performance of Albacore for the US with Different Moving Averages} \label{tab:results_us_popALL1}
      \vspace{-0.3cm}}
    \setlength{\tabcolsep}{0.69em} 
    \setlength\extrarowheight{2.9pt}
    \begin{tabular}{l| rrrrrrrrr | rrrrrrrrr } 
      \toprule \toprule 
      \addlinespace[2pt]
      &  \multicolumn{9}{c}{2010m1-2019m12} &  \multicolumn{9}{c}{2020m1-2023m12}  \\
      \cmidrule(lr){2-10} \cmidrule(lr){11-19}
      \multicolumn{1}{r|}{$h \rightarrow$}&  \multicolumn{1}{c}{$1$} & &  \multicolumn{1}{c}{$3$} & & \multicolumn{1}{c}{$6$} & & \multicolumn{1}{c}{$12$} & & \multicolumn{1}{c}{$24$}&  \multicolumn{1}{c}{$1$} & &  \multicolumn{1}{c}{$3$} & & \multicolumn{1}{c}{$6$} & & \multicolumn{1}{c}{$12$} & & \multicolumn{1}{c}{$24$}  \\
      \midrule
       \addlinespace[5pt] 
      \rowcolor{gray!15} 
      \multicolumn{9}{l}{Albacore$_\text{\tiny comps}$ Level 2 ($K = 15$)}   & & &&&&& & & &  \cellcolor{gray!15}  \\ \addlinespace[2pt]
      MoM  & 1.06 &  & 1.13 &  & 1.23 &  & 1.36 &  & 1.38 & 0.92 &  & 0.97 &  & 0.82 &  & 0.74 &  & 0.96 \\
      QoQ & 1.02 &  & 1.02 &  & 1.08 &  & 1.13 &  & 1.12 & 0.94 &  & 0.89 &  & 0.85 &  & 0.63 &  & 0.88 \\
      6Mo6M& 1.06 &  & 1.09 &  & 1.12 &  & 1.12 &  & 1.05 & 0.95 &  & 0.94 &  & 0.85 &  & 0.58 &  & 0.87 \\  
      YoY& 1.05 &  & 1.02 &  & 1.08 &  & 1.08 &  & 0.97 & 1.01 &  & 1.03 &  & 0.91 &  & 0.73 &  & 1.01 \\ 
      \midrule
      \addlinespace[5pt] 
      \rowcolor{gray!15} 
       \multicolumn{9}{l}{Albacore$_\text{\tiny comps}$ Level 3 ($K = 50$)}  & & &&&&& & & &  \cellcolor{gray!15}  \\ \addlinespace[2pt]
        MoM  & 1.03 &  & 1.07 &  & 1.25 &  & 1.25 &  & 1.30 & 0.91 &  & 0.92 &  & 0.81 &  & 0.75 &  & 0.91 \\ 
       QoQ & 1.02 &  & 1.05 &  & 1.15 &  & 1.14 &  & 1.06 & 0.91 &  & 0.87 &  & 0.78 &  & 0.75 &  & 1.01 \\
       6Mo6M & 1.06 &  & 1.07 &  & 1.08 &  & 1.14 &  & 1.03 & 0.89 &  & 0.91 &  & 0.88 &  & 0.77 &  & 1.10 \\ 
       YoY& 1.05 &  & 1.03 &  & 1.05 &  & 1.05 &  & 0.85 & 1.01 &  & 1.05 &  & 0.94 &  & 0.88 &  & 1.15 \\ 
      \midrule
\addlinespace[5pt] 
\rowcolor{gray!15} 
\multicolumn{9}{l}{Albacore$_\text{\tiny comps}$ Level 6 ($K = 215$)}  & & &&&&& & & &  \cellcolor{gray!15}  \\ \addlinespace[2pt]
MoM  & 1.06 &  & 1.21 &  & 1.36 &  & 1.26 &  & 1.26 & 0.94 &  & 0.96 &  & 0.90 &  & 0.72 &  & 0.92 \\
QoQ & 1.09 &  & 1.03 &  & 1.15 &  & 1.19 &  & 1.16 & 0.96 &  & 0.91 &  & 0.82 &  & 0.71 &  & 0.99 \\ 
6Mo6M & 1.05 &  & 1.02 &  & 1.14 &  & 1.26 &  & 1.14 & 0.97 &  & 0.97 &  & 0.68 &  & 0.74 &  & 1.04 \\  
YoY& 1.05 &  & 1.01 &  & 1.17 &  & 1.10 &  & 1.00 & 0.90 &  & 0.96 &  & 0.97 &  & 0.88 &  & 1.19 \\  
\addlinespace[5pt] 
\specialrule{1.3pt}{1pt}{1pt} 
\addlinespace[5pt] 
\rowcolor{gray!15} 
\multicolumn{9}{l}{Albacore$_\text{\tiny ranks}$ Level 2 ($K= 15$)}   & & &&&&& & & &\cellcolor{gray!15}  \\ \addlinespace[2pt]
MoM  & 1.03 &  & 1.01 &  & 1.07 &  & 1.08 &  & 1.03 & 0.87 &  & 0.80 &  & 0.61 &  &  \textbf{\color{ForestGreen}0.56}&  & 0.71 \\
QoQ  & 1.01 &  & 0.97 &  & 0.99 &  & 0.98 &  & 0.87 & 0.84 &  & 0.74 &  &  \textbf{\color{ForestGreen}0.57} &  & 0.59 &  & 0.69 \\
6Mo6M& 1.02 &  & 1.00 &  & 0.99 &  & 0.93 &  & 0.83 & 0.83 &  &  \textbf{\color{ForestGreen}0.71 }&  & 0.67 &  & 0.64 &  & 0.74 \\
YoY& 1.02 &  & 0.99 &  & 0.95 &  & 0.83 &  & 0.73 & 0.87 &  & 0.81 &  & 0.70 &  & 0.68 &  & 0.81 \\ 
\midrule
\addlinespace[5pt] 
\rowcolor{gray!15} 
\multicolumn{9}{l}{Albacore$_\text{\tiny ranks}$ Level 3 ($K = 50$)}  & & &&&&& & & &  \cellcolor{gray!15}  \\ \addlinespace[2pt]
MoM  & 0.98 &  & 0.93 &  & 0.95 &  & 0.91 &  & 0.80 & 0.85 &  & 0.78 &  & 0.63 &  & 0.58 &  & 0.65 \\ 
QoQ &  \textbf{\color{ForestGreen}0.97} &  & 0.92 &  & 0.93 &  & 0.87 &  & 0.73 &  \textbf{\color{ForestGreen}0.81} &  & 0.76 &  & 0.61 &  &  \textbf{\color{ForestGreen}0.56} &  &  \textbf{\color{ForestGreen}0.63} \\ 
6Mo6M & 1.00 &  & 0.96 &  & 0.94 &  & 0.85 &  & 0.72 & 0.86 &  & 0.78 &  & 0.65 &  & 0.60 &  & 0.64 \\  
YoY& 1.01 &  & 0.96 &  & 0.90 &  &  \textbf{\color{ForestGreen}0.76} &  &  \textbf{\color{ForestGreen}0.66}& 0.88 &  & 0.81 &  & 0.65 &  & 0.61 &  & 0.71 \\  
\midrule
\addlinespace[5pt] 
\rowcolor{gray!15} 
\multicolumn{9}{l}{Albacore$_\text{\tiny ranks}$ Level 6 ($K = 215$)}  & & &&&&& & & &  \cellcolor{gray!15}  \\ \addlinespace[2pt]
MoM  & 1.01 &  & 0.92 &  &  \textbf{\color{ForestGreen}0.88}&  & 0.85 &  & 0.80 & 0.80 &  & 0.81 &  & 0.70 &  & 0.62 &  & 0.68 \\
QoQ & 0.98 &  &  \textbf{\color{ForestGreen}0.91} &  &  \textbf{\color{ForestGreen}0.88}&  & 0.84 &  & 0.77 & 0.86 &  & 0.82 &  & 0.70 &  & 0.62 &  & 0.67 \\ 
6Mo6M  & 0.99 &  & 0.93 &  & 0.93 &  & 0.85 &  & 0.75 & 0.88 &  & 0.86 &  & 0.73 &  & 0.62 &  & 0.65 \\ 
YoY& 1.00 &  & 0.96 &  & 0.93 &  & 0.85 &  & 0.70 & 0.94 &  & 0.90 &  & 0.75 &  & 0.66 &  &  \textbf{\color{ForestGreen}0.63}\\ 
      \bottomrule \bottomrule
\end{tabular} 
\begin{tablenotes}[para,flushleft]
\scriptsize 
\textit{Notes}: The table presents root mean square error (RMSE) relative to $\boldsymbol{X}_t^{\text{bm}} = [\text{PCE}_t \phantom{.}  \text{PCEcore}_t \phantom{.}  \text{PCEtrim}_t  ]$  with intercept. The transformations of the regressors are: month-over-month (MoM), quarter-over-quarter (QoQ), 6-months-over-6-months (6Mo6M), year-over-year (YoY). Numbers in \textbf{bold} indicate the best model for each \textit{level} and each \textit{horizon} in each of the out-of-sample periods. Numbers highlighted in {\color{ForestGreen} green} show the best model per \textit{horizon} and out-of-sample period \textit{across levels}.
\end{tablenotes}
\end{threeparttable}
\end{table}

\begin{table}[t!]
  \footnotesize
  \centering
  \begin{threeparttable}
    \caption{\normalsize {Forecasting Performance of Benchmarks for the US with Different Moving Averages} \label{tab:results_us_popALL2}
      \vspace{-0.3cm}}
    \setlength{\tabcolsep}{0.69em} 
    \setlength\extrarowheight{2.9pt}
    \begin{tabular}{l| rrrrrrrrr | rrrrrrrrr } 
      \toprule \toprule 
      \addlinespace[2pt]
      &  \multicolumn{9}{c}{2010m1-2019m12} &  \multicolumn{9}{c}{2020m1-2023m12}  \\
      \cmidrule(lr){2-10} \cmidrule(lr){11-19}
      \multicolumn{1}{r|}{$h \rightarrow$}&  \multicolumn{1}{c}{$1$} & &  \multicolumn{1}{c}{$3$} & & \multicolumn{1}{c}{$6$} & & \multicolumn{1}{c}{$12$} & & \multicolumn{1}{c}{$24$}&  \multicolumn{1}{c}{$1$} & &  \multicolumn{1}{c}{$3$} & & \multicolumn{1}{c}{$6$} & & \multicolumn{1}{c}{$12$} & & \multicolumn{1}{c}{$24$}  \\
      \midrule
      \addlinespace[5pt] 
      \rowcolor{gray!15}
    $\boldsymbol{X}_t^{\text{bm}}$ & & & & & & & &&&&& & & & & & & \cellcolor{gray!15}  \\ \addlinespace[2pt]
    MoM  & 0.96 &  & 0.97 &  & 0.97 &  & 0.98 &  & 1.00 & 0.94 &  & 1.12 &  & 1.07 &  & 0.99 &  & 1.03 \\
    QoQ& 1.00 &  &  1.00 &  &  1.00 &  & 1.00&  &  1.00& 1.00&  &  1.00 &  &  1.00 &  &  1.00 &  & 1.00 \\ 
    6Mo6M  & 1.03 &  & 1.02 &  & 1.02 &  & 1.01 &  & 0.98 & 0.97 &  & 1.01 &  & 1.00 &  & 1.01 &  & 1.00 \\ 
    YoY& 1.03 &  & 1.01 &  & 1.01 &  & 0.98 &  & 0.97 & 1.05 &  & 1.19 &  & 1.13 &  & 0.96 &  & 1.03 \\
      \midrule 
      \rowcolor{gray!15}
      $\boldsymbol{X}_t^{\text{bm}}$, $\phantom{..}$($w_0=0$) & & & & & & & &&&&& & & & & & & \cellcolor{gray!15}  \\ \addlinespace[2pt]
      MoM  & 0.97 &  & 1.00 &  & 1.04 &  & 1.09 &  & 1.03 & 0.87 &  & 0.95 &  & 0.86 &  & 0.84 &  & 0.94 \\ 
      QoQ& 1.00 &  & 0.99 &  & 1.03 &  & 1.03 &  & 0.96 & 0.91 &  & 0.89 &  & 0.79 &  & 0.80 &  & 0.93 \\ 
      6Mo6M  & 1.04 &  & 1.02 &  & 1.02 &  & 1.01 &  & 0.90 & 0.86 &  & 0.84 &  & 0.81 &  & 0.80 &  & 0.93 \\ 
      YoY& 1.02 &  & 0.99 &  & 0.99 &  & 0.93 &  & 0.82 & 0.92 &  & 0.96 &  & 0.90 &  & 0.86 &  & 0.99 \\ 
        \midrule 
      \rowcolor{gray!15}
      $\boldsymbol{X}_t^{\text{bm+}}$ & & & & & & & &&&&& & & & & & & \cellcolor{gray!15}  \\ \addlinespace[2pt]
      MoM  & 0.96 &  & 0.97 &  & 0.98 &  & 0.99 &  & 1.01 & 0.94 &  & 1.18 &  & 1.11 &  & 1.01 &  & 1.05 \\
      QoQ& 1.01 &  & 1.04 &  & 1.04 &  & 1.04 &  & 1.02 & 1.37 &  & 1.48 &  & 1.18 &  & 1.14 &  & 1.09 \\ 
      6Mo6M   & 1.05 &  & 1.06 &  & 1.07 &  & 1.09 &  & 1.01 & 1.13 &  & 1.19 &  & 1.08 &  & 1.03 &  & 1.08 \\
      YoY& 0.98 &  & 0.96 &  & 0.97 &  & 1.02 &  & 1.00 & 0.89 &  & 0.86 &  & 0.84 &  & 0.98 &  & 1.05 \\
        \midrule 
      \rowcolor{gray!15}
      $\boldsymbol{X}_t^{\text{bm+}}$, ($w_0=0$) & & & & & & & &&&&& & & & & & & \cellcolor{gray!15}  \\ \addlinespace[2pt]
      MoM  & 0.97 &  & 1.00 &  & 1.05 &  & 1.09 &  & 1.03 & 0.87 &  & 1.01 &  & 0.88 &  & 0.86 &  & 0.95 \\ 
      QoQ& 1.02 &  & 1.04 &  & 1.07 &  & 1.07 &  & 0.97 & 1.10 &  & 1.15 &  & 0.86 &  & 0.87 &  & 0.96 \\ 
      6Mo6M  & 1.07 &  & 1.06 &  & 1.07 &  & 1.05 &  & 0.91 & 0.95 &  & 0.88 &  & 0.78 &  & 0.79 &  & 0.95 \\ 
      YoY & 0.99 &  & 0.96 &  & 0.97 &  & 0.97 &  & 0.85 & 0.83 &  & 0.79 &  & 0.69 &  & 0.74 &  & 0.88 \\ 
      \bottomrule \bottomrule
    \end{tabular} 
    \begin{tablenotes}[para,flushleft]
      \scriptsize 
      \textit{Notes}: The table presents root mean square error (RMSE) relative to $\boldsymbol{X}_t^{\text{bm}} = [\text{PCE}_t \phantom{.}  \text{PCEcore}_t \phantom{.}  \text{PCEtrim}_t  ]$  with intercept. The remaining benchmarks are: $\boldsymbol{X}_t^{\text{bm}} = [\text{PCE}_t \phantom{.}  \text{PCEcore}_t \phantom{.}  \text{PCEtrim}_t  ]$ without an intercept (i.e., $w_0=0$), $\boldsymbol{X}_t^{\text{bm+}}$ with and without an intercept. The transformations of the regressors are: month-over-month (MoM), quarter-over-quarter (QoQ), 6-months-over-6-months (6Mo6M), year-over-year (YoY). Numbers in \textbf{bold} indicate the best model for each \textit{level} and each \textit{horizon} in each of the out-of-sample periods. Numbers highlighted in {\color{ForestGreen} green} show the best model per \textit{horizon} and out-of-sample period \textit{across levels}.
    \end{tablenotes}
  \end{threeparttable}
\end{table}

\clearpage

\subsection{Embracing Sparsity?} \label{sec:sparse}

\noindent The main analysis focuses on the most disaggregated level as it often yields the best prediction results, and it is this high-dimensional space in which our penalized regression encounters the largest sandbox to explore non-trivial predictability patterns (i.e., items are not forced into groups ex-ante).  However,  Section \ref{sec:usresults}'s forecasting results show that highly aggregated regressions can also provide impressive results, especially for the periods covering the inflation surge (see Table \ref{tab:results_us_pop3}).   This is achieved without the need to think deeply about  the regularization scheme and its strength.  The grouping itself can be interpreted as  regularization in the form of constraining the weights of all the subcomponent to equal the weight of a higher-level component, respectively shrink them towards the official PCE weights in lockstep.  While the performance and practicality of such regressions is desirable,  the allocation of components they provide, can map into rather unrealistic consumption weights.  Still,  their surprising performance warrants an investigation.


\begin{figure}[h!]
\vspace{0.2cm}
  \caption{\normalsize{\acc for Aggregated Components}} \label{fig:albacore_us_sparse}
  \begin{center}
    \vspace{-0.7cm}
    \begin{subfigure}[t]{0.5\textwidth}
      \centering
      \includegraphics[width=\textwidth, trim = 0mm 0mm 0mm 0mm, clip]{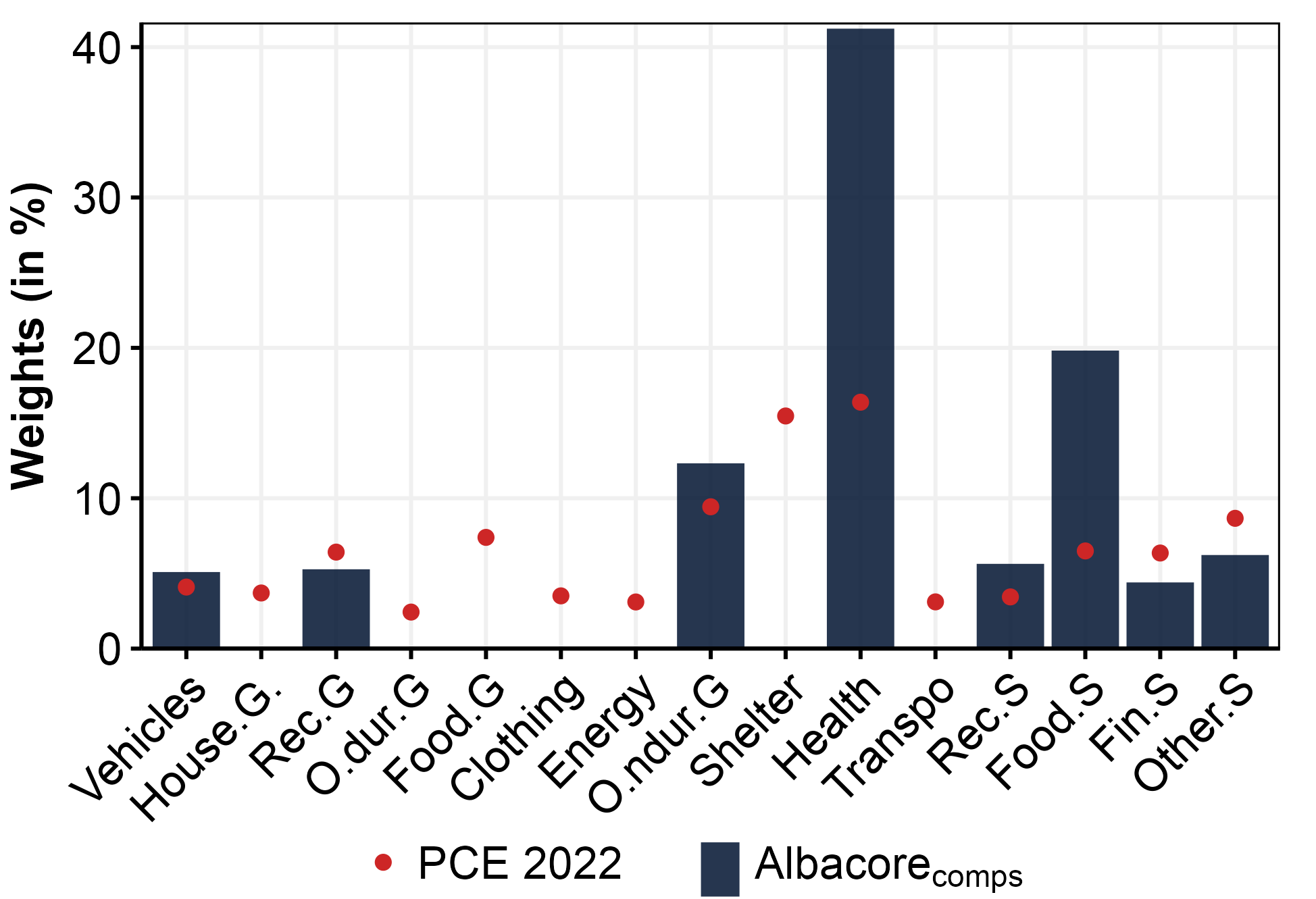}
      \vspace{0.08cm}
      \caption{Comparison of Component Weights}
    \end{subfigure}%
    \begin{subfigure}[t]{0.515\textwidth}
      \centering
      \includegraphics[width=\textwidth, trim = 0mm -8mm 0mm 0mm, clip]{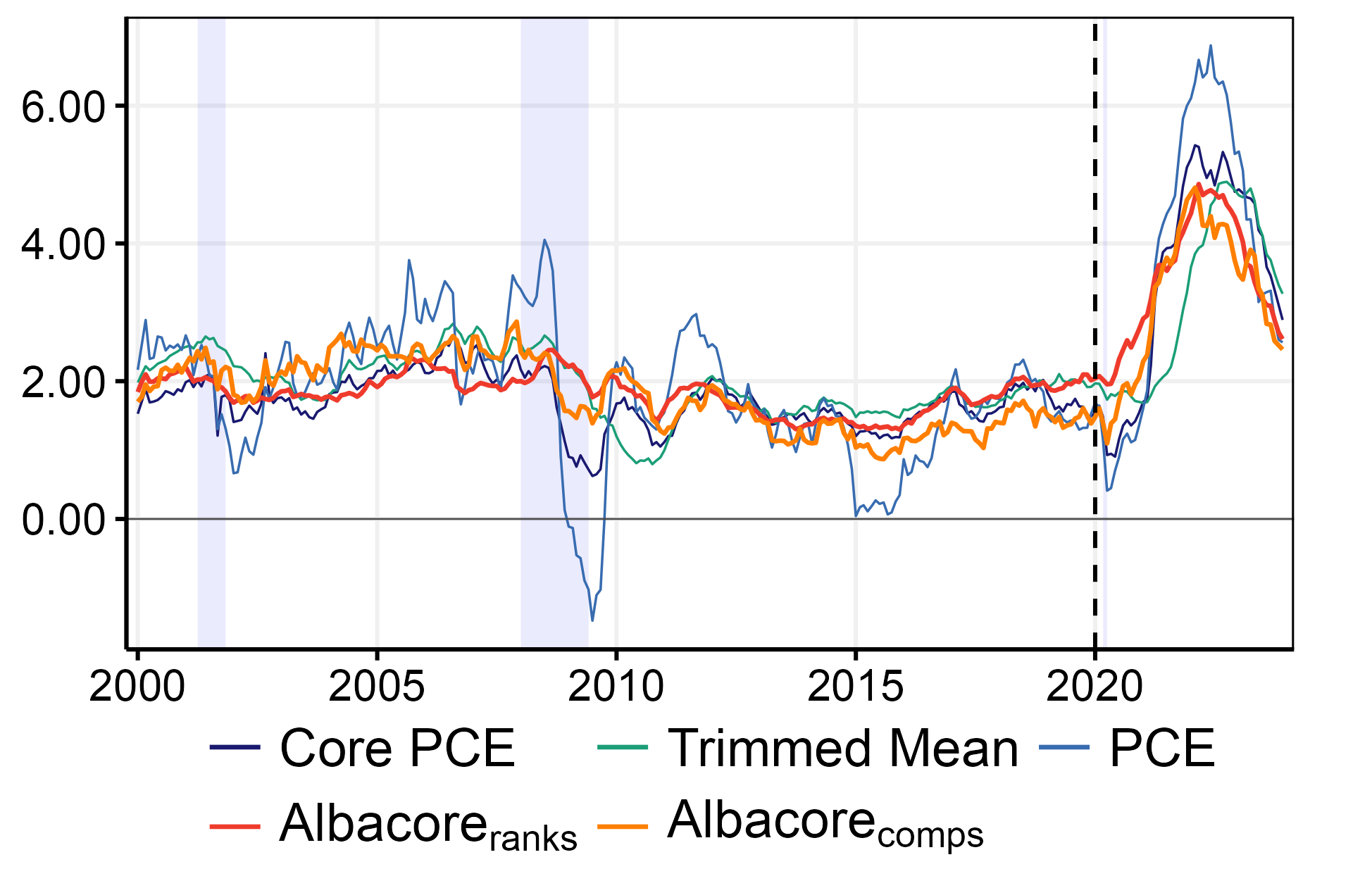}
      \vspace{0.08cm}
      \caption{Time Series Comparison}
    \end{subfigure}%

  \end{center}

    \begin{threeparttable}
    \centering
    \begin{minipage}{\textwidth}
    \vspace*{-0.7cm}
      \begin{tablenotes}[para,flushleft]
    \setlength{\lineskip}{0.2ex}
    \notsotiny 
  {\textit{Notes}: We show Albacore$_\text{comps}$ in level 2 and Albacore$_\text{ranks}$ in level 3. Results are based on models trained from 2001m1 to 2019m12.}
    \end{tablenotes}
  \end{minipage}
  \end{threeparttable}
\end{figure}

As can be seen in Figure \ref{fig:albacore_us_sparse} (right panel), when assembling components at their highest level of aggregation (level 2), \acc  features high predictive qualities. In line with \acr, it provides early signs of the inflation surge following the outbreak of the Covid-19 pandemic. In contrast to officially reported inflation measures, \acc shows a rather small and short-lived dip in 2020m4, which turns into a fast and strong run-up thereafter. It peaks in 2022m2, indicating the onset of the disinflationary process early. While other core inflation measures remain elevated throughout 2022, \acc in level 2 declines rather fast and lands at 2.5\% in 2023m12.

The left panel reveals that \acc gives a rather sparse solution in this case. Most weight gets assigned to items related to health (40\%), followed by food services, and other nondurable goods.  7 out of 15 components are entirely excluded from the aggregate.  Those include volatile items such as food goods and energy, as well as lagging indicators such as shelter, and slow-growing ones such as durable goods and clothing. Clearly, the resulting consumption basket is hardly representative of a plausible consumer. Moreover, relying on a few components makes \acc highly dependent on developments in certain sectors and sector-specific shocks may thus overly influence its dynamics, raising concerns about the generalization of such an allocation. Nonetheless, its forecasting performance is undeniable, and as such,  is a price pressure indicator worth keeping an eye on.

\subsection{Albacore for Canada} \label{sec:canresults}

\begin{figure}[t!]
  \caption{\normalsize{Albacore for Canada}} \label{fig:albacore_can_h12}
  \begin{center}
    
    \begin{subfigure}[t]{\textwidth}
      \vspace*{-0.7cm}
      \centering
      \includegraphics[width=0.98\textwidth, trim = 0mm -10mm 0mm 0mm, clip]{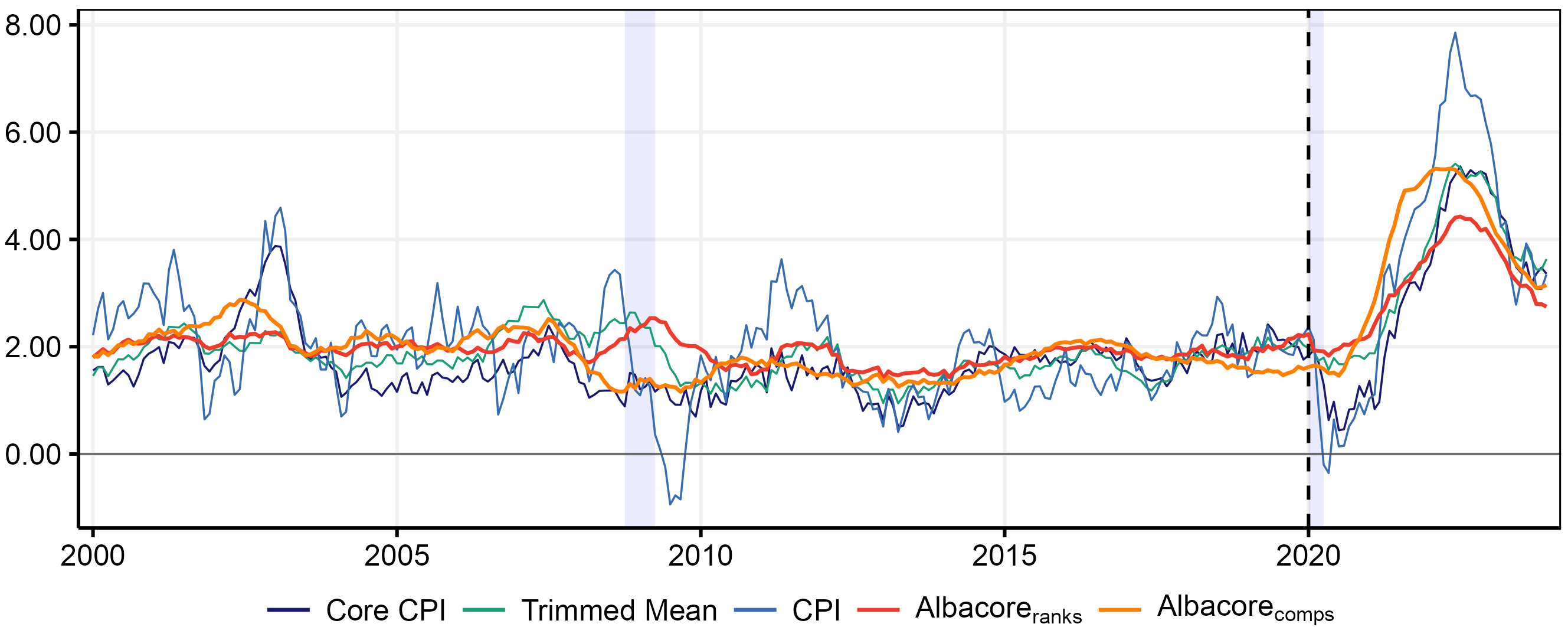}
    \end{subfigure}%
    
    \vspace*{-0.3cm}
    \begin{subfigure}[t]{0.51\textwidth}
      \centering
      \includegraphics[width=\textwidth, trim = 1mm 0mm 0mm -5mm, clip]{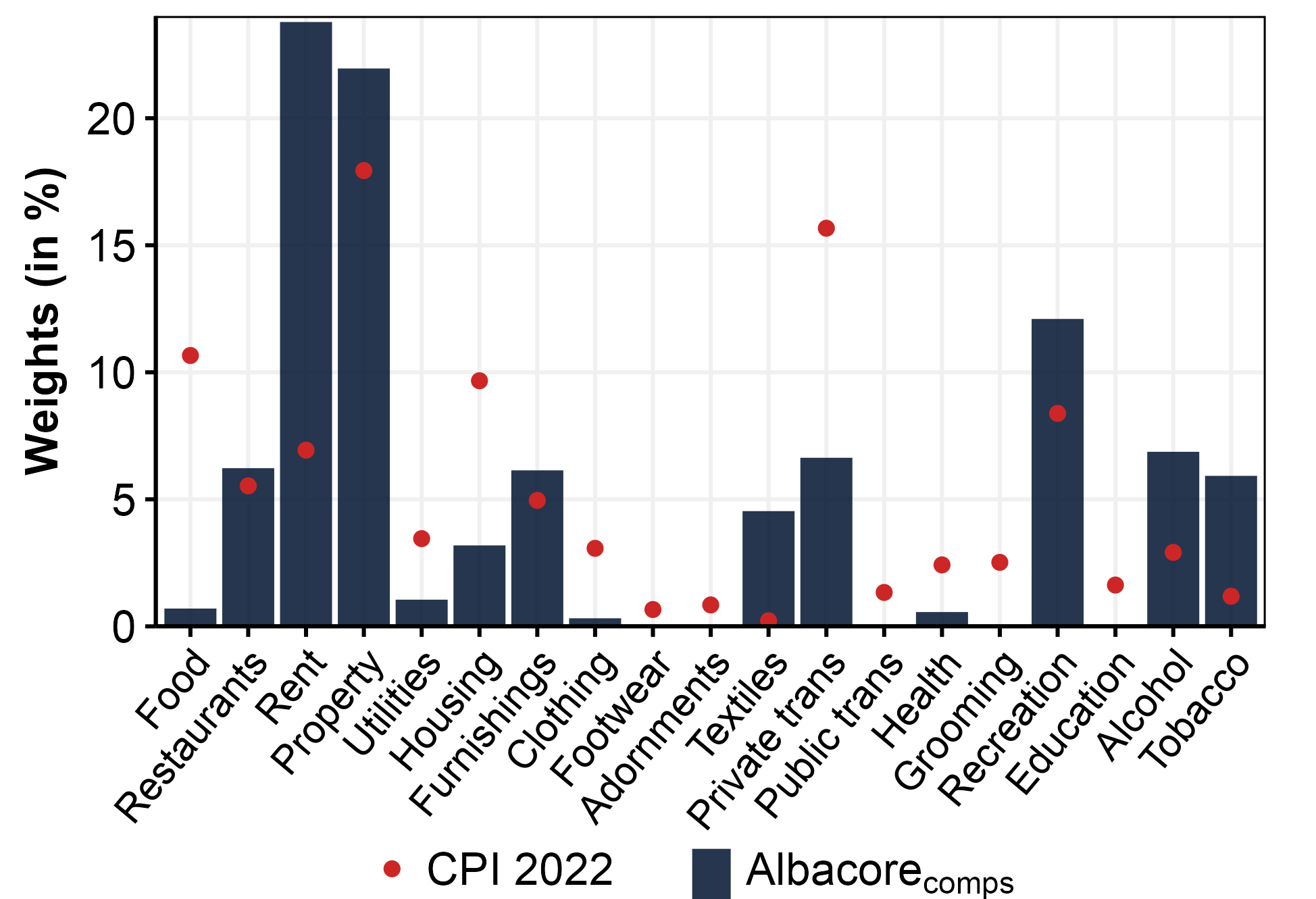}
      \vspace*{0.07cm}
      \caption{Comparison in Component Weights}\label{fig:albacore_can_h12_comps}
    \end{subfigure}%
    \begin{subfigure}[t]{0.515\textwidth}
      \centering
      \includegraphics[width=\textwidth, trim = 0mm -19mm -12mm 0mm, clip]{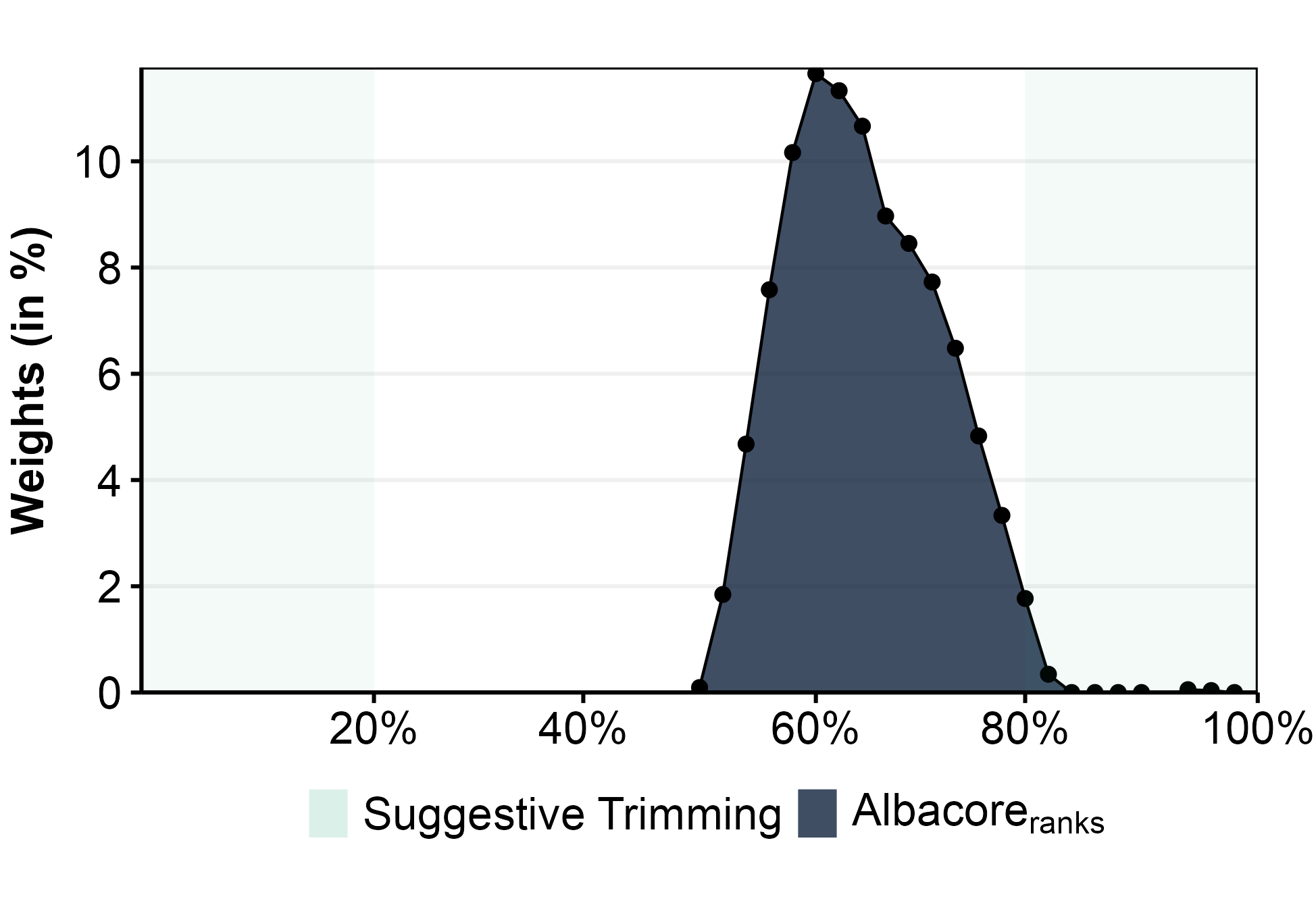}
      \vspace*{0.07cm}
      \caption{Comparison in Rank Weights}\label{fig:albacore_can_h12_comps_ranks}
    \end{subfigure}%
  \end{center}

  \begin{threeparttable}
    \centering
    \begin{minipage}{\textwidth}
    \vspace*{-0.7cm}
      \begin{tablenotes}[para,flushleft]
    \setlength{\lineskip}{0.2ex}
    \notsotiny 
  {\textit{Notes}: For more details we refer to Figure \ref{fig:albacore_us_h12}. Country-specific changes can be summarized as follows: For Canada, we present Albacore in level 4. For the trimmed mean, we choose the CPI-trim. The green shaded area marks the excluded share of monthly price changes as is suggested by the trimmed mean before reweighting (20\% of the lower tail and 20\% of the upper tail). Details on the labeling can be found in Appendix \ref{sec:data}.}
    \end{tablenotes}
  \end{minipage}
  \end{threeparttable}
\end{figure}

In addition to the analysis on the US and the EA, we introduce Albacore for Canada. Following our main procedure, we present its forecasting performance in Table \ref{tab:results_can_pop3}, and provide detailed insights on the proposed measure in Figure \ref{fig:albacore_can_h12}.  The main challenge with the Canadian data is its variable quality,  especially when looking at intra-year disaggregated price series. We use  CPI data taken from Statistics Canada in level 3, 4, and 5, implying a combination of 19, 49, and 87 components, respectively.  We remove 11 components in level 4 and 26 components in level 5 from the Canadian price data based on the fact that these items experience a substantial number of zero monthly growth rates, which may lead to model instabilites and includes little information to learn from. \textcolor{black}{Competitors for the Canadian Albacore are the officially reported CPI rate, CPI excluding energy and food (i.e., core CPI), CPI-trim, CPI-median and CPI x8.} 

Consistent with the findings in Section \ref{sec:emp}, convex combinations of existing core inflation measures demonstrate competitive predictive performances. Yet, Albacore is either broadly in line with the benchmarks (for pre-Covid observations) or superior (for the post-Covid sample), with increasing accuracy for higher-order forecasts. In the Canadian case, we find that the inclusion of an intercept in the pre-Covid sample is crucial, making it difficult for alternative models to yield improvements. When abstracting from benchmarks with intercept, we find consistent gains of Albacore over the benchmarks, with \acr topping the list for all horizons. Notably, these findings reverse for the post-pandemic era. Models with intercept give inferior results, and \acc emerges as the best performer. It displays superior predictive accuracy across all horizons with highest gains for the medium term ($h \in \{6,12\}$) and level 4.

\begin{table}[h!]
\vspace{0.2cm}
  \footnotesize
  \centering
  \begin{threeparttable}
    \caption{\normalsize {Forecasting Performance of Albacore for Canada} \label{tab:results_can_pop3}
      \vspace{-0.3cm}}
    \setlength{\tabcolsep}{0.4em} 
    \setlength\extrarowheight{2.5pt}
    \begin{tabular}{l| rrrrrrrrr | rrrrrrrrr } 
      \toprule \toprule
      \addlinespace[2pt]
      &  \multicolumn{9}{c}{2010m1-2019m12} &  \multicolumn{9}{c}{2020m1-2023m12}  \\
      \cmidrule(lr){2-10} \cmidrule(lr){11-19}
      \multicolumn{1}{r|}{$h \rightarrow$}&  \multicolumn{1}{c}{$1$} & &  \multicolumn{1}{c}{$3$} & & \multicolumn{1}{c}{$6$} & & \multicolumn{1}{c}{$12$} & & \multicolumn{1}{c}{$24$}&  \multicolumn{1}{c}{$1$} & &  \multicolumn{1}{c}{$3$} & & \multicolumn{1}{c}{$6$} & & \multicolumn{1}{c}{$12$} & & \multicolumn{1}{c}{$24$}  \\
      \midrule
      \addlinespace[5pt] 
      \rowcolor{gray!15} 
      \multicolumn{6}{l}{Level 3 ($K = 19$)}   & & &&&&& & & & & & & \cellcolor{gray!15}  \\ \addlinespace[2pt]
      Albacore$_\text{\tiny comps}$ & 1.02 &  & 1.04 &  & \textbf{1.05} &  & \textbf{1.13} &  & 1.37 & \textbf{0.83} &  & \textbf{0.73} &  & \textbf{0.68} &  & \textbf{0.63} &  & 0.85 \\
      Albacore$_\text{\tiny ranks}$& \textbf{1.01} &  & \textbf{1.01} &  & \textbf{1.05} &  & \textbf{1.13} &  & \textbf{1.15} & 0.84 &  & 0.78 &  & 0.74 &  & 0.74 &  & \textbf{0.78} \\ 
      \midrule 
      \rowcolor{gray!15} 
      \multicolumn{6}{l}{Level 4 ($K = 49$)}  & & &&&&& & & & & & & \cellcolor{gray!15}  \\ \addlinespace[2pt]
      Albacore$_\text{\tiny comps}$ & 1.05 &  & 1.10 &  & 1.12 &  & 1.23 &  & 1.24 & {\color{ForestGreen} \textbf{0.82}} &  & {\color{ForestGreen} \textbf{0.69}} &  & {\color{ForestGreen} \textbf{0.54}} &  & {\color{ForestGreen} \textbf{0.47}} &  & \textbf{0.72} \\ 
      Albacore$_\text{\tiny ranks}$& {\color{ForestGreen} \textbf{1.00}} &  & {\color{ForestGreen} \textbf{1.00}} &  & \textbf{1.07} &  & \textbf{1.15} &  & \textbf{1.18} & 0.89 &  & 0.82 &  & 0.68 &  & 0.67 &  & 0.76 \\  
      \midrule 
      \rowcolor{gray!15}
      \multicolumn{6}{l}{Level 5 ($K = 87$)}  & & &&&&& & & & & & & \cellcolor{gray!15}  \\ \addlinespace[2pt]
      Albacore$_\text{\tiny comps}$ & 1.02 &  & 1.09 &  & 1.10 &  & 1.22 &  & \textbf{1.24} & {\color{ForestGreen} \textbf{0.82}} &  & \textbf{0.73} &  & \textbf{0.69} &  & \textbf{0.55} &  & {\color{ForestGreen} \textbf{0.71}} \\  
      Albacore$_\text{\tiny ranks}$ & {\color{ForestGreen} \textbf{1.00}} &  & \textbf{1.01} &  & \textbf{1.05} &  & \textbf{1.12} &  & 1.27 & 0.90 &  & 0.83 &  & 0.76 &  & 0.71 &  & 0.74 \\ 
      \midrule 
      \rowcolor{gray!15}
      ${\mathbf{Benchmarks}}$  & & & & & & & &&&&& & & & & & & \cellcolor{gray!15}  \\ \addlinespace[2pt]
      $\boldsymbol{X}_t^{\text{bm}}$, $\phantom{..}$($w_0=0$) & 1.03 &  & 1.06 &  & 1.10 &  & 1.34 &  & 1.51 & 0.83 &  & 0.77 &  & 0.72 &  & 0.78 &  & 0.86 \\
      $\boldsymbol{X}_t^{\text{bm+}}$ & {\color{ForestGreen} \textbf{1.00}} &  & {\color{ForestGreen} \textbf{1.00}} &  & {\color{ForestGreen} \textbf{1.00}} &  & {\color{ForestGreen} \textbf{1.00}} &  & {\color{ForestGreen} \textbf{1.00}} & 1.00 &  & 1.00 &  & 1.00 &  & 1.00 &  & 1.00 \\  
      $\boldsymbol{X}_t^{\text{bm+}}$, ($w_0=0$) & 1.01 &  & 1.04 &  & 1.08 &  & 1.24 &  & 1.35 & 0.84 &  & 0.77 &  & 0.74 &  & 0.80 &  & 0.85 \\ 

      \bottomrule \bottomrule
    \end{tabular}
    \begin{tablenotes}[para,flushleft]
      \scriptsize 
      \textit{Notes}: The table presents root mean square error (RMSE) relative to $\boldsymbol{X}_t^{\text{bm}} = [\text{CPI}_t \phantom{.}  \text{CPIcore}_t \phantom{.}  \text{CPItrim}_t  ]$  with intercept. The remaining benchmarks are: $\boldsymbol{X}_t^{\text{bm}} = [\text{CPI}_t \phantom{.}  \text{CPIcore}_t \phantom{.}  \text{CPItrim}_t  ]$ without an intercept (i.e., $w_0=0$), $\boldsymbol{X}_t^{\text{bm+}}$ with and without an intercept. Numbers in \textbf{bold} indicate the best model for each \textit{level} and each \textit{horizon} in each of the out-of-sample periods. Numbers highlighted in {\color{ForestGreen} green} show the best model per \textit{horizon} and out-of-sample period \textit{across levels}.
    \end{tablenotes}
  \end{threeparttable}
\end{table}


Based on these results, our detailed analysis focuses on $h=12$ and level 4 (estimated using data up to December 2019). In-sample, all core inflation measures show similar trends, closely tracking the 2\% target. Deviations of \acc and core CPI from trimming-based approaches can be observed during the GFC, with the former downward trending already in 2008 followed by the trimmed mean and \acr in early-/mid-2009. Moreover, our proposed measures exhibit smooth trajectories compared to their well-established counterparts. Out-of-sample, both \acr and \acc demonstrate forward-looking qualities. 
\acc takes the lead in showing robust inflationary tendencies (already in 2020m7), as well as in indicating an easing of the strong upward pressures at the beginning of 2022. While benchmarks remain elevated until the end of 2022, \acc reaches the turning point in 2022m7 and starts descending thereafter. 
Notably, \acr for Canada is less estranged with trimmed mean inflation than in the US. Yet, it moves above target in the course of 2020, when officially reported series are still well below. We attribute these dynamics to \acr's familiar asymmetric trimming. As shown in Figure \ref{fig:albacore_can_h12_comps_ranks}, it gives a rather sparse solution favoring the 60$^{th}$ percentile, similar to our European results. After continuing its upward trajectory until mid-2022, \acr decreases from 2022m10 onwards.

At the end of the horizon, all measures point toward a slowdown of the disinflationary process. While rebounds for headline CPI are mainly attributable to positive base effects in gasoline prices, the main driver for core is shelter prices. Similarly, \acc is affected by strong upward pressures from rents as it allocates a sizable amount of weight to the corresponding component (see Figure \ref{fig:albacore_can_h12_comps}). Unlike in the US, shelter prices in Canada are more timely in tracking inflation dynamics. Rental properties typically come with amenities like parking, appliances, heating, and air conditioning. As such, rent is influenced by a broad range of factors, including energy costs as well as the prevailing interest rate environment, and offers a comprehensive measure of household needs \citep{StatcanRentMore,macisaac2023rent}. Since \acc assigns negligible weight to food and energy components, upweighting rent helps to buffer against the high volatility associated with utility costs without entirely eliminating their influence.  Lastly,  the transformed series for rent (after the application of various treatments,  see Appendix \ref{sec:data}) is particularly smooth, \textcolor{black}{close to its behavior before the methodological change of 2019.} Given the level of noise in the Canadian data,  rents may capture a reliable low-frequency movement that is obscured in other series.  Given this and many other challenges with the Canadian data,  caution should be exerted in drawing more sweeping economic conclusions.



\clearpage

\subsection{Properties} \label{sec:prop}



\begin{table}[h]
\centering
\tiny
\begin{threeparttable}[h]
\caption{Properties of Underlying Inflation Measures for the US\label{tab:prop_us}} 
\begin{tabular}{lcccccccccccc}
\toprule
\multicolumn{1}{c}{\bfseries Inflation series}&\multicolumn{1}{c}{\bfseries }&\multicolumn{2}{c}{\bfseries Bias}&\multicolumn{1}{c}{\bfseries }&\multicolumn{2}{c}{\bfseries  Volatility}&\multicolumn{1}{c}{\bfseries }&\multicolumn{2}{c}{\bfseries Coefficient of variation}&\multicolumn{1}{c}{\bfseries }&\multicolumn{2}{c}{\bfseries Lead/lag correlation}\tabularnewline
\cline{1-1} \cline{3-4} \cline{6-7} \cline{9-10} \cline{12-13}
\multicolumn{1}{c}{}&\multicolumn{1}{c}{}&\multicolumn{1}{c}{Full-sample}&\multicolumn{1}{c}{Pre-Covid}&\multicolumn{1}{c}{}&\multicolumn{1}{c}{Full-sample}&\multicolumn{1}{c}{Pre-Covid}&\multicolumn{1}{c}{}&\multicolumn{1}{c}{Full-sample}&\multicolumn{1}{c}{Pre-Covid}&\multicolumn{1}{c}{}&\multicolumn{1}{c}{Full-sample}&\multicolumn{1}{c}{Pre-Covid}\tabularnewline
\midrule
\textbf{3Mo3M} &&&&&&&&&&&\tabularnewline
   Core PCE&   &   -0.12&   \cellcolor{ForestGreen!15} -0.09&   &   0.57&   0.36&   &   0.59&   0.39&   &   \cellcolor{ForestGreen!15} 0&   \cellcolor{ForestGreen!15} 0\tabularnewline
   Trimmed Mean (FedDallas)&   &   \cellcolor{ForestGreen!15} 0.02&    0.14&   &   \cellcolor{ForestGreen!15} 0.42&   \cellcolor{ForestGreen!15} 0.29&   &   \cellcolor{ForestGreen!15} 0.41&   \cellcolor{ForestGreen!15} 0.27&   &   -8&  \cellcolor{ForestGreen!15} 0\tabularnewline
   Core excl. housing&   &    2.35&    2.37&   &   5.45&   1.83&   &   2.52&   0.80&   &   \cellcolor{ForestGreen!15} 1&   \cellcolor{ForestGreen!15} 2\tabularnewline
   Median CPI (FedCleveland)&   &    0.59&    0.59&   &   0.61&   0.38&   &   0.46&   0.29&   &   -8&   -9\tabularnewline
   Trimmed Mean (FedCleveland)&   &    0.33&    0.31&   &   0.63&   0.40&   &   0.53&   0.34&   &   \cellcolor{ForestGreen!15} 0&   \cellcolor{ForestGreen!15} 0\tabularnewline
   Sticky Core (FedAtlanta)&   &    0.51&    0.56&   &   0.57&   0.37&   &   \cellcolor{ForestGreen!15} 0.45&   \cellcolor{ForestGreen!15} 0.28&   &   -9&   -11\tabularnewline
   Albacore$_\text{ranks}$&   &   \cellcolor{ForestGreen!15} -0.03&   \cellcolor{ForestGreen!15} 0.03&   &   \cellcolor{ForestGreen!15} 0.40&   \cellcolor{ForestGreen!15} 0.22&   &   \cellcolor{ForestGreen!15} 0.40&   \cellcolor{ForestGreen!15} 0.21&   & \cellcolor{ForestGreen!15}  0& \cellcolor{ForestGreen!15}  0\tabularnewline
   Albacore$_\text{comps}$&   &   \cellcolor{ForestGreen!15} -0.08&   \cellcolor{ForestGreen!15} 0.00&   &   \cellcolor{ForestGreen!15} 0.48&   \cellcolor{ForestGreen!15} 0.34&   &   0.49&   0.34&   &   \cellcolor{ForestGreen!15}0&   \cellcolor{ForestGreen!15}0\tabularnewline
\midrule
\textbf{YoY} &&&&&&&&&&&\tabularnewline
   Core PCE&   &   -0.13&   -0.11&   &   0.69&   \cellcolor{ForestGreen!15} 0.39&   &   0.49&   \cellcolor{ForestGreen!15} 0.22&   &   \cellcolor{ForestGreen!15} 0&   \cellcolor{ForestGreen!15}0\tabularnewline
   Trimmed Mean (FedDallas)&   &   \cellcolor{ForestGreen!15} 0.01&    0.13&   &   \cellcolor{ForestGreen!15} 0.56&  \cellcolor{ForestGreen!15} 0.46&   &   \cellcolor{ForestGreen!15} 0.37&  \cellcolor{ForestGreen!15} 0.23&   &   -7&   -9\tabularnewline
   Core excl. housing&   &    2.42&    2.43&   &   2.93&   2.03&   &   0.91&   0.46&   &   \cellcolor{ForestGreen!15} 2&   \cellcolor{ForestGreen!15} 3\tabularnewline
   Median CPI (FedCleveland)&   &    0.55&    0.56&   &   0.82&   0.61&   &   0.43&   0.25&   &   -8&   -11\tabularnewline
   Trimmed Mean (FedCleveland)&   &    0.30&    0.29&   &   0.82&   0.55&   &   0.47&   0.25&   &   -4&   -6\tabularnewline
   Sticky Core (FedAtlanta)&   &    0.48&    0.53&   &   0.74&   0.59&   &   \cellcolor{ForestGreen!15} 0.40&   0.24&   &   -8&   -12\tabularnewline
   Albacore$_\text{ranks}$&   &   \cellcolor{ForestGreen!15} -0.04&   \cellcolor{ForestGreen!15} 0.01&   &   \cellcolor{ForestGreen!15} 0.54&   \cellcolor{ForestGreen!15} 0.28&   &   \cellcolor{ForestGreen!15} 0.36&   \cellcolor{ForestGreen!15} 0.15&   &  \cellcolor{ForestGreen!15} 0&   -1\tabularnewline
   Albacore$_\text{comps}$&   &   \cellcolor{ForestGreen!15}-0.09&   \cellcolor{ForestGreen!15} 0.00&   &   \cellcolor{ForestGreen!15} 0.60&   0.50&   &   0.41&   0.26&   &  \cellcolor{ForestGreen!15} 0&   \cellcolor{ForestGreen!15} 2\tabularnewline
\bottomrule
\end{tabular}
\begin{tablenotes}[para,flushleft]
      \tiny 
      \textit{Notes}: Cells shaded in {\color{ForestGreen} green} indicate the three best performing models for each criterion. The evaluation is based on the sample ranging from 2000$m1$ to 2023$m12$ for the full sample and from 2000$m1$ to 2019$m12$ for the pre-Covid case. Bias is determined as the difference between the long-run average of the respective series and PCE headline. Volatility refers to the standard deviation of each series relative to the standard deviation of headline. Coefficient of variation is the ratio between standard deviation and long-run average of each measure. Positive (negative) numbers for the lead/lag correlation refer to leads (lags) with the highest cross-correlation between each measure and PCE headline.
    \end{tablenotes}
  \end{threeparttable}
\end{table}

\vspace*{2cm}
\begin{table}[h]
\centering
\tiny
\begin{threeparttable}[h]
\caption{Properties of Underlying Inflation Measures for the Euro Area\label{tab:prop_ea}} 
\begin{tabular}{lcccccccccccc}
\toprule
\multicolumn{1}{c}{\bfseries Inflation series}&\multicolumn{1}{c}{\bfseries }&\multicolumn{2}{c}{\bfseries Bias}&\multicolumn{1}{c}{\bfseries }&\multicolumn{2}{c}{\bfseries  Volatility}&\multicolumn{1}{c}{\bfseries }&\multicolumn{2}{c}{\bfseries Coefficient of variation}&\multicolumn{1}{c}{\bfseries }&\multicolumn{2}{c}{\bfseries Lead/lag correlation}\tabularnewline
\cline{1-1} \cline{3-4} \cline{6-7} \cline{9-10} \cline{12-13}
\multicolumn{1}{c}{}&\multicolumn{1}{c}{}&\multicolumn{1}{c}{Full-sample}&\multicolumn{1}{c}{Pre-Covid}&\multicolumn{1}{c}{}&\multicolumn{1}{c}{Full-sample}&\multicolumn{1}{c}{Pre-Covid}&\multicolumn{1}{c}{}&\multicolumn{1}{c}{Full-sample}&\multicolumn{1}{c}{Pre-Covid}&\multicolumn{1}{c}{}&\multicolumn{1}{c}{Full-sample}&\multicolumn{1}{c}{Pre-Covid}\tabularnewline
\midrule
\textbf{PoP} &&&&&&&&&&&\tabularnewline
   HICPX&   &   -0.46&   -0.29&   & \cellcolor{ForestGreen!15}  0.51&  \cellcolor{ForestGreen!15} 0.40&   &   0.70&   0.40&   &   -6&  \cellcolor{ForestGreen!15} 0\tabularnewline
   Trimmed Mean 30\%&   &  \cellcolor{ForestGreen!15} -0.21&  \cellcolor{ForestGreen!15} -0.07&   &   0.65&   0.55&   &   0.77&   0.48&   &   \cellcolor{ForestGreen!15} 0&  \cellcolor{ForestGreen!15} 0\tabularnewline
   HICP excl. energy&   &   \cellcolor{ForestGreen!15} -0.19&   -0.11&   &   0.67&   0.49&   &   0.78&   0.44&   &   -6&   \cellcolor{ForestGreen!15} 0\tabularnewline
   HICP excl. energy \& unpr food&   &   -0.25&   -0.15&   &   0.64&   0.44&   &   0.77&   0.40&   &   -6&  \cellcolor{ForestGreen!15} 0\tabularnewline
   Supercore&   &   -0.24&   -0.10&   &   0.55&   \cellcolor{ForestGreen!15} 0.36&   &   0.68&   0.33&   &   -6&   -2\tabularnewline
   PCCI excl. energy&   &    5.29&    4.89&   &   1.44&   1.00&   &   \cellcolor{ForestGreen!15} 0.43&   \cellcolor{ForestGreen!15} 0.21&   &   -5&  \cellcolor{ForestGreen!15} -1\tabularnewline
   PCCI&   &    5.92&    5.43&   &   1.68&   1.24&   &   \cellcolor{ForestGreen!15} 0.47&   \cellcolor{ForestGreen!15} 0.24&   &  \cellcolor{ForestGreen!15} -1&  \cellcolor{ForestGreen!15} 0\tabularnewline
   Albacore$_\text{ranks}$&   &   \cellcolor{ForestGreen!15} -0.17&   \cellcolor{ForestGreen!15} -0.01&   &   \cellcolor{ForestGreen!15} 0.46&   \cellcolor{ForestGreen!15} 0.26&   &   \cellcolor{ForestGreen!15} 0.56&   \cellcolor{ForestGreen!15} 0.23&   &  \cellcolor{ForestGreen!15} -2&   \cellcolor{ForestGreen!15}-1\tabularnewline
   Albacore$_\text{comps}$&   &   -0.26&   \cellcolor{ForestGreen!15} -0.06&   &   \cellcolor{ForestGreen!15} 0.49&   0.42&   &   0.62&   0.38&   &   \cellcolor{ForestGreen!15} 0&  \cellcolor{ForestGreen!15} 0\tabularnewline
\midrule
\textbf{YoY} &&&&&&&&&&&\tabularnewline
   HICPX&   &   -0.48&   -0.29&   &   0.55&   0.46&   &   0.63&   0.31&   &   -5&   -6\tabularnewline
   Trimmed Mean 30\%&   &   -0.23&   -0.07&   &   0.70&   0.66&   &   0.70&   0.39&   &  \cellcolor{ForestGreen!15} -3&   -3\tabularnewline
   HICP excl. energy&   &  \cellcolor{ForestGreen!15} -0.22&   -0.13&   &   0.73&   0.59&   &   0.73&   0.36&   &   -4&   -2\tabularnewline
   HICP excl. energy \& unpr food&   &   -0.26&   -0.15&   &   0.70&   0.56&   &   0.71&   0.34&   &   -4&   -2\tabularnewline
   Supercore&   &   -0.27&   -0.11&   &   0.61&   0.46&   &   0.64&   0.29&   &   -4&   -5\tabularnewline
   PCCI excl. energy&   &   -0.25&  \cellcolor{ForestGreen!15} -0.04&   &   \cellcolor{ForestGreen!15} 0.44&   \cellcolor{ForestGreen!15} 0.36&   &   \cellcolor{ForestGreen!15} 0.44&   \cellcolor{ForestGreen!15} 0.21&   &   \cellcolor{ForestGreen!15} 0&   \cellcolor{ForestGreen!15} 0\tabularnewline
   PCCI&   &   \cellcolor{ForestGreen!15} -0.09&    0.10&   &   \cellcolor{ForestGreen!15} 0.51&   \cellcolor{ForestGreen!15} 0.45&   &   \cellcolor{ForestGreen!15} 0.48&   \cellcolor{ForestGreen!15} 0.24&   &   \cellcolor{ForestGreen!15} 2&   \cellcolor{ForestGreen!15} 2\tabularnewline
   Albacore$_\text{ranks}$&   &  \cellcolor{ForestGreen!15} -0.18&   \cellcolor{ForestGreen!15} -0.01&   &   \cellcolor{ForestGreen!15} 0.52&   \cellcolor{ForestGreen!15} 0.34&   &   \cellcolor{ForestGreen!15} 0.53&   \cellcolor{ForestGreen!15} 0.21&   &  \cellcolor{ForestGreen!15} -3&   -5\tabularnewline
   Albacore$_\text{comps}$&   &   -0.28&  \cellcolor{ForestGreen!15} -0.06&   &   0.54&   0.53&   &   0.58&   0.33&   &  \cellcolor{ForestGreen!15} -3&   \cellcolor{ForestGreen!15} 1\tabularnewline
   \bottomrule
\end{tabular}
\begin{tablenotes}[para,flushleft]
      \tiny 
      \textit{Notes}: For more details we refer to Table \ref{tab:prop_us}. Values are relative to HICP.
    \end{tablenotes}
  \end{threeparttable}
\end{table}

\clearpage

\subsection{Data Wrangling and Labeling of Subcomponents}\label{sec:data} 

We employ the Census X-13ARIMA-SEATS Seasonal Adjustement Program provided in the R package \texttt{seasonal} \citep{sax2018seasonal} to extract seasonal variation present in the EA and Canadian components data. For the Canadian price data, we include an outlier detection step after removing seasonal patterns to account for sudden, one-time price changes that might obscure the underlying trend. These outliers may arise from various factors, such as changes in taxes (e.g., the tobacco tax in Canada), methodological shifts, and variations in sampling methods; particularly Statistics Canada's efforts to update its sampling process in recent years. The latter leads to high volatility in those series that switched from being updated yearly (or infrequently) to monthly, and structural breaks in others (e.g., observed for rent in 2019\footnote{Details on the new approach to estimate the CPI's rent component can be found \href{https://www150.statcan.gc.ca/n1/pub/62f0014m/62f0014m2019002-eng.htm}{here}.}). For cases of heavy irregularities, we substitute the erratic high-level component with its lower-level counterpart. This mainly concerns series with a significant number of zero monthly growth rates. For example, we find more than 17 components at level 4 exhibiting more than 5 cumulative years of zero monthly growth rates between 2000 and 2019. Services related to household furnishings and equipment show 14.75 years of cumulative zeros over a 20 year span. As a result, we remove 11 components at level 4, and 26 components at level 5.

\vspace{0.5cm}
\begin{table}[h]
\centering
\footnotesize
\begin{threeparttable}[h]
\caption{Labeling of Subcomponents in the US (Level 2) \label{tab:USnaming}} 
\begin{tabular}{lllc}
\toprule
\multicolumn{1}{l}{\bfseries Abbreviation}&\multicolumn{1}{l}{\bfseries Code}&\multicolumn{1}{l}{\bfseries Label}&\multicolumn{1}{c}{\bfseries Aggregate} \tabularnewline
\midrule
   Vehicles & DMOTRG & Motor vehicles and parts & Goods \tabularnewline
   House.G & DFDHRG & Furnishings and durable household equipment & Goods \tabularnewline
   Rec.G & DREQRG & Recreational goods and vehicles & Goods \tabularnewline
   O.dur.G & DODGRG & Other durable goods & Goods \tabularnewline
   Food.G  & DFXARG & Food and beverages purchased for off-premises consumption & Food \tabularnewline
   Cothing & DCLORG & Clothing and footwear & Goods \tabularnewline
   Energy & DGOERG & Gasoline and other energy goods & Energy \tabularnewline
   O.ndur.G & DONGRG & Other nondurable goods & Goods \tabularnewline
   Shelter & DHUTRG & Housing and utilities & Shelter \tabularnewline
   Health & DHLCRG & Health care & Other \tabularnewline
   Transpo & DTRSRG & Transportation services & Other \tabularnewline
   Rec.S & DRCARG & Recreation services & Services \tabularnewline
   Food.S & DFSARG & Food services and accommodations & Services \tabularnewline
   Fin.S & DIFSRG & Financial services and insurance & Services \tabularnewline
   Other.S & DOTSRG & Other services & Services \tabularnewline
   \bottomrule
\end{tabular}
\begin{tablenotes}[para,flushleft]
      \scriptsize 
      \textit{Notes}: Labeling and grouping is based on the BEA's National Income and Products Account (NIPA) table 2.4.4 (Price Indexes for Personal Consumption Expenditures by Type of Product). Weights to aggregate lower to higher levels of aggregation are taken from the NIPA table 2.4.6 (Real Personal Consumption Expenditures by Type of Product).
    \end{tablenotes}
  \end{threeparttable}
\end{table}


\begin{table} 
\centering
\footnotesize
\begin{threeparttable}[h]
\caption{Labeling of Subcomponents in the EA (Level 2) \label{tab:EAnaming}} 
\begin{tabular}{lllc}
\toprule
\multicolumn{1}{l}{\bfseries Abbreviation}&\multicolumn{1}{l}{\bfseries Code}&\multicolumn{1}{l}{\bfseries Label}&\multicolumn{1}{c}{\bfseries Aggregate} \tabularnewline
\midrule
   Food & CP01 & Food and non-alcoholic beverages & Food \tabularnewline
   Alcohol & CP02 & Alcoholic beverages, tobacco and narcotics & Food  \tabularnewline
   Clothing & CP03 & Clothing and footwear & NEIG / Services \tabularnewline
   Housing &  CP04 (ex. CP045) & Housing and water & NEIG / Services \tabularnewline
   Furnishings & CP05 & Furnishings, household equipment  & NEIG / Services \tabularnewline
   &&and routine household maintenance& \tabularnewline
   Health & CP06 & Health & NEIG / Services \tabularnewline
   Transpo & CP07 (ex. CP0722)  & Transport  & NEIG / Services \tabularnewline
   Comm. & CP08 & Communication & NEIG / Services \tabularnewline
   Recrea. & CP09 & Recreation and culture  & NEIG / Services\tabularnewline
   Education & CP10 & Education  & Services \tabularnewline
   Restaurants & CP11 & Restaurants and hotels & Services  \tabularnewline
   Other & CP12 & Miscellaneous goods and services & NEIG / Services  \tabularnewline
   Energy & CP045, CP0722 & Energy & Energy  \tabularnewline
   \bottomrule
\end{tabular}
\begin{tablenotes}[para,flushleft]
      \scriptsize 
      \textit{Notes}: Labeling is based on the COICOP classifciation taken from Eurostat. Level 2 refers to the main divisions, i.e., two-digits level. The aggregates are classified as is suggested by Eurostat's ``Special Aggregates'', i.e., Energy, NEIG - Non-Energy Industrial Goods, Services, Food (including alcohol and tobacco). The exact allocation of disaggregated components and further details are available in RAMON (\href{http://ec.europa.eu/eurostat/ramon}{\texttt{http://ec.europa.eu/eurostat/ramon}}). Weights to aggregate lower to higher levels of aggregation are based on expenditure shares and taken from Eurostat (\texttt{prc\_hicp\_inw}).
    \end{tablenotes}
  \end{threeparttable}
\end{table}

\begin{table} 
  \centering
  \footnotesize
  \begin{threeparttable}[h]
    \caption{Labeling of Subcomponents in Canada (Level 3) \label{tab:CANnaming}} 
    \begin{tabular}{lll}
      \toprule
      \multicolumn{1}{l}{\bfseries Abbreviation}&\multicolumn{1}{l}{\bfseries Code}&\multicolumn{1}{l}{\bfseries Label} \tabularnewline
      \midrule
      Food & v41690975 &Food purchased from stores  \tabularnewline
      Restaurants & v41691046 & Food purchased from restaurants  \tabularnewline
      Rent & v41691051 & Rented accommodation \tabularnewline
      Property & v41691055 & Owned accommodation \tabularnewline
      Utilities & v41691062 & Water, fuel and electricity \tabularnewline
      Housing & v41691068 & Household operations \tabularnewline
      Furnishing & v41691087  & Household furnishings and equipment \tabularnewline
      Clothing & v41691109 & Clothing \tabularnewline
      Footwear & v41691113 & Footwear\tabularnewline
      Adornments & v41691118 & Clothing accessories, watches and jewellery \tabularnewline
      Textiles & v41691123 & Clothing material, notions and services  \tabularnewline
      Private trans & v41691129 & Private transportation \tabularnewline
      Public trans & v41691146 & Public transportation \tabularnewline
      Health & v41691154 & Health care \tabularnewline
      Grooming & v41691163 & Personal care \tabularnewline
      Recreation & v41691171 & Recreation \tabularnewline
      Education & v41691197 & Education and reading \tabularnewline
      Alcohol & v41691207 & Alcoholic beverages \tabularnewline
      Tobacco & v41691216 & Tobacco products and smokers' supplies \tabularnewline
      \bottomrule
    \end{tabular}
    \begin{tablenotes}[para,flushleft]
      \scriptsize 
      \textit{Notes}: Labeling is based on the Consumer Price Index survey classifciation taken from Statistic Canada. Level 3 refers to the main divisions, i.e., major group. We excluded v1043024072, Recreational cannabis (201812=100), since the component was introduced in December 2018 (i.e., legalisation of cannabis). The exact allocation of disaggregated components and further details are available at Statistic Canada's website (\href{https://www150.statcan.gc.ca/t1/tbl1/en/tv.action?pid=1810000701}{\texttt{Basket weights of the Consumer Price Index}}). 
    \end{tablenotes}
  \end{threeparttable}
\end{table}

\end{document}